\documentstyle[sprocl,psfig,epsf]{article}

\def\be{\begin{equation}}
\def\ee{\end{equation}}
\def\bea{\begin{eqnarray}}
\def\eea{\end{eqnarray}}
\newcommand{\beq}{\begin{equation}}
\newcommand{\eeq}{\end{equation}}
\newcommand{\beqa}{\begin{eqnarray}}
\newcommand{\eeqa}{\end{eqnarray}}
\newcommand{\dfrac}{\displaystyle \frac}

\newcommand{\ve}{\varepsilon}
\newcommand{\krig}[1]{\stackrel{\circ}{#1}}
\newcommand{\barr}[1]{\not\mathrel #1}

\begin{document}

\title{CHIRAL QCD: BARYON DYNAMICS~\footnote{Essay for the {\em Festschrift
in honor of Boris Ioffe}, to appear in the ``Encyclopedia of Analytic QCD,''
edited by M. Shifman, to be published by World Scientific.}
}

\author{ ULF-G. MEI{\ss}NER }

\address{Forschungszentrum J\"ulich, Institut f\"ur Kernphysik (Theorie)\\
D-52425 J\"ulich, Germany\\
Electronic address: Ulf-G.Meissner@fz-juelich.de}


\maketitle\abstracts{
I review the consequences of the chiral symmetry of QCD for the
structure and dynamics of the low--lying baryons, with particular
emphasis on the nucleon.
}
  
\vspace{-5.5cm}

\hfill {\small {\tt FZJ-IKP(TH)-2000-17}}

\vspace{4.5cm}

\section{A short guide through this long essay}

There are many approaches which relate baryon structure and properties to 
QCD in the non--perturbative regime, i.e. at energies, where the running
coupling constant $\alpha_S (Q^2)$
is sufficiently large to render conventional perturbation theory
doubtful. In this regime, quarks and gluons are confined within hadrons.
In addition, the low--energy dynamics is severely constrained by the 
spontaneously and explicitly broken chiral symmetry of QCD. To account for these facts,
many  models have been invented which emphasize a certain aspect of the
underlying theory. To pick one example, let me just point out that 
the bag model describes baryons made of bagged 
(confined) quarks in a simple and appealing picture, with the confinement due to the
difference of the pressure in the vacuum and in the baryon state.  
However, in such a framework one has very little control over the
approximations involved. Another type of approaches can be considered  
more ambitious in that they only try to implement basic principles 
underlying QCD at the expense of some parameters which have to be determined from
data and  which parameterize our ignorance of
going directly from QCD to the baryon states. One of these approaches are
the sum rules, championned by Boris Ioffe and his collaborators for baryons.
Here, the non--perturbative vacuum structure is given in terms of a string
of quark, gluon and mixed condensates, which are matched to the
operator product expansion of the n--point function under consideration.
A detailed discussion of this framework can be found in the contribution
by Colangelo and Khodjamirian to this {\em Festschrift}. 
In this essay, I will be concerned with
a very different approach. In a certain sense it is more general than
QCD since it solely explores the consequences of the spontaneously and 
explicitly broken chiral symmetry for the structure and 
dynamics of the low--lying baryons. The underlying framework is an effective
field theory based on asymptotically observable states, here the Goldstone
bosons and the matter fields. The basic ideas of the Goldstone boson dynamics
are discussed in detail by Heiri Leutwyler in this book, so I will focus
on the picture of baryon structure at low energies as it follows from
the chiral QCD dynamics. This picture completely masks the quark and
gluon degrees of freedom clearly seen at high energies, which is not that
surprising since in many physical systems collective excitations govern the
low energy dynamics. The Goldstone bosons of QCD can be thought off as
such collective excitations and the baryons essentially as sources surrounded
by a cloud of these. As I will show, this meson cloud picture has far--reaching
consequences for the baryon structure and is well established by now through
a series of precision experiments. In any case, this non--perturbative regime
of QCD is most challenging to theorists and it is therefore not surprising 
that the progress is slow but steady. Still, the field has matured to a
point that truely {\em quantitative} tests of chiral symmetry breaking can
be performed, as I will detail in the following sections. 
A basic discussion of hadron effective field 
theory\footnote{The terminus ``hadron effective field theory'' means
nothing else but baryon chiral perturbation theory.} is given in 
in section~\ref{sec:EFT}. Many applications for the
two-- and three--flavor cases are discussed in sections~\ref{sec:SU2} and
\ref{sec:SU3}, respectively. No attempt to achieve completeness is made
and only material available before July 2000 was included.
Before embarking on any of these issues in
detail, I will give a very pedestrian type of introduction in
section~\ref{sec:foot} trying to
outline the basic ideas and the emerging picture in a fairly descriptive
way. This should allow the beginner to catch the essential concepts without
being overwhelmed by heavy machinery. The reader more interested in
results or derivations might want to skip this section.
More precisely, a short review of the chiral symmetry of QCD is given in
section~\ref{sec:QCD}, followed by a discussion on how one can derive
dynamics from symmetry in section~\ref{sec:dynamics} and some remarks
on the physical picture emerging for the structure of the baryons in
section~\ref{sec:baryon}.

\section{From symmetry to dynamics}
\label{sec:foot}

\subsection{Some introductory remarks}
The spectrum of the low--lying mesons reveals some interesting features.
Starting at the mass of the rho meson, $M_\rho \simeq 770\,$MeV, a whole
zoo of scalar, vector, pseudoscalar and axial states is observed. These
states come in multiplets (octets or nonets) organized in terms of a 
flavor SU(3). The average mass of these lowest multiplets is $M_M \simeq 1\,$GeV.
Below, there are nine {\em pseudoscalar} states which are suspiciously light, 
in particular the three pions ($\pi^+, \pi^0, \pi^-$), $M_\pi/M_M \simeq 1/7$,
the kaons ($K^+,K^-,K^0,\overline{K^0}$) and the eta with $M_K \simeq M_\eta \simeq
M_M/2$. As explained below, these nine particles can be considered  
as the Pseudo--Goldstone bosons related to the {\em spontaneous breakdown of the chiral
symmetry of QCD}. Due to Goldstone's theorem,
their interactions vanish as the momentum transfer goes to zero. In fact,
this statement does not only hold for interactions between Goldstone
bosons, like e.g. for low energy elastic pion--pion scattering, but also for
Goldstone bosons interacting with ``matter'' fields. Here, by matter I
denote any particle which is not a Goldstone boson, in particular also
the nucleon. As a consequence, it follows that the pion--nucleon
interaction at low energies has to be weak and thus paves the way
for a systematic expansion of such low energy processes in terms of
small external momenta (or derivatives). Clearly, the mass scale
of these matter fields is not related to chiral symmetry and thus
sets a natural boundary for such type of approach. This is exactly
the situation which allows one to make use of the concept of effective
field theory, namely that one has a {\em scale separation}. In the
simplified world of exactly massless up, down and strange quarks, 
the Goldstone bosons would be massless and the strong interactions
could be described by a theory with a mass gap, the height of this gap given by the
lowest resonance mass. More generally, one
can identify the scale of chiral symmetry breaking to be of the
order of 1~GeV. The small expansion parameters to be dealt with are
momenta divided by this scale and quark (meson) masses divided by it. 
This leads  in a very natural
way  to chiral perturbation theory and a plethora of predictions
for reactions involving pions, kaons, etas and also external fields
like e.g. photons. These considerations can be extended in a fairly
straightforward manner to the ground state octet of baryons, since
these can not further decay. Stated differently, one can extend the
effective field theory to include these massive degrees of freedom
and still have a systematic power counting. This can most easily
be understood in terms of  relativistic quantum mechanics. By
construction, external momenta impinging on such a baryon are small
compared to its mass because $m_B \simeq 1\,$GeV, which is close to the scale
of chiral symmetry breaking. Therefore, one is dealing with a Dirac
equation for a very heavy fermion, which can be treated in a
systematic fashion by means of a Foldy--Wouthuysen transformation.
This allows to shuffle the large fermion mass into a string of
local operators with increasing powers in the inverse of the mass. By the
same argument, it becomes clear why the treatment of excited states is
in general out of the realm of such a systematic expansion. When an
excited state decays, the energy release is most often too large (one
exception will be discussed below).
I will now try to explain these concepts in simple terms, without any
mathematical rigor.

\subsection{Chiral symmetry of QCD}
\label{sec:QCD}

It is instructive to describe chiral symmetry using only elementary 
mathematics.\cite{john} For that, consider an idealized world in which the up, down and
strange quarks are massless. This is the so--called {\em chiral limit} of
QCD. For each quark, there is a right--handed (RH) helicity state with the
spin parallel to the momentum  and a left--handed (LH) helicity state with the
spin antiparallel to the momentum. All interactions in the QCD Lagrangian
are the same for left and right helicity and, furthermore, do  not flip
helicity chiefly because gluons couple to color and not to flavor.
 Therefore, a left--handed massless quark will always remain
left--handed and similarly, a right--handed massless state stays
right--handed. Thus, one has two separate worlds, a LH world and a RH world.
These two worlds do not communicate. Since each flavor is massless and
has the same QCD couplings, there exists a separate flavor SU(3) in
each world, i.e. we can perform independent SU(3) rotations on the
right-- and the left--handed quarks. This global symmetry is called
SU(3)$_{\rm L}\times$SU(3)$_{\rm R}$, the {\it chiral symmetry}. 
Let us now get closer to the physical situation and add a common
mass to the quarks. Clearly, one can no longer maintain the separate
invariances under left-- and right--handed rotations. For a massive
particle with a given handedness, one can always perform a boost to a
frame moving in the other direction, such that the particle changes
say from being left-- to right--handed. Therefore, by simple kinematics,
the two worlds are no longer separated and one has no longer two 
independent SU(3) symmetries. However, there remains an invariance
under common L+R rotations, with the corresponding SU(3)$_{\rm V}$
symmetry (since the sum of left-- and right--handed generators leads
to a vector). If for some reason the mass is small, the  SU(3)$_{\rm
L}\times$SU(3)$_{\rm R}$ symmetry could be an {\it approximate}
symmetry allowing to treat the mass term as a perturbation. In the
real world, all light quark masses are different, so the SU(3)$_{\rm
  V}$ is really broken into a direct product of three U(1)
symmetries.  However, to the  extent that the mass differences are
small, one can consider the SU(3)$_{\rm V}$ as an approximate symmetry
and treat these mass differences as perturbations. To see this in more
detail consider the triplet of quark fields
\beq
q = \left( u,d,s \right)^T~,
\eeq  
where $'T'$ means transposed, and the left and right projection operators
\beq
P_L = {1\over 2} (1+ \gamma_5) ~, \quad
P_R = {1\over 2} (1- \gamma_5) ~, 
\eeq
with
\beq
P_L^2 = P_L~, \quad  P_R^2 = P_R~, \quad  P_L \, P_R = 0 ~, \quad  
P_L + P_R = I~, 
\eeq
with $I$ the unit matrix such that
\beq
q_L = P_L \,q ~, \quad  q_R = P_R \,q ~, \quad q= q_L + q_R~.
\eeq
For massless particles, the $P_{L,R}$ project out the helicity
of the particle. For massive particles, the $P_{L,R}$ are still
projection operators but do not give exactly the helicity. Therefore,
one introduces the word {\em chirality} (or handedness). In terms
of the chiral quarks, the Dirac part of the QCD Lagrangian takes the form
\beqa
{\cal L}_{\rm QCD, fermions} &=& \bar{q} (i D\!\!\!\!/ - m)q \nonumber \\
         &=& \bar{q}_L \, i D\!\!\!\!/ \, q_L + \bar{q}_R \, i D\!\!\!\!/
         \, q_R +   \bar{q}_L \, m \, q_R +  \bar{q}_R \, m \, q_L~,
\eeqa
which shows the advocated feature that for $m = 0$, the left-- and
right--handed worlds decouple  so that the QCD Lagrangian is invariant
under $q_L \to L q_L $,  $q_R \to R q_R $ for $L,R \in$~SU(3)$_{L,R}$.
The mass term mixes the different
chiralities as explained before. Since such flavor transformations do
not affect the gluon fields, there was no need of spelling out
explicitly how these spin--1 fields are hidden in the gauge covariant derivative
$D_\mu$ and have their own  Yang--Mills Lagrangian. Since SU(3) has 8
generators, Noether's theorem tells us that the massless theory should have 16
conserved currents and charges, eight constructed from bilinears of the
left--handed fermions and eight from the right--handed ones. Since we
can also combine L+R to give a vector and L-R to give an
axial--vector, one can equally well talk of eight conserved vector and
eight conserved axial--vector currents. The vector currents and
charges are the usual SU(3)$_{\rm V}$ = SU(3)$_{\rm flavor}$ ones which are the basis
of the hadron multiplet structure (the ``eightfold way''). With the additional eight axial
generators, one would expect larger multiplets. This can be seen as
follows. Performing an axial rotation on a single particle state leads
to a different state with exactly the same mass but of opposite parity 
because $[{\rm H_{QCD}},Q_5^a] = 0$, where $Q_5^a$ is any one of the eight axial generators.
Such a doubling of hadron states is {\em not} observed in nature, e.g. there
exists no state with the same mass as the proton but opposite
parity. Consequently, this larger symmetry must be hidden. This
phenomenon is also called {\em spontaneous symmetry breaking}.
It is important to realize that although the full symmetry is not
present in the spectrum, it still allows one to make powerful {\em predictions}.

\subsection{Dynamical consequences of chiral symmetry}
\label{sec:dynamics}

\begin{figure}[htb]
\hskip 1.5in
\epsfysize=1.5in
\epsffile{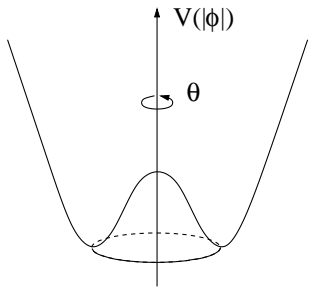}
\vspace{-0.1cm}
 \caption{
The mexican hat potential ${\cal V}(\phi)$ for the scalar field $\phi$.
          \label{fig:hat} }
\end{figure}

The most important and also most difficult step in the understanding
of dynamical symmetry breaking is why one can make predictions at all.
Such predictions will be at the heart of this article. So let us be
more general and consider a Lagrangian which is invariant under some 
symmetry but allow for a continuous family of ground states. These are
related to each other by the symmetry but are assumed not to be
invariant under it individually. This can be illustrated in some classic examples.
First, consider a field theory of a complex scalar field $\phi$,
\beq
{\cal L} = |\partial_\mu \phi|^2 - {\cal V} (|\phi |)~,
\eeq 
and ${\cal V}(|\phi |)$ is assumed to have the mexican hat (wine
bottle) shape as depicted in Fig~\ref{fig:hat}. The rotational symmetry becomes
manifest if one decomposes the field $\phi$ in polar coordinates,
\beq
\phi = \frac{1}{\sqrt{2}} \rho {\rm e}^{i\theta} ~, \quad 
\partial_\mu\phi = \frac{1}{\sqrt{2}}  {\rm e}^{i\theta} (\partial_\mu
\rho + i \rho \partial_\mu \theta )~, 
\eeq
clearly exhibiting the symmetry of ${\cal L}$ under $\theta \to \theta
+ \alpha$, i.e. rotations. In the rim of the mexican hat, we have our family of ground
states and these are obviously connected by rotations around the
symmetry axis of the potential. There are
different particle excitations possible, one is along the rim of the
mexican head (the $\theta$ variable) and the other orthogonal to it,
rolling up and down the hills (the $\rho$ variable). 
\noindent
Spontaneous
symmetry breaking happens if we now choose exactly one possible ground
state, for example $\rho (x) = \rho_0$, $\theta = 0$. This clearly
breaks the rotational symmetry, or better, hides it.  As an immediate
consequence of this, there will be massless excitations (particles) in
the theory, according to a theorem due to Goldstone.\cite{Goldstone} This
is not hard to see conceptually in our example. For that, we expand
around the selected minimum, i.e. set $\rho = \rho_0 + \chi$. In these
new variables, the Lagrangain takes the form
\beq
{\cal L} = {1\over 2} (\partial_\mu \chi)^2 
        + {1\over 2} \rho_0^2 (\partial_\mu \theta)^2  
- {\cal V} \left(\frac{\rho_0}{\sqrt{2}}\right) -{1\over 2} \chi^2 {\cal
    V}'' \left(\frac{\rho_0}{\sqrt{2}}\right) + \ldots  ~,
\eeq
where the ellipsis denotes terms of higher order in the fields. From
the curvature of the potential we can read off the particle masses,
\beq
m_\chi^2 =  {\cal  V}'' \left(\frac{\rho_0}{\sqrt{2}}\right)~, 
\quad m_\theta^2 = 0~.
\eeq
This means that there are massless excitations in the $\theta$
direction, which are the advocated Goldstone boson modes. In contrast
to these, the $\chi$ modes are our everyday massive particles.
Another classic example is the ferromagnet. The magnet Hamiltonian is
rotationally invariant but in a given domain below the Curie
temperature, all spins are aligned in a certain direction and thus
one has only an invariance with respect to rotations around this
axis. In mathematical terms, the O(3) of the Lagrangian is broken down
to O(2) of the ground state. The Goldstone boson excitations in this
example are spin waves related to rotations of the whole system (the
infinitely extended magnetic domain) which have infinite
wavelength. Therefore, the corresponding energy is zero since
$E \sim c/\lambda \sim cp$. Thus the symmetry transformation of rotating
the whole domain in fact corresponds to the excitation of a massless particle.

\medskip\noindent
In QCD, the situation is similar but can not be explained so easily.
In the chiral limit, we can freely rotate between the light quark
flavors, and we have a continuous family of vacua with different
compositions of left-- and right--handed paired quarks and antiquarks.
The physical vacuum itself corresponds to one particular combination
since it does not have this invariance as can be deduced from the
non--vanishing of the matrix--elements (see e.g.~\cite{VW,LS})
\beq
\langle 0 | \bar{q}q| 0 \rangle = 
\langle 0 | \bar{q}_L q_R + \bar{q}_R q_L| 0 \rangle~.  
\eeq
This operator is clearly not invariant under the
SU(3)$_{\rm L}\times$SU(3)$_{\rm R}$ and thus must vanish if the
vacuum would be invariant. It has been established theoretically 
that this operator is indeed non--zero in case of spontaneous symmetry
breaking. In that case, the pions, kaons and eta would be the
Goldstone bosons with their small but finite mass related to the
small current quark masses. Still, the question remains why such
a hidden symmetry can be used to make predictions? Consider first
the example of a conventionally realized symmetry such as isospin.
To a very good approximation, the proton and the neutron can be
considered  an iso--doublet and are thus related by an SU(2) 
transformation. Consequently, the couplings of protons and neutrons
must be equal up to computable Clebsch--Gordan coefficients, such as
$g_{\pi^0 pp} = g_{\pi^0 nn }$. In the case of a broken symmetry,
matters are different. As argued before, there is no such state as a 
proton with the opposite parity to our everyday proton. Instead, under
an axial transformation the proton transforms into {\em a proton plus
a zero energy pion} (to leading order). Such pions are also called ``soft pions''. 
Because of Goldstone's theorem, this state has the same
energy as the pure proton state. This can be generalized to some
arbitrary state $|\psi\rangle$, i.e. the symmetry relates
$|\psi\rangle$ to $|\psi + \pi(p_\mu = 0)\rangle$
and gives a relation between their couplings. Stated differently,
the symmetry relates processes with different numbers of pions,
such as $K\to 3\pi$ to $K\to 2\pi$ or $K \to \pi\pi e \nu$
to $\pi\pi  \to \pi\pi$.  One can also extend such considerations
to systems with more pions, e.g. scattering processes like $i + \pi^a \to 
f + \pi^b$. All this can be expressed more formally in terms of 
``soft pion theorems''. A famous example is the prediction for the S--wave
scattering lengths for pions scattering off some target with isospin 
$I_t$ (this target not being a pion),\cite{weinscatt,tomozawa}
\beq
a = -L \, \left( 1 + {M_\pi \over m_t }\right)^{-1} \,
[I(I+1) -I_t(I_t+1) -2]~,
\eeq
where $I$ is the total isospin, and $L$ is a universal length introduced
by Weinberg, $L = M_\pi /(8\pi F_\pi^2) \simeq 0.09/ M_\pi$, 
expressed in terms of the pion mass $M_\pi$  and the pion decay constant
$F_\pi$. The
precise meaning of these low energy theorems will be discussed in 
section~\ref{sec:LET}. In modern terminology, one simply uses an effective
Lagrangian to calculate such processes, not only to leading order but also
{\em systematically including corrections}. This will be the theme elaborated on
in the rest of the essay.

\subsection{Chiral dynamics with baryons}
\label{sec:baryon}

So far, I have given some arguments why pions, kaons and the eta play a
special role in the strong interactions at low energies. We already know
that their couplings to matter fields are of derivative nature. A particular
type of matter fields are the low--lying baryon octet states
($p, n, \Sigma^+, \Sigma^0,\Sigma^-, \Lambda^0, \Xi^0, \Xi^-)$. Because these
can not further decay due to baryon number conservation, they also play
a special role which will be explored in the following. To be specific,
consider the two--flavor case, i.e. nucleons chirally coupled to pions. We
had already drawn the analogy to the Dirac equation for a heavy particle
which is subject to external probes like pions or photons. When these external
momenta are small compared to the nucleon mass, it is obvious that we can
think of the following idealized picture: The nucleon consists of some a
priori structureless spin--1/2 field surrounded by a could of pions, which
dress up the ``bare'' nucleon. An extreme case of such a picture is the
static source model of the nucleon, for a very illustrative discussion see
Ref.7. 
It also relates directly to the meson cloud model developed
many decades ago (see e.g.~\cite{Pauli}), in which e.g. 
the neutron can be pictured as a proton
surrounded by a negatively charged pion. This very naive approach indeed
gives the qualitative correct description of the charge distribution in the
neutron. Such pictures are, of course, to be taken with a grain of salt, but
might be helpful in getting some first grasp of the certainly more complicated
reality. What we really will be doing is to formulate an effective Lagrangian
of pions, nucleons and external sources and use it to calculate the properties
of the nucleon dictated by chiral symmetry in a systematic fashion using
standard quantum field theoretical methods.
Before formulating such a procedure in mathematical terms, let me briefly discuss
one very often stated misconception. The nucleon certainly is an extended
object, with a charge radius of about 0.85~fm as measured e.g. in electron scattering.
So how can it possibly make sense to consider a local field theory for such 
particles? First, one should be aware that  the pion itself  is not that small
an object, $r_\pi \simeq 0.65\,$fm. But since the pion field can be related
to the divergence of the axial current, one might circumvent this by simply
formulating the whole approach in terms of currents, that is current algebra,
i.e. exploring once more the special character of this particular
hadron. Still, the finite extension of the pion offers the clue to understand the
nucleon in the low energy domain. 
The pion size can be understood simply by vector meson dominance,
$r_\pi = \sqrt{6}/{M_\rho} \simeq 0.63\,$fm. The small difference to the
physical radius  is due to uncorrelated pions in the cloud surrounding the pion.
In a similar manner, any nucleon size can be thought of as due to the pion
cloud and  heavier excitations, the latter being represented by some
local pion--pion or pion--nucleon contact interactions. That such a scheme
works will be demonstrated below, in fact, it works because it can be derived
solely from the transformation properties of matter under the spontaneously
broken chiral symmetry and is based on a systematic power counting. This is 
also the major difference to all the various models implementing chiral
symmetry. While these can give some (mostly) qualitative insight, one in
general has no way of controlling the approximations one is forced to do.
Chiral perturbation theory is a {\em model--independent} approach and 
a direct consequence of the symmetries of QCD, formulated most
elegantly in a hierarchy of chiral Ward identities. 

\section{Effective field theory with matter fields}
\label{sec:EFT}
\setcounter{equation}{0}

\subsection{General remarks}
\label{sec:general}
Chiral perturbation theory (CHPT) is the effective field theory of the
Standard Model (SM) at low
energies in the hadronic sector. Since as an EFT it contains all terms allowed
by the symmetries of the underlying theory,\cite{wein79} it should be viewed 
as a direct consequence of the SM itself. The two main assumptions
underlying CHPT are that (i) the masses of the light quarks u, d (and possibly s) can be
treated as perturbations (i.e., they are small compared to a typical 
hadronic scale of 1 GeV),  and  that  (ii) in the limit of zero quark
masses, the chiral symmetry is
spontaneously broken  to its vectorial subgroup. The resulting Goldstone
bosons are the pseudoscalar mesons (pions, kaons and eta).
CHPT is a systematic low--energy expansion around the 
chiral limit.\cite{wein79} \cite{GL1} \cite{GL2} \cite{leut} It is a
well--defined quantum field theory although it has to be renormalized
order by order. Beyond leading order, one has to include loop diagrams to
 restore unitarity perturbatively. Furthermore, Green functions calculated in
CHPT at a given order contain certain parameters that are not constrained by 
the symmetries, the so--called low--energy constants (LECs).
At each order in the chiral expansion, those LECs have to be
determined  from phenomenology (or can be estimated with some
model dependent assumptions).  These issues are discussed in more
detail by Leutwyler in this book. I will now turn to the inclusion of
matter fields. Here, ``matter'' are {\em all} fields that have a
nonvanishing mass in the chiral limit, such as the vector mesons, the
ground--state baryon octet and decuplet states, and so on.

\medskip\noindent
Since in the remainder of this section  the basic principles and tools underlying baryon
chiral perturbation theory are presented, I consider it helpful to summarize
already here the contents of the various subsections:
\begin{itemize}
\item[3.2] The general structure of the Lagrangian
  is discussed. The notions of tree and loop graphs as well as
  of low--energy constants and power counting  are introduced. 
\item[3.3] It is shown how one can construct the
  most general effective Lagrangian of pions, nucleons and external
  fields. This is somewhat technical and might be skipped during a first
  reading.
\item[3.4] The scale set by the nucleon mass is of comparable size as
  the scale of chiral symmetry breaking. It can therefore not be
  treated perturbatively. One way of avoiding this scale is the
  heavy baryon projection, in which the nucleon mass is transfered
  from the propagator to a string of interaction vertices with
  increasing powers in the {\em inverse} of the mass.
\item[3.5] The form of the effective Lagrangian beyond leading order
  is now made explicit. The differences between the relativistic and
  the heavy baryon approach are discussed.
\item[3.6] The problem of the nucleon mass scale in loop graphs is
  discussed and it is shown how one can set up symmetry preserving
  regularization schemes. One is based on the heavy fermion approach and another
  one starts from relavistic spin-1/2 fields, employing a clever
  separation of any one loop graph into ``soft'' and ``hard'' parts.
\item[3.7] Some remarks about renormalization in effective field
  theories are made.
\item[3.8] The principle underlying the systematics of effective
  field theory is {\em power counting}. The explicit formula for the
  so--called chiral dimension for any process and any diagram is derived.
\item[3.9] Beyond next--to--leading order, coupling constants not
  fixed by chiral symmetry appear. The physics behind these
  low--energy constants is discussed.
\item[3.10] Isospin breaking can be due to quark mass differences or
  virtual photons. It is shown how to modify the machinery to deal
  with such effects.
\item[3.11] A proper defintion of ``low energy theorems'' in the
  framework of chiral perturbation theory is given. Some examples
  are discussed and possible loopholes of less systematic
  definitions are spelled out.
\item[3.12] It is shown that one can also include resonances in the
  effective field theory in a systematic way provided one accepts certain assumptions.  
\end{itemize}
I will now discuss these issues in more detail.

\subsection{Structure of the effective Lagrangian}
\label{sec:structure}
First, I discuss the general structure of the effective Lagrangian,
built from the Goldstone--boson octet,
\begin{equation}\label{matphi}
\Phi = \sqrt{2} \left( 
\begin{array}{ccc}
\frac{1}{\sqrt{2}} \pi^0 + \frac{1}{\sqrt{6}} \eta & \pi^+ & K^+ \\ 
\pi^- & -\frac{1}{\sqrt{2}} \pi^0 + \frac{1}{\sqrt{6}} \eta & K^0 \\ 
K^- & \bar{K}^0 & -\frac{2}{\sqrt{6}} \eta
\end{array}
\right)~,
\end{equation}
and the octet of the low--lying spin--1/2 baryon states,
\begin{equation}\label{octet}
B = \left( 
\begin{array}{ccc}
\frac{1}{\sqrt{2}}\Sigma^0 + \frac{1}{\sqrt{6}}\Lambda & \Sigma^+ & p \\ 
\Sigma^- & - \frac{1}{\sqrt{2}}\Sigma^0 + \frac{1}{\sqrt{6}}\Lambda & n
\\ 
\Xi^- & \Xi^0 & -\frac{2}{\sqrt{6}}\Lambda
\end{array}
\right)~.
\end{equation}
The Goldstone boson fields are parameterized in some highly non--linear
fashion in a matrix--valued function $U(x)$, whose explicit form is
not needed. In writing down the diagonal matrix elements in
Eq.(\ref{matphi}), I have been sloppy. In fact, there is a small
mixing between the neutral pion and the eta which is proportional
to the quark mass difference $m_d-m_u$. It was pointed out by Ioffe
and Shifman\cite{IoSh} already 20 years ago that the decays 
$\psi ' \to J/\psi + \pi^0 \, (\eta )$ can be used to pin down the
light quark mass ratio $m_d/m_u$. I come back to such isospin
violating terms later. Goldstone's theorem requires decoupling of these bosons
from the matter field as the momentum (energy) transfer vanishes. This
can most easily be achieved by requiring a {\em non-linear realization} of the
chiral symmetry, leading naturally to derivative
couplings. Such a non--linear realization is also in agreement with the
spectrum of strongly interacting particles.
 Consequently, the chiral effective meson--baryon  (MB) Lagrangian
consists of a string of terms with increasing chiral
dimension,\footnote{Note that this chiral dimension has nothing to do
  with the canonical field dimension.}
\beq\label{Lagr}
{\cal L}_{\rm eff} =  {\cal L}_{\rm MB}^{(1)} + {\cal L}_{\rm MB}^{(2)} + 
{\cal L}_{\rm MB}^{(3)} + {\cal L}_{\rm MB}^{(4)} + {\cal L}_{\rm M}^{(2)}
+ {\cal L}_{\rm M}^{(4)} + \ldots~,
\eeq 
which are constructed in harmony with general principles like Lorentz
invariance and also with the underlying continuous and discrete
symmetries. The superscripts refer to the chiral dimension, i.e. a
term of order $p^l$ contains $n$ derivatives and $m$ powers of meson masses
subject to the constraint $l = m+n$, with $l,m,n$ integers
(this is the standard scenario of a large quark--antiquark condensate).
In what follows, we will denote any small momentum or meson mass by $p$,
small with respect to the scale of symmetry breaking, $\Lambda_\chi \simeq
1$~GeV. The meson Lagrangian ${\cal L}_{\rm MB}^{(2,4,\ldots)}$
also has to be considered since at one loop
order, one has to renormalize the external legs and so on. For the details
concerning the Goldstone--boson interactions, I refer to Leutwyler's
contribution. Let me come back to the meson--baryon system.
 That the lowest order terms are of dimension one can be
seen as follows. As just argued, Goldstone's theorem requires
derivative couplings and thus the lowest order interaction terms must
be of dimension one (an explicit construction for the two--flavor case
is given below). But: matter fields have mass and the
canonical mass term $m_B \langle\bar{B} B\rangle$,
is clearly of dimension zero. Here and in what follows, 
$\left\langle \ldots\right\rangle $ stands for the flavor trace.
However, there is also the kinetic term $i\langle \bar{B}
\barr{\partial} B\rangle$ and the
operator $(i\barr \partial -m_B) B$ is obviously of dimension one since
the time derivative  also gives the baryon mass. The
Lagrangian is a tool to calculate matrix--elements and transition
currents in a systematic expansion in external momenta and quark
(meson) masses (the complications due to the baryon mass scale will
be discussed later). To be more specific, let me give the lowest
order Lagrangian for the two-- and the three--flavor case,
\beqa\label{Lpin12}
{\rm SU(2)} &:&  {\cal L}_{\pi N}^{(1)}  = \bar{\Psi}
 \left( i \gamma_\mu D^\mu - \krig{m} + \frac{1}{2}\krig{g}_A \gamma^\mu
   \gamma_5 u_\mu \right) \Psi~, \,\,\, \Psi  = \left( \begin{array}{c}
                 p \\ n
\end{array} \right)~,    \\
{\rm SU(3)} &:&  {\cal L}_{\rm MB}^{(1)}  = \langle \bar{B} 
i \gamma_\mu D^\mu B \rangle - \krig{m}_B \langle \bar{B}{B} \rangle
\nonumber \\
&& \qquad\qquad + \frac{D}{2} \langle \bar{B}
\gamma^\mu \gamma_5 \{ u_\mu, B\} \rangle 
+ \frac{F}{2} \langle \bar{B}\gamma^\mu 
\gamma_5 [ u_\mu, B] \rangle~.\label{Lpin13}
\eeqa
The precise definition of the covariant derivatives and the axial--vector
$u_\mu$ are given in the next section.
Here, the superscript '$\circ$' denotes quantities in the chiral limit, i.e.
\beq
Q = ~\stackrel{\circ}{Q} [ 1 + {\cal O}(m_{\rm quark})]
\eeq
(with the exception of $M$ which is
the leading term in the quark mass expansion of the pion mass and $F$
which is the leading term in the expansion of the pion deacy constant),
$g_A$ is the axial--vector
coupling constant measured in neutron $\beta$--decay, $g_A =
1.267$, and $m$ ($m_B$) denotes the nucleon (average baryon octet) mass.
For the three flavor case, one has of course two axial
couplings, the $F$ and $D$ couplings. I remark that the
baryon mass splittings only arise at next--to--leading (second) order.
Note that
the symbol $F$ is used for two different objects, but no confusion
can arise. From now on, I will be somewhat sloppy and not specifically
denote quantities in the chiral limit.
From the Lagrangian one calculates tree and
Goldstone boson loop graphs according to a systematic power counting,
which is at the heart of any effective field theory. The exact form
of the power counting rules is derived below, here I only state that
in any scheme which ,,properly'' accounts for the baryon mass (which
will also be discussed later), the following structure emerges:
\smallskip
\renewcommand{\arraystretch}{1.2}
\begin{center}
\begin{tabular}{|c|c|c|l|}
\hline
Order & Trees from & Loops with insertions from \\ 
\hline
$p$    &  ${\cal L}_{\rm MB}^{(1)}$  &  none \\
$p^2$  &  ${\cal L}_{\rm MB}^{(1)} + {\cal L}_{\rm MB}^{(2)}$ & none
\\
$p^3$  &  ${\cal L}_{\rm MB}^{(1)} + {\cal L}_{\rm MB}^{(2)} +{\cal L}_{\rm MB}^{(3)} $ 
& ${\cal L}_{\rm MB}^{(1)}$ \\
$p^4$  &  ${\cal L}_{\rm MB}^{(1)} + {\cal L}_{\rm MB}^{(2)} +{\cal
  L}_{\rm MB}^{(3)}  + {\cal L}_{\rm MB}^{(4)}$ 
& ${\cal L}_{\rm MB}^{(1)} + {\cal L}_{\rm MB}^{(2)} $ \\
\hline
\end{tabular}
\end{center}
\smallskip
Note that the local operators comprising the various terms
of ${\cal L}_{\rm MB}^{(i)}$  with dimension two (or higher), 
i.e. $i\ge 2$, are
accompanied by coupling constants not fixed by chiral symmetry,
the so--called {\em low}--{\em energy} {\em constants} (LECs). These must be pinned down
by a fit to data. The lowest order term in Eq.(\ref{Lagr}) is unique
in that it only contains parameters related to chiral symmetry, like
the weak pion decay constant, $F_\pi$, the pion mass, $M_\pi$, or the weak axial
baryon--meson couplings $F$ and $D$ (or $g_A = F+D$ in the two--flavor case). 
The special role of $g_A$ becomes most transparent in the context of the celebrated
Goldberger--Treiman relation,\cite{GTR}
\beq
g_A = g_{\pi NN} {F_\pi \over m_N} + {\cal O}(M_\pi^2)~,
\eeq
which relates the strong pion--nucleon coupling constant $g_{\pi NN}$
to properties of the weak axial current. This remarkable relation
later gave birth to the concept of the partially conserved
axial--vector current and was a cornerstone of current algebra. More
precisely, it is exact in the chiral limit and fulfilled in nature
within a few percent.
The baryon mass is somewhat special and will be discussed in more detail
in subsequent paragraphs.
The loop graphs are, of course, required by unitarity - tree graphs
are always real and thus can not give any absorptive contributions. In
CHPT, unitarity is not fulfilled exactly but perturbatively, in the
sense that there are always higher order corrections not considered (in
some cases, these can be large and one has to go to high
orders or needs to perform some type of resummation). Of course, such loop
graphs are in general divergent, but that is not a problem since to
the same order, one has by construction all possible counterterm structures.
In fact, most of the low--energy constants, denoted $C_i$ here, decompose into
an infinite and a renormalized finite piece, 
\beq\label{LECform}
C_i = C_i^r (\lambda) + C_i^\infty~,
\eeq
with $\lambda$ some regularization scale. This scale dependence is of course
balanced by the corresponding one of the loop graphs since physical observables
${\cal O}$ are renormalization group equation (RGE) invariant,
\beq
{d \over d\lambda} {\cal O} = 0~.
\eeq
This also means that in general it is not meaningful to separate the
loop from the tree graph contribution since such a separation depends on the choice
of the scale $\lambda$. Exceptions from this statement are finite loop
contributions or the  non--analytic chiral limit behaviour of certain
observables. This latter type of phenomenon is related to the fact that
in the chiral limit of vanishing Goldstone boson masses, loop graphs can
acquire IR singularities, some examples will be given in later sections.

\subsection{Construction of the chiral pion--nucleon Lagrangian}
\label{sec:constru}
This paragraph is somewhat technical, but I consider it important
to show explicitely how the most general effective Lagrangian can be
calculated to a given order. The reader not so much interested in these
details is invited to skip this section. It follows essentially
Ref.15, 
in which more details can be found.
To construct the effective Lagrangian, one first has to 
collect the basic  building blocks which are consistent with the
pertinent symmetries. I will restrict myself to the two--flavor case.
The underlying Lagrangian is that of QCD with massless $u
$ and $d$ quarks, coupled to external hermitian $2\times2$ matrix--valued fields
$v_{\mu},$ $a_{\mu},$ $s$ and $p$ (vector, axial--vector, scalar and
pseudoscalar, respectively)~\cite{GL1}
\begin{equation}
{\cal{L}}={\cal{L}}_{\rm{QCD}}^{0}+\bar q\gamma^{\mu}\left(  v_{\mu
}+\gamma_{5}a_{\mu}\right)  q-\bar q(s-i\gamma_{5}p)q\, ,
\qquad q=\left(\begin{array}{c} u \\d \end{array} \right)~.
\end{equation}
With suitably transforming
external fields, this Lagrangian is locally ${\rm SU(2)_{L}}\times
{\rm SU(2)_{R}}\times
{\rm U(1)_{V}}$ invariant. The chiral ${\rm SU(2)_{L}}\times {\rm
  SU(2)_{R}}$ symmetry of QCD is 
spontaneously broken down to its vectorial subgroup, ${\rm SU(2)_V}$. To
simplify life further, we disregard the isoscalar axial currents as
well as the winding number density.\footnote{Note that isoscalar axial
currents only play a role in the discussion of the so--called spin
content of the nucleon. That topic can only be addressed properly in  a
three flavor scheme. For a lucid discussion of the $\theta$ term,
we refer to Ref.16.
} 
Explicit chiral symmetry breaking, i.e. the
nonvanishing $u$ and $d$ current quark mass, is taken into account by setting
$s={\cal M}={\rm diag}(m_{u},m_{d}).$ 
On the level of the effective field theory, the spontaneously broken
chiral symmetry is non-linearly realized in terms of the pion and the nucleon
fields.\cite{CCWZ,weinnon} The pion fields $\Phi,$ being coordinates of the chiral
coset space, are naturally represented by elements $u(\Phi)$ of this coset
space. The most convenient choice of fields for the construction of the
effective Lagrangian is given by
\begin{eqnarray}\label{BB}
u_{\mu} & = & i\{u^{\dagger}(\partial_{\mu}-ir_{\mu})u-u(\partial_{\mu}
-i\ell_{\mu})u^{\dagger}\} ~,\nonumber \\
\chi_{\pm} & = & u^{\dagger}\chi u^{\dagger}\pm u\chi^{\dagger}u~, \quad 
\chi  =  2B(s+ip)~,
\nonumber\\
F^{\pm}_{\mu\nu} & = & u^{\dagger}F^{R}_{\mu\nu}u \pm uF^{L}_{\mu\nu}u^{\dagger}  
~, \nonumber \\
F_{R}^{\mu\nu} & = & \partial^{\mu}r^{\nu}-\partial^{\nu}r^{\mu}-i[r^{\mu
},r^{\nu}]~, \quad r_{\mu} = v_{\mu}+a_{\mu}~, \nonumber \\
F_{L}^{\mu\nu} & = & \partial^{\mu}\ell^{\nu}-\partial^{\nu}\ell^{\mu}
-i[\ell^{\mu},\ell^{\nu}]~, \quad \ell_{\mu}  =  v_{\mu}-a_{\mu}~,
\end{eqnarray}
and $B$ is the parameter of the meson Lagrangian of $O(p^{2})$ related
to the strength of the quark--antiquark condensate.\cite{GL1} We work
here in the standard framework, $B \gg F_\pi$.
For a discussion of the generalized scenario ($B \simeq F_\pi$) in the
presence of matter fields, see e.g. Ref.19.
The reason why the fields in Eq.(\ref{BB}) are so convenient is that they all
transform in the same way under chiral transformations, namely as
\begin{equation}
X\stackrel{g}{\rightarrow}h(g,\Phi)Xh^{-1}(g,\Phi)~,\label{hXh}
\end{equation}
where $g$ is an element of  ${\rm SU(2)_{L}}\times {\rm SU(2)_{R}}$ and 
the so--called compensator $h(g,\Phi)$ defines a non--linear realization
of the chiral symmetry. The compensator $h(g,\Phi)$ depends on $g$ and $\Phi$
in a complicated way, but since the nucleon field $\Psi$ transforms as
\begin{equation}
\Psi\stackrel{g}{\rightarrow}h(g,\Phi)\Psi~,\qquad\quad\bar{\Psi}\stackrel
{g}{\rightarrow}\bar{\Psi}h^{-1}(g,\Phi)\label{hPsi}%
\end{equation}
one can easily construct invariants of the form $\bar{\Psi}O\Psi$ without
explicit knowledge of the compensator. Moreover, for the covariant derivative
defined by
\begin{equation}%
D_{\mu}  =  \partial_{\mu}+\Gamma_{\mu}~, \qquad
\Gamma_{\mu} =  \frac{1}{2}\{u^{\dagger}(\partial_{\mu}-ir_{\mu
})u+u(\partial_{\mu}-i\ell_{\mu})u^{\dagger}\}~,
\label{covD}%
\end{equation}
it follows that also $\left[  D_{\mu},X\right]  ,$ 
$\left[  D_{\mu},\left[  D_{\nu},X\right]
\right]  ,$ etc. transform according to Eq.(\ref{hXh}) and $\ D_{\mu}\Psi,$
$D_{\mu}D_{\nu}\Psi,$ etc. according to Eq.(\ref{hPsi}). This allows for a simple
construction of invariants containing these derivatives.
The definition of the covariant derivative Eq.(\ref{covD}) implies two important
relations. The first one is the so--called curvature relation
\begin{equation}
\left[  D_{\mu},D_{\nu}\right]  =\frac{1}{4}\left[  u_{\mu},u_{\nu}\right]
-\frac{i}{2}F_{\mu\nu}^+~,   \label{curv}%
\end{equation}
which allows one to consider products of covariant derivatives only in the
completely symmetrized form (and ignore all the other possibilities). The
second one is 
\begin{equation}
\left[  D_{\mu},u_{\nu}\right]  -\left[  D_{\nu},u_{\mu}\right]  =F_{\mu\nu}^-~,
   \label{Du anti}%
\end{equation}
which in a similar way allows one to consider covariant derivatives of $u_{\mu}$
only in the explicitly symmetrized form in terms of the tensor $h_{\mu\nu}$,
\begin{equation}
h_{\mu\nu}=\left[  D_{\mu},u_{\nu}\right]  +\left[  D_{\nu},u_{\mu}\right]~.
\end{equation}
For the construction of terms, as well as for further phenomenological
applications, it is advantageous to treat isosinglet and isotriplet components
of the external fields separately. We therefore define
\begin{equation}
\widetilde{X}=X-\frac12\langle X\rangle~.
\end{equation}
It is thus convenient to 
work with the following set of fields: $u_\mu,$ $\widetilde{\chi}_{\pm},$ $\langle
\chi_{\pm}\rangle,$ $\widetilde{F}_{\pm},$ $\langle F_{+}\rangle$ (the trace
$\langle F_{-}\rangle$ is zero because we have omitted the isoscalar
axial current.)
To construct an hermitian effective Lagrangian, which is not only chiral, but
also parity (P) and charge conjugation (C) invariant, 
one needs to know the transformation properties of the
fields under space inversion, charge conjugation and hermitian conjugation.
For the fields under consideration (and their covariant derivatives), hermitian
conjugation  is equal to $\pm$itself and charge conjugation amounts to $\pm
$transposed, where the signs are given (together with the parity) in 
Table~\ref{signs1}. This table also contains the chiral dimensions of the fields.
This is necessary for a systematic construction of the chiral
effective Lagrangian, order by order. Covariant derivatives acting on pion or
external fields count as quantities of first chiral order.
\begin{table}[h] \centering
\caption{Chiral dimension and transformation properties 
of the basic fields and the covariant derivative acting on the
pion and the external fields.\label{signs1}}%
\begin{tabular}
[c]{|l|c|c|c|c|c|c|}\hline
& $u_{\mu}$ & $\chi_{+}$ & $\chi_{-}$ & $F_{\mu\nu}^{+}$ & $F_{\mu\nu}^{-}$ &
$D_{\mu}$\\\hline
chiral dimension & $1$ & $2$ & $2$ & $2$ & $2$ & $1$\\
parity & $-$ & $+$ & $-$ & $+$ & $-$ & $+$\\
charge conjugation & $+$ & $+$ & $+$ & $-$ & $+$ & $+$\\
hermitian conjugation & $+$ & $+$ & $-$ & $+$ & $+$ & $+$\\\hline
\end{tabular}
\end{table}%
\noindent
In what follows, an analogous information for Clifford algebra
elements, the metric $g_{\mu\nu}$  and the totally antisymmetric 
(Levi-Civita) tensor $\varepsilon_{\lambda\mu\nu\rho}$ (for $d=4)$
together with the covariant derivative acting on the nucleon fields will
be needed (for a more detailed discussion, see Ref.20).
In this case, hermitian conjugate equals $\pm\gamma^{0}%
\,$(itself)$\,\gamma^{0}$ and charge conjugation amounts to $\pm$transposed,
where the signs are given (together with parity and chiral dimension) in
Table \ref{signs2}. The covariant derivative acting on nucleon fields counts as a
quantity of zeroth chiral order, since the time component of the
derivative gives the nucleon energy, which cannot be considered a small
quantity. However, the combination $\left(i\barr{D}-m\right)  \Psi$ is of first
chiral order. The minus sign for the charge and hermitian conjugation of
$D_{\mu}\Psi$ as well as the chiral dimension of $\gamma_5$
in Table \ref{signs2} is formal.\cite{FMMS}
\begin{table}[h] \centering
\caption{Transformation properties and chiral dimension
of the elements of the Clifford algebra together with the
metric and Levi-Civita tensors as well as the covariant derivative acting
on the nucleon field.\label{signs2}}%
\begin{tabular}
[c]{|l|l|l|l|l|l|l|l|}\hline
& $\gamma_{5}$ & $\gamma_{\mu}$ & $\gamma_{\mu}\gamma_{5}$ & $\sigma_{\mu\nu}$%
& $g_{\mu\nu}$ & $\varepsilon_{\lambda\mu\nu\rho}$ & $D_{\mu}\Psi$\\\hline
chiral dimension & \multicolumn{1}{|c|}{$1$} & \multicolumn{1}{|c|}{$0$} &
\multicolumn{1}{|c|}{$0$} & \multicolumn{1}{|c|}{$0$} &
\multicolumn{1}{|c|}{$0$} & \multicolumn{1}{|c|}{$0$} &
\multicolumn{1}{|c|}{$0$}\\
parity & \multicolumn{1}{|c|}{$-$} & \multicolumn{1}{|c|}{$+$} &
\multicolumn{1}{|c|}{$-$} & \multicolumn{1}{|c|}{$+$} &
\multicolumn{1}{|c|}{$+$} & \multicolumn{1}{|c|}{$-$} &
\multicolumn{1}{|c|}{$+$}\\
charge conjugation & \multicolumn{1}{|c|}{$+$} & \multicolumn{1}{|c|}{$-$} &
\multicolumn{1}{|c|}{$+$} & \multicolumn{1}{|c|}{$-$} &
\multicolumn{1}{|c|}{$+$} & \multicolumn{1}{|c|}{$+$} &
\multicolumn{1}{|c|}{$-$}\\
hermitian conjugation & \multicolumn{1}{|c|}{$-$} & \multicolumn{1}{|c|}{$+$} &
\multicolumn{1}{|c|}{$+$} & \multicolumn{1}{|c|}{$+$} &
\multicolumn{1}{|c|}{$+$} & \multicolumn{1}{|c|}{$+$} &
\multicolumn{1}{|c|}{$-$}\\\hline
\end{tabular}
\end{table}
We are now in the position to combine these building blocks to form
invariant monomials. Any invariant monomial in the effective $\pi$N
Lagrangian  is of the generic form
\begin{equation}
\bar{\Psi}A^{\mu\nu\ldots}\Theta_{\mu\nu\ldots}\Psi+\mathrm{h.c.}~.%
\label{monom}
\end{equation}
Here, the quantity $A^{\mu\nu\ldots}$ is a product 
of pion and/or external fields and their
covariant derivatives. $\Theta_{\mu\nu\ldots}$, on the other hand, is a product
of a Clifford algebra element $\Gamma_{\mu\nu\ldots}$ and a totally
symmetrized product of $n$ covariant derivatives acting on nucleon fields,
$D_{\alpha\beta\ldots\omega}^{n}=\left\{  D_{\alpha},\left\{  D_{\beta
},\left\{  \ldots,D_{\omega}\right\}  \right\}  \right\}  $,%
\begin{equation}
\Theta_{\mu\nu\ldots\alpha\beta\ldots}=\Gamma_{\mu\nu\ldots}D_{\alpha
\beta\ldots}^{n}~.%
\end{equation}
The Clifford algebra elements are understood to be  expanded in the standard basis
($1$, $\gamma_{5}$, $\gamma_{\mu}$, $\gamma_{\mu}\gamma_{5}$,
$\sigma_{\mu\nu}$) and all
the metric and Levi-Civita tensors are included in $\Gamma_{\mu\nu\ldots}$.
Note that Levi-Civita tensors may have some upper indices, which are, however,
contracted with other indices within $\Theta_{\mu\nu\ldots}$.
The structure given in Eq.(\ref{monom}) is not the most general Lorentz invariant
structure, it  already obeys some of the restrictions dictated by 
chiral symmetry. The curvature relation Eq.(\ref{curv}) manifests itself in the
fact that we only consider symmetrized products of covariant derivatives acting on $\Psi$. Another
feature of Eq.(\ref{monom}) is that, except for the $\varepsilon$-tensors, no
two indices of $\Theta_{\mu\nu\ldots}$ are contracted with each other. The
reason for this lies in the fact that at a given chiral order, $\barr{D}\Psi$ can 
always be replaced by $-im\Psi$, since their difference $\left(\barr{D}+im\right)
\Psi$ is of higher order. One can therefore ignore $D_{\mu}\Psi$ contracted
with $\gamma^{\mu},$ $\gamma^{5}\gamma^{\mu}$ and also with $\sigma
^{\lambda\mu}=i\gamma^{\lambda}\gamma^{\mu}-ig^{\lambda\mu}$. The last
relation explains also why $g^{\lambda\mu}\{D_{\lambda},D_{\mu}\}\Psi$ can be
ignored. To get the complete list of terms contributing to
$A^{\mu\nu\ldots}$, one writes down all 
possible products of the pionic and external fields and covariant derivatives
thereof. Every index of $A^{\mu\nu\ldots}$ increases the chiral order by one,
therefore the overall number of indices of $A^{\mu\nu\ldots}$ (as well as
lower indices of $\Theta_{\mu\nu\ldots}$) is constrained by the chiral order
under consideration. Since the matrix fields do not commute, one has to take
all the possible orderings. To get $A^{\mu\nu\ldots}$ with simple
transformation properties under  charge and hermitian conjugation, all the
products are rewritten in terms of commutators and anticommutators. In the
SU(2) case, this is equivalent to decomposing each product of any two
terms into isoscalar and isovector parts, since the only nontrivial product is
that of two traceless matrices, and the isoscalar (isovector) part of this product
is given by an anticommutator (commutator) of the matrices. After such a
rearrangement of products has been performed, the following relations hold:
\begin{equation}
A^{\dagger}=(-1)^{h_{A}}A~,\qquad A^{c}=(-1)^{c_{A}}A^{T}~,%
\end{equation}
where $(-1)^{h_A}$ and $(-1)^{c_A}$ are determined by the signs from 
Table~\ref{signs1} (as products of factors $\pm1$ for every field and covariant
derivative, and an extra factor $-1$ for every commutator). For the Clifford
algebra elements one has similar relations,
\begin{equation}
\Gamma^{\dagger}=(-1)^{h_{\Gamma}}\gamma^{0}\Gamma\gamma^{0}~,\qquad
\Gamma^{c}=(-1)^{c_{\Gamma}}\Gamma^{T}~,%
\end{equation}
where $h_{\Gamma}$ and $c_{\Gamma}$ are determined by the signs from 
Table~\ref{signs2}. The monomial Eq.(\ref{monom}) can now be written as
\begin{equation}
\bar{\Psi}A^{\mu\nu\ldots\alpha\beta\ldots}\Gamma_{\mu\nu\ldots}%
D_{\alpha\beta\ldots}^{n}\Psi+(-1)^{h_{A}+h_{\Gamma}}\bar{\Psi
}\overleftarrow{D}_{\alpha\beta\ldots}^{n}\Gamma_{\mu\nu\ldots}A^{\mu\nu
\ldots\alpha\beta\ldots}\Psi~.\label{monom2}%
\end{equation}
After elimination of total derivatives ($\overleftarrow{D}^{n}\rightarrow
(-1)^{n}D^{n}$) and subsequent use of the Leibniz rule, one obtains (modulo
higher order terms with derivatives acting on $A^{\mu\nu\ldots}$)
\begin{equation}
\bar{\Psi}A^{\mu\nu\ldots\alpha\beta\ldots}\Gamma_{\mu\nu\ldots}%
D_{\alpha\beta\ldots}^{n}\Psi+(-1)^{h_{A}+h_{\Gamma}+n}\bar{\Psi}%
A^{\mu\nu\ldots\alpha\beta\ldots}\Gamma_{\mu\nu\ldots}D_{\alpha\beta\ldots
}^{n}\Psi~.\label{monom3}%
\end{equation}
We see that to a given chiral order, the second term in Eq.(\ref{monom2}) either
doubles or cancels the first one. If the later is true, i.e. if $h_{A}%
+h_{\Gamma}+n$ is odd, this term only contributes at
higher orders and can thus be ignored. Consider now charge conjugation acting on the 
remaining  terms of the type given in Eq.(\ref{monom3}) (again
modulo higher order terms, with derivatives acting on $A^{\mu\nu\ldots}$)
\begin{equation}
2\bar{\Psi}A^{\mu\nu\ldots\alpha\beta\ldots}\Gamma_{\mu\nu\ldots
}D_{\alpha\beta\ldots}^{n}\Psi+(-1)^{c_{A}+c_{\Gamma}+n}2\bar{\Psi}%
A^{\mu\nu\ldots\alpha\beta\ldots}\Gamma_{\mu\nu\ldots}D_{\alpha\beta\ldots
}^{n} \Psi~.
\end{equation}
By the same reasoning as before, the terms with odd
$c_{A}+c_{\Gamma}+n$  are to be discarded.
The formal minus sign for the charge and hermitian conjugation of $D_{\mu}%
\Psi$ in Table~\ref{signs2}  takes care of this in a simple way. With
this convention, for any $\Theta_{\mu\nu\ldots}$, the two numbers $h_{\Theta}$
and $c_{\Theta}$ are determined by the entries in  Table~\ref{signs2} 
($h_{\Theta}=$ $h_{\Gamma}+n$, $c_{\Theta}=$ $c_{\Gamma}+n$) 
and invariant monomials Eq.(\ref{monom}) of a given chiral order 
are obtained if and only if
\begin{equation}
(-1)^{h_{A}+h_{\Theta}}=1~,\qquad (-1)^{c_{A}+c_{\Theta}}=1~.\label{h c}%
\end{equation}
The list of invariant monomials generated by the complete lists of $A^{\mu
\nu\ldots}$ and $\Theta_{\mu\nu\ldots}$ together with the condition
(\ref{h c}) is still overcomplete. It contains linearly dependent terms,
which can be reduced to a minimal set by use of various identities.
First of all, there are general identities, like the cyclic property
of the trace  or Schouten's identity
$\varepsilon^{\lambda\mu\nu\rho}a^{\tau}+\varepsilon^{\mu\nu\rho\tau}
a^{\lambda}+\varepsilon^{\nu\rho\tau\lambda}a^{\mu}+\varepsilon^{\rho
\tau\lambda\mu}a^{\nu}+\varepsilon^{\tau\lambda\mu\nu}a^{\rho}=0$.
Another general identity, frequently used in the construction of 
chiral Lagrangians, is provided by the Cayley-Hamilton theorem. 
For 2$\times$2 matrices $a$ and $b$, this theorem just implies
$\left\{  a,b\right\}  =a\left\langle b\right\rangle +\left\langle
a\right\rangle b+\left\langle ab\right\rangle -\left\langle a\right\rangle
\left\langle b\right\rangle$.
This was already accounted for by the separate treatment of the traces
and  the traceless
matrices. However, there are other nontrivial identities among products of
traceless matrices. Let us explicitly mention one such identity, which turns
out to reduce the number of independent terms containing four $u$ fields. 
It reads
\begin{equation}
\{  \widetilde{a},\widetilde{b}\,\}  [  \widetilde{a}%
,\widetilde{c}\,]  - \{  \widetilde{a},\widetilde{c}\, \}  
[\widetilde{a},\widetilde{b}\, ]  = \{  \widetilde{a},\widetilde
{a}\, \}  [  \widetilde{b},\widetilde{c}\, ]  -\widetilde
{a} \{  \widetilde{a}, [  \widetilde{b},\widetilde{c}\, ]\, \}~,
\end{equation}
where $\widetilde{a},\widetilde{b},\widetilde{c}$ are traceless 2$\times$2 matrices.
Another set of identities is provided by the curvature relation Eq.(\ref{curv}).
In connection with the Bianchi identity for covariant derivatives,
\begin{equation}
\left[  D_{\lambda},\left[  D_{\mu},D_{\nu}\right]  \right]  +\mathrm{cyclic}%
=0~,
\end{equation}
where ``$\mathrm{cyclic}$'' stands for cyclic permutations, it entails
\begin{equation}
\left[  D_{\lambda},F_{\mu\nu}^+ \right]  +\mathrm{cyclic}=\frac{i}{2}\left[
u_{\lambda},F_{\mu\nu}^- \right]  +\mathrm{cyclic}\label{cyclic Fp}~,%
\end{equation}
where we have used the Leibniz rule and Eq.(\ref{Du anti}) on the
right--hand--side. On the other hand,
when combined with Eq.(\ref{Du anti}), the curvature relation gives
\begin{equation}
\left[  D_{\lambda},F_{\mu\nu}^- \right]  +\mathrm{cyclic}=\frac{i}{2}\left[
u_{\lambda},F_{\mu\nu}^+ \right]  +\mathrm{cyclic}~,\label{cyclic Fm}%
\end{equation}
where we have used the Jacobi identity $[[u_{\lambda},u_{\mu}],u_{\nu
}]+\mathrm{cyclic}=0$ on the right--hand--side.
These relations can be used for the elimination of
(some) terms which contain $\left[  D_{\lambda},F^\pm_{\mu\nu}\right]  $.
Yet another set of identities is based on the equations of motion (EOM) deduced
from the lowest order $\pi\pi$ 
\begin{equation}
{\cal L}_{\pi\pi}^{(2)} = {F^2 \over 4} \langle D_\mu U D^\mu
U^\dagger + \chi^\dagger U + U^\dagger \chi \rangle \to 
\left[  D_{\mu},u^{\mu}\right]  =\frac{i}{2}\widetilde{\chi}_{-}~,
\label{pi eom}%
\end{equation}%
and $\pi$N Lagrangians, see Eq.(\ref{Lpin12}), 
\begin{equation}
\left(i\barr{D} -m+\frac{1}{2}{g}_{A}\barr{u}\gamma^{5}\right)  \Psi  = 
0 =
\bar{\Psi}\left(  i\overleftarrow{\barr{D}}+m-\frac{1}{2}{g}_A
\barr{u}\gamma^{5}\right)~.
\label{Psi eom}%
\end{equation}
One can directly use these EOM or (equivalently) perform specific field
redefinitions --- both techniques yield the same result. 
The pion EOM is used to
get rid of all the terms containing $h_{\;\mu}^{\mu}$, as well as $[D^{\mu
},h_{\mu\nu}]$, which can be eliminated using Eq.(\ref{pi eom}) together with
Eqs.(\ref{curv},\ref{Du anti}). 
The main effect of the nucleon EOM is
a remarkable restriction of the structure of $\Theta_{\mu\nu\ldots}$.
Here, I only give two
specific relations based on partial integrations and the nucleon EOM 
used in the reduction from the overcomplete to the
minimal set of terms, see Refs.15,20.
Two specific relations of this type are
\begin{eqnarray} 
\bar{\Psi} A^\mu i\,D_\mu \Psi + {\rm h.c.} & \doteq & 
2 m\,\bar{\Psi} \gamma_\mu A^\mu \Psi~,
\label{R1} \nonumber
\\[0.3em]
\bar{\Psi} A^{\mu\nu} D_\nu D_\mu \Psi + {\rm h.c.} & \doteq &
-m\left(\bar{\Psi} \gamma_\mu A^{\mu\nu} i\,D_\nu \Psi + {\rm h.c.} \right)~.
\label{R2}  
\end{eqnarray}
Here, the symbol $\doteq$ means equal up to terms of higher order.
With the techniques given in this section one is now able to
construct the minimal pion--nucleon Lagrangian to a given order.

\subsection{Heavy baryon projection}
\label{sec:heavy}
The so--called heavy baryon projection is interesting for various
reasons. First, it is modelled after heavy quark effective field
theory and was the first scheme including baryons that allowed for
a consistent power counting, see below. Second, it has some
resemblance to the static source model mentioned in section~\ref{sec:baryon}
and thus lends  to some ``intuitive'' physical interpretation. Let me first
discuss the issue of power counting.
Clearly, the appearance of the baryon mass scale in the lowest order
meson--baryon Lagrangian  causes
trouble. To be precise, if one calculates the baryon self--energy 
to one loop, one encounters terms of dimension 
  zero using  standard dimensional regularization,\cite{GSS}
\beq\label{mkrig}
 {\cal L}_{\rm MB}^{(0)} =
\bar{c}_0 \, \bar{B} B \, , \quad \, 
\bar{c}_0 \sim \left(\frac{{m}}{F}\right)^2
\frac{1}{d-4} + \ldots \quad , \label{massdiv} \eeq 
where the ellipsis stands for terms which are finite as $d \to 4$.
Such terms clearly make it difficult to organize the chiral expansion
in a straightforward and simple manner. They can be avoided if
the additional mass scale ${m}_B \sim 1$ GeV can be eliminated from the
lowest order effective Lagrangian (more precisely, what appears in 
${\cal L}_{\rm MB}^{(1)}$ is the baryon mass in the chiral limit.
For the following discussion, I will ignore 
this).\footnote{Another possibility to avoid
this complication will be discussed in section~\ref{sec:reg}.} 
Notice here the difference to the
pion case - there the mass vanishes as the quark masses are sent to
zero. Consider now  the mass of the baryon large compared to
the typical external momenta transferred by pions or photons (or any
other external source) and write
the baryon four--momentum as \cite{JM}~\cite{BKKM} 
\beq p_\mu = m_B \, v_\mu + \ell_\mu \, , \quad p^2 = m^2_B \, , 
\quad v \cdot \ell \ll m \, , \eeq
with $v_\mu$ is the baryon four--velocity (in the
rest--frame, we have $v_\mu =( 1 , \vec 0 \, )$), and
$l_\mu$ is a small residual momentum. In that
case, we can decompose the baryon field $B$ into velocity
eigenstates 
\beq B (x) = \exp [ -i {m}_B v \cdot x ] \, [ H(x) + h(x) ] \eeq
with 
\beq \barr v \, H = H \,\, , \quad  \barr  v \, h = -h \,\, , \eeq
or in terms of velocity projection operators
\beq 
P_v^+ H = H \, , \, P_v^- h = h \, , \quad P_v^\pm =
\frac{1}{2}(1 \pm \barr v \,) \, , \quad P_v^+ + P_v^- = 1\,  . 
\eeq
One now eliminates the 'small' component $h(x)$ either by using the
equations of motion or path--integral methods.
The Dirac equation for the velocity--dependent
baryon field $H = H_v$ (I will often suppress the label '$v$') 
takes the form $i v \cdot \partial H_v = 0$ to lowest
order in $1/m_B$. This allows for a consistent chiral counting as described
below and the effective meson--baryon Lagrangian takes the form:  
\beq {\cal L}_{\rm MB}^{(1)}  = \langle \bar{H} i v_\mu D^\mu H \rangle
+ D \langle \bar{H} S^\mu \{u_\mu, H\} \rangle
+ F \langle \bar{H} S^\mu  [u_\mu, H ] \rangle
+ {\cal O}\left(\frac{1}{m_B} \right) \, , \label{lagr} \eeq
with $S_\mu$ the covariant spin--operator
\beq S_\mu = \frac{i}{2} \gamma_5 \sigma_{\mu \nu} v^\nu \, , \, 
S \cdot v = 0 \, , \, \lbrace S_\mu , S_\nu \rbrace = \frac{1}{2} \left(
v_\mu v_\nu - g_{\mu \nu} \right) \, , \, [S_\mu , S_\nu] = i
\epsilon_{\mu \nu \gamma \delta} v^\gamma S^\delta \,
 \, , \label{spin}\eeq
in the convention $\epsilon^{0123} = -1$. There is one subtlety to be
discussed here. In the calculation of loop graphs, divergences appear
and one needs to regularize and renormalize these. That is done most
easily in dimensional regularization since it naturally preserves the
underlying symmetries. However, the totally antisymmetric Levi--Civita
tensor is ill-defined in $d \neq 4$ space--time dimensions. One
therefore has to be careful with the spin algebra. In essence,
one has to give a prescription how to uniquely fix the finite pieces.
The mostly used convention to do this consists in only using the
anticommutator to simplify products of spin matrices and only
taking into account that the commutator is antisymmetric under
interchange of the indices. Furthermore, $S^2$ can be uniquely
extended to $d$ dimensions via $S^2 = (1-d)/4$. With that in mind,
two important observations can be made. The lowest order meson--baryon
Lagrangian does not
contain the baryon mass term any more and also, all Dirac matrices
can be expressed as combinations of $v_\mu$ and $S_\mu$,\cite{JM}
\begin{displaymath} \bar H \gamma_\mu H = v_\mu \bar H H \, , \, 
\bar H \gamma_5 H = 0 \, , \, \bar H \gamma_\mu \gamma_5 H =
2 \bar H S_\mu H \, \, ,  \end{displaymath}
\beq \label{spinga}
\bar H \sigma_{\mu \nu} H = 2 \epsilon_{\mu \nu \gamma
  \delta} v^\gamma \bar H S^\delta H \, , \, \bar H \gamma_5 
\sigma_{\mu \nu} H = 2i \bar H (v_\mu S_\nu - v_\nu S_\mu) H \, \, ,
\eeq
to leading order in $1/m_B$. More precisely, 
this means e.g. $\bar H \gamma_5 H = {\cal O}(1/m_B)$. 
 We read off the baryon propagator,
\beq S_B (\omega ) = \frac{i}{\omega + i \eta} \, , \quad \omega = v
\cdot \ell \, , \quad \eta > 0\, \, . \label{prop} \eeq
The Fourier transform of Eq.(\ref{prop}) gives the space--time 
representation of the heavy baryon propoagator. Its explicit form
${\tilde S}(t, \vec{r}\,) = \Theta(t) \, \delta^{(3)} ( \vec{r}\,)$
illustrates very clearly that the field $H$ represents an (infinitely
heavy) static source.
A list of Feynman rules for the heavy baryon approach can be 
found in Ref.24. 
It is also instructive to consider the transition from the relativistic
fermion propagator to its counterpart in the heavy fermion limit.
Starting with $S_B (p) = i / (\barr{p} -m_B)$, one can project out
the light field component,
\beqa
S_B &=& P_v^+ \, i \, {\barr{p} -m_B \over p^2 -m^2}\, P_v^+ = i\,
{p \cdot v +m_B \over p^2 - m_B^2} \, P_v^+ = i \,
{2m + v \cdot \ell \over 2m_B  v \cdot \ell + \ell^2} \, P_v^+ 
\nonumber \\
&& \simeq {i \over  v \cdot \ell + \ell^2/2m_B - ( v \cdot \ell)^2/2m_B
+ {\cal O}(1/m_B^2)} \, P_v^+ \nonumber \\
&&\simeq  {i\, P_v^+  \over  v \cdot \ell} +
 {\cal O}({1\over m_B}) \,\,\, , 
\eeqa
which in fact shows that one can include the kinetic energy
corrections already in the propagator. So far, 
the arguments have been fairly hand--waving.
A more direct approach starting from the path integral of the
relativistic theory allows one to easily construct the
complete $1/m_B$ expansion systematically, in particular one easily
generates terms with fixed coefficients as demanded by Lorentz invariance
and lets one perform matching to the relativistic theory.
This is discussed in appendix~A, see also.\cite{BKKM,EM}
On the other hand, one can stay entirely within the heavy fermion approach
outlined so far and use reparameterization invariance to impose the
strictures from Galilean invariance,\cite{LM} a short discussion is
given in appendix~B. It is also important
to stress that simple concepts like wavefunction renormalization are
somewhat tricky in such type of approach, I refer the interested
reader to Ref.27. 
The explicit power counting based on such
ideas is given in section~\ref{sec:count}.

\subsection{Explicit form of the Lagrangian}

Here, I will discuss the explicit form of the effective Lagrangian for
the two--flavor case, in its relativistic as well as the heavy baryon
formulation. As we have already seen,
at lowest order, the effective $\pi$N Lagrangian is given in terms
of two parameters, the nucleon mass  and the axial--vector coupling
constant (in the chiral limit).
At second order, seven independent terms with LECs appear, so that
the relativistic Lagrangian reads (the explicit form of the various
operators $O_i^{(2)}$ is given in Table~\ref{dim2})
\begin{equation}
{\cal L}_{\pi
  N}^{(2)}=\sum_{i=1}^{7}c_{i}\bar{\Psi}O_{i}^{(2)}\Psi~.
\end{equation}
The LECs $c_i$ are finite because loops only start to contribute at
third order.
The HB projection is straightforward. We work here with the standard
form (see e.g. Refs.23,24)
and do not transform away the
$(v\cdot D)^2$ term from the kinetic energy (as it was done e.g. in
Ref.25), 
\begin{equation}
{\cal{L}}_{\pi N}^{(2)}   =\frac{1}{2m}\bar{N}_{v}\left(
\left(  v\cdot D\right)  ^{2}-D^{2}-ig_{A}\left\{  S\cdot D,v\cdot u\right\}
\right)  N_{v} +\sum_{i=1}^{7}\widehat{c}_{i}\bar{N}_{v}
\widehat{O}_{i}^{(2)} N_{v}~.
\end{equation}
The first three terms are indeed the fixed coefficients terms alluded
to in the previous section. The first two  simply give the
kinetic energy corrections to the leading order propagator. The third
term can not have a free coefficient because it generates the
low--energy theorem for neutral pion production off protons (as
detailed in \cite{BKKM}).
Again, the monomials $\widehat{O}_{i}^{(2)}$ are listed in
Table~\ref{dim2} together with the $1/m$ corrections, which some
of these operators receive (this splitting is done mostly to have
an easier handle on estimating LECs via resonance
saturation as will be discussed below).
\renewcommand{\arraystretch}{1.2}
\begin{table}[htb] \centering
\caption{Independent dimension two operators for the relativistic and
  the HB Lagrangian. The $1/m$ corrections to these operators in the 
  HB formulation are also displayed.\label{dim2}}%
\begin{tabular}
[c]{|c|c|c|c|}\hline
$i$ & $O_{i}^{(2)}$ & $\widehat{O}_{i}^{(2)}$ & $2 m (\widehat{c}_{i}-c_{i})%
$\\\hline
\multicolumn{1}{|l|}{$1$} & \multicolumn{1}{|l|}{$\left\langle \chi
_{+}\right\rangle $} & \multicolumn{1}{|l|}{$\left\langle \chi_{+}%
\right\rangle $} & \multicolumn{1}{|l|}{$0$}\\
\multicolumn{1}{|l|}{$2$} & \multicolumn{1}{|l|}{$-\frac{1}{8m^{2}%
}\left\langle u_{\mu}u_{\nu}\right\rangle D^{\mu\nu}+{\rm h.c.}$} &
\multicolumn{1}{|l|}{$\frac{1}{2}\left\langle \left(  v\cdot u\right)
^{2}\right\rangle $} & \multicolumn{1}{|l|}{$-\frac{1}{4} g_{A}^{2}$}\\
\multicolumn{1}{|l|}{$3$} & \multicolumn{1}{|l|}{$\frac{1}{2}\left\langle
u\cdot u\right\rangle $} & \multicolumn{1}{|l|}{$\frac{1}{2}\left\langle
u\cdot u\right\rangle $} & \multicolumn{1}{|l|}{$0$}\\
\multicolumn{1}{|l|}{$4$} & \multicolumn{1}{|l|}{$\frac{i}{4}\left[  u_{\mu
},u_{\nu}\right]  \sigma^{\mu\nu}$} & \multicolumn{1}{|l|}{$\frac{1}{2}\left[
S^{\mu},S^{\nu}\right]  \left[  u_{\mu},u_{\nu}\right]  $} &
\multicolumn{1}{|l|}{$\frac{1}{2}$}\\
\multicolumn{1}{|l|}{$5$} & \multicolumn{1}{|l|}{$\widetilde{\chi}_{+}$} &
\multicolumn{1}{|l|}{$\widetilde{\chi}_{+}$} & \multicolumn{1}{|l|}{$0$}\\
\multicolumn{1}{|l|}{$6$} & \multicolumn{1}{|l|}{$\frac{1}{8m}F_{\mu\nu}%
^{+}\sigma^{\mu\nu}$} & \multicolumn{1}{|l|}{$-\frac{i}{4m}\left[  S^{\mu
},S^{\nu}\right]  F_{\mu\nu}^{+}$} & \multicolumn{1}{|l|}{$2 m $}\\
\multicolumn{1}{|l|}{$7$} & \multicolumn{1}{|l|}{$\frac{1}{8m}\left\langle
F_{\mu\nu}^{+}\right\rangle \sigma^{\mu\nu}$} & \multicolumn{1}{|l|}{$-\frac
{i}{4m}\left[  S^{\mu},S^{\nu}\right]  \left\langle F_{\mu\nu}^{+}%
\right\rangle $} & \multicolumn{1}{|l|}{$0$}\\\hline
\end{tabular}
\end{table}

\noindent
At third order, one has 23 independent terms. We follow the notation of
Ref.20
(see also~\cite{EM}),
\begin{equation}
{\cal L}_{\pi N}^{(3)}=\sum_{i=1}^{23}d_{i}\, \bar{\Psi}\, 
O_{i}^{(3)}\, \Psi~.
\end{equation}
The LECs $d_i$ decompose into a renormalized scale--dependent and an
infinite (also scale--dependent) part in the standard manner, 
\beq\label{diren} d_i =
d_i^r (\lambda ) + \frac{\kappa_i}{(4\pi F)^2} L(\lambda )~, \quad
 L(\lambda) = \frac{\lambda^{d-4}}{(4\pi)^2}\biggl\{ \frac{1}{d-4} - 
 \frac{1}{2} [ \log(4\pi) + 1 - \gamma] \biggr\}~,
\eeq
with $\lambda$ the scale of dimensional regularization, 
$\gamma = 0.57722$ the Euler--Mascheroni constant, 
and the $\beta$--functions $\kappa_i$ were first given by Ecker.\cite{GE}
The HB projection including the $1/m$ corrections reads
\begin{equation}
{\cal{L}}_{\pi N}^{(3)} = 
\sum_{i=1}^{23}\widehat{d}_{i}\bar{N}_{v}\widehat{O}_{i}^{(3)} N_{v}
+ \bar{N}_{v}\, \widehat{O}_{\rm fixed}^{(3)}\, N_{v}
+ \bar{N}_{v}\, \widehat{O}_{\rm div}^{(3)}\, N_{v}~,
\end{equation}
with the corresponding operators and $1/m$ corrections together with  
terms with fixed coefficients, i.e. the monomials
${\cal O}_{\rm fixed}^{(3)}$  collected in Ref.15. 
In addition, there are eight terms just needed for the
renormalization, see again Ref.20. 

\smallskip
\noindent
At fourth order, there are 118 independent dimension four operators, which
are in principle measurable.
Four of these are special in the sense
that they are pure contact interactions of the nucleons with the external
sources that have no pion matrix elements. For example, two of these operators
contribute to the scalar nucleon form factor but not to
pion--nucleon scattering. 
The relativistic dimension four Lagrangian takes the form
\begin{equation}
{\cal L}_{\pi N}^{(4)}=\sum_{i=1}^{118} e_{i} \, \bar{\Psi} 
\,O_{i}^{(4)}\, \Psi~,
\end{equation}
and the monomials $O_{i}^{(4)}$ can be found in Ref.15. 
It should be stressed that many of the operators 
contribute only to very exotic processes, like three or four pion
production induced by photons or pions. Moreover, for a given
process, it often happens that some of these operators appear in
certain linear combinations. Note also that operators with a separate
$\langle \chi_+ \rangle$ simply amount to a quark mass renormalization
of the corresponding dimension two operator.
For example, in the case of elastic pion--nucleon scattering, one has
4, 5, and 4 independent LECs (or combinations) thereof at second, third
and fourth order, respectively, which is a much smaller number than
the total number of
independent terms. Consequently, the often cited folkore that CHPT
becomes useless beyond a certain order because the number of LECs increases
drastically is not really correct.
The HB projection leads to a much more complicated Lagrangian,
\begin{eqnarray}\label{L4HB}
{{\cal L}}_{\pi N}^{(4)} &=&\bar{N}_{v}\left( \sum_{i=1}^{118}
\hat{e}_{i}\widehat{O}_{i}^{(4)}+\sum_{i=1}^{23}e_{i}^{\prime
}W_{i}+\sum_{i=1}^{67}\left( e_{i}^{\prime \prime }X_{i}^{\lambda
}D_{\lambda }+{\rm h.c.}\right) \right. \nonumber  \\
&&\left. +\sum_{i=1}^{23}\left( e_{i}^{\prime \prime \prime }D_{\mu
}Y_{i}^{\mu \nu }D_\nu +{\rm h.c.}\right) +\sum_{i=1}^{4}\left(
e_{i}^{iv}Z_{i}^{\lambda \mu \nu }D_{\lambda }D_{\mu }D_{\nu }+{\rm h.c.}%
\right) \right.    \nonumber\\
&&\left. +\frac{1}{8m^{3}}\left( v\cdot DD_{\mu }D^{\mu }v\cdot D-\left(
v\cdot D\right) ^{4} \right) +  \widehat{O}_{\rm div}^{(4)}\right) N_{v}~,
\end{eqnarray}
where the first sum contains the 118 low--energy constants
in the basis of the heavy nucleon fields. Various additional terms appear:
First, there are the leading $1/m$ corrections to (most of) the 118
dimension four operators. These contribute to the difference in the
LECs $e_i$ and $\hat{e}_i$. Furthermore,
we have additional terms with fixed coefficients. They can
be most compactly represented by counting the number of covariant
derivatives acting on the nucleon fields.
There are two other fixed coefficient terms which are listed in the last
line of Eq.(\ref{L4HB}). All these terms are spelled out in Ref.15. 

\subsection{Loops and regularization methods}
\label{sec:reg}
In the meson sector, one can use standard dimensional
regularization for consistently separating the finite from the
infinite pieces of any loop diagram. The corresponding low--energy constants
have the the form as given in Eq.(\ref{LECform}) and the infinite parts
at one loop can be written as:
\beq
 C_i^\infty \sim \beta_i \, L(\lambda)~,
\eeq
with $L(\lambda)$ as defined in Eq.(\ref{diren}), 
and $\beta_i$ the pertinent $\beta$--function. The renormalized parts of the LECs 
then satisfy the RGE 
\beq
\lambda \frac{d}{d\lambda} C_i^r (\lambda) = - \beta_i~.
\eeq
Here, $C_i$ is a generic name for any LEC. In actual calculations to
be discussed, other symbols are used to characterize these couplings.
\begin{figure}[htb]
\centerline{
\psfig{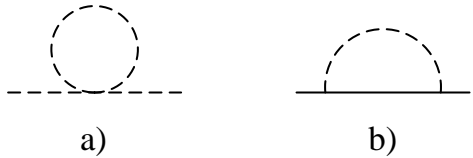}
}
\caption{One loop corrections to the a) pion and b) the nucleon mass.
\label{fig:tase}}
\end{figure}  
\noindent
For example, in the isospin limit ($m_u=m_d$) the quark mass expansion of the pion
mass takes the form 
\beq
M_\pi^2 = M^2 \biggl\{ 1 - \frac{M^2}{32\pi^2 F_\pi^2} \bar{\ell}_3
\biggr\} + {\cal O}(M^6)~,
\eeq
where the renormalized coupling $\bar{\ell}_3$ depends logarithmically
on the quark mass, $M^2 d\bar{\ell}_3/dM^2 = -1$. The infinite contribution
of the pion tadpole $\sim 1/(d-4)$, see graph a) in Fig.\ref{fig:tase},
has been absorbed in the infinite part of this LEC. In this scheme,
there is a consistent power counting since the only mass scale
(i.e. the pion mass) vanishes in the chiral limit (this power counting
is derived in the next section). If one uses the same method in the
baryon case, one encounters the scale problem already alluded to. More
precisely, the nucleon mass shift calculated from the self--energy diagram
(see graph b) in Fig.\ref{fig:tase}),
treating the nucleon field relativistically   based on standard
dimensional regularization, can be expressed via~\cite{GSS} 
\beq
\delta m_N = \frac{3\krig{g}_A^2 \krig{m}^2}{32\pi^2 F^2} \, \krig{m} \,
\biggl\{ \bar{c}_0 + \bar{c}_1 z - \pi z^{3/2} - \frac{1}{2} z^2 \ln z
+ \sum_{\nu = 4}^\infty a_\nu z^{\nu/2} \biggr\}~,
\eeq
with $z = M_\pi^2/m^2$ and the $\bar{c}_{0,1}$ are LECs, compare
also Eq.(\ref{mkrig}). Although only shown for the nucleon mass here,
a similar type of expansion holds for the whole ground--state octet.
This expansion is very different from the one for the pion mass. This
difference is due to the fact that the nucleon mass does not vanish in
the chiral limit and thus introduces a new mass scale apart from the
one set by the quark masses. Therefore, any power of the quark masses 
can be generated by chiral loops in the nucleon (baryon) case, whereas in the
meson case a loop order corresponds to a definite number of quark mass
insertions. This is the reason why one has resorted to the heavy mass
expansion in the nucleon case. Since in that case the nucleon (baryon) mass is
transformed from the propagator into a string of vertices with
increasing powers of $1/m_B$, a consistent power counting emerges (as
detailed in the next section). However, this method has the disadvantage
that certain type of diagrams are at odds with strictures from analyticity.
The best example is the so--called triangle graph, which enters e.g. the
scalar form factor or the isovector electromagnetic form factors of the
nucleon. This diagram has its threshold at $t_0= 4M_\pi^2$ but also
a singularity on the second Riemann sheet, at $t_c =  4M_\pi^2 - M_\pi^4/
m^2 = 3.98M_\pi^2$, i.e. very close to the threshold. To leading order
in the heavy baryon approach, this singularity coalesces with the
threshold and thus causes problems (a more detailed discussion can
be found e.g. in Refs.29,30).  
In a fully relativistic treatment,
such constraints from analyticity are automatically fulfilled. It was recently
argued in~\cite{EllTa} that relativistic one--loop integrals can be separated
into ``soft''' and ``hard'' parts. While for the former the power counting
as in HBCHPT applies, the contributions from the latter can be absorbed in 
certain  LECs.
In this way, one can combine the advantages of both methods. A more formal
and rigorous implementation of such a program is due to Becher and 
Leutwyler.\cite{BL} They call their method ``infrared regularization''. Any
dimensionally regularized
one--loop integral $H$ is split into an infrared singular and a regular part
by a particular choice of Feynman parameterization. Consider first the
regular part, called $R$. If one chirally  expands these terms, one generates
polynomials in momenta and quark masses. Consequently, to any order, $R$ can
be absorbed in the LECs of the effective Lagrangian.  On the other hand, the
infrared (IR) singular part $I$ has the same analytical properties as the full
integral $H$ in the low--energy region and its chiral expansion leads to the
non--trivial momentum and quark--mass dependences of CHPT, like e.g. the
chiral logs or fractional powers of the quark masses.
To be specific, consider the self--energy diagram b) of Fig.\ref{fig:tase} in
$d$ dimensions,
\beq
H(p^2) = \frac{1}{i} \int \frac{d^dk}{(2\pi)^d} \frac{1}{M_\pi^2 -k^2}
{1\over m^2 -(p-k)^2}~.
\eeq
At threshold, $p^2 = s_0 = (M_\pi+m)^2$, this gives
\beq
H(s_0) = c(d)\frac{M_\pi^{d-3} + m^{d-3}}{M_\pi + m} = I+R~,
\eeq 
with $c(d)$ some constant depending on the dimensionality of space--time.
The IR singular piece $I$ is characterized by fractional powers in the 
pion mass (for a non--integer dimension $d$) and generated by the
momenta of order $M_\pi$. For these soft
contributions, the power counting is fine. On the other hand, the IR regular
part $R$ is characterized by integer powers in the pion mass and generated
by internal momenta of the order of the nucleon mass (the large mass scale).
These are the terms which lead to the violation of the power counting in
the standard dimensional regularization discussed above. For the self--energy
integral, this splitting can be achieved in the following way
\beqa
H &=& \int  \frac{d^dk}{(2\pi)^d} {1 \over AB} 
= \int_0^1 dz \int  \frac{d^dk}{(2\pi)^d} {1 \over [(1-z)A+zB]^2}
\nonumber \\
 &=& \biggl\{ \int_0^\infty - \int_1^\infty \biggr\} dz
 \int  \frac{d^dk}{(2\pi)^d} {1 \over [(1-z)A+zB]^2} = I + R~,
\eeqa
with $A=M_\pi^2-k^2-i\epsilon$, $B=m^2 -(p-k)^2 -i\epsilon$ and
$\epsilon \to 0^+$. Any general
one--loop diagram with arbitrary many insertions from external sources
can be brought into this form by combining the propagators to a single
pion and a single nucleon propagator. It was also shown
that this procedure leads to a unique, i.e.
process--independent result, in accordance with the chiral Ward
identities of QCD.  This is essentially based on the fact that terms
with fractional versus integer powers in the pion mass must be separately
chirally symmetric. Consequently, the transition from any one--loop graph $H$
to its IR singular piece $I$ defines a symmetry--preserving regularization.
However, at present it is not known how to generalize this method to higher
loop orders. Also, as will be discussed later, its phenomenological
consequences  have so far not been explored in great detail. It is,
however, expected that this approach will be applicable in a larger energy
range than the heavy baryon approach.

\subsection{Renormalization}

\begin{figure}[htb]
\hskip 1.5in
\epsfysize=2.6cm
\epsffile{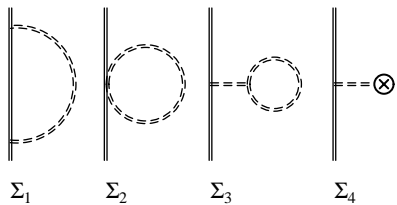}
\caption{Contributions to the one--loop generating functional
($\Sigma_1$,  $\Sigma_2$, $\Sigma_3$) and the tree level mesonic
generating functional ($\Sigma_4$) at order $\hbar$. The solid
(dashed) lines denote the baryon (meson) propagator in the presence
of external fields. The circle--cross in $\Sigma_4$ denotes the
counter terms from ${\cal L}_M^{(4)}$. The contributions
$\Sigma_{1,2}$ are called irreducible, whereas $\Sigma_{3,4}$ are
reducible.
\label{fig:ren}}
\end{figure}

In effective field theories, renormalizability to all orders is not
an issue due to the parametric suppression of higher order loop
graphs. Still, to obtain meaningful and unique results at a given
order, one has to perform standard field theoretic renormalization.
This can be done either by considering the specific divergences for
a process under consideration and combining these with appropriate
contact terms to obtain a finite result, or to systematically work
out the divergence structure of the generating functional to a given
order in the chiral expansion. Without going into mathematical
details, let me make some remarks concerning the one loop approximation.
To be specific, I consider the heavy baryon formalism. 
In Fig.~\ref{fig:ren} the various contributions to the 
one--loop generating functional together with the tree level generating 
functional at order $\hbar$ are shown. The solid (dashed) double lines 
represent the baryon (meson) propagator in the presence of external fields.
Only if one ensures that the field definitions underlying the meson
and the baryon--meson Lagrangian match, the divergences are entirely given
by the irreducible self--energy ($\Sigma_1$) and the tadpole ($\Sigma_2$)
graphs. The explicit calculations to extract the divergences 
from $\Sigma_{1,2}$ are given  in~\cite{GE} for SU(2) to third
order, in~\cite{mue:mei} for SU(3) to third order and in~\cite{MMS98}
for SU(2) to fourth order employing standard heat kernel techniques.
In the baryon sector, this technique is somewhat ackward due to the
different treatment of the bosonic and the fermionic propagators.
Therefore, in Ref.35  
a super--heat--kernel technique inspired
from supersymmetry methods was proposed and used in~\cite{Neu} for the
chiral pion--nucleon Lagrangian to third order. The so constructed 
divergences in the generating functional are simple poles in $1/(d-4)$.
These can be renormalized by introducing the appropriate local
counterterm Lagrangian to that order with corresponding LECs
of the form Eq.(\ref{LECform}). In that way, one constructs in general
a fairly long list of terms, i.e.
complete set for the renormalization with off--shell baryons.  
As long as one is only interested in Greens functions
with on--shell baryons, the number of terms can be reduced considerably
by making use of the baryon equations of motion. Also, many of these terms 
involve processes with three or more mesons. Such terms rarely show
up in actual calculations. For more details on the procedure,
the reader should consult in particular Ref.33. 

\medskip\noindent
Consider now renormalization within the relativistic approach based on IR
regularization. To leading order, the infrared singular parts coincide with 
the heavy baryon
expansion, in particular the infinite parts of loop integrals are the same. 
Therefore, the $\beta$--functions for low energy constants which absorb
these infinities are identical. However, infrared singular parts of relativistic
loop integrals also contain infinite parts which are suppressed by powers
of $M_\pi /m_N$, which hence cannot be absorbed as long as one only 
introduces counterterms to a finite order: exact renormalization only works up to the order at
which one works, higher order divergences have to be removed by hand.
Closely related to this problem is the one of the new mass scale $\lambda$ which 
one has to introduce in the progress of regularization and renormalization.
In dimensional regularization and related schemes, 
loop diagrams depend logarithmically on $\lambda$.
This $\log \lambda$ dependence is compensated for by running coupling constants,
the running behavior determined by the corresponding $\beta$--functions. 
In the same way as the contact terms cannot consistently absorb higher order
divergences, their $\beta$--functions cannot compensate for scale dependence
which is suppressed by powers of $\mu$. 
In order to avoid this unphysical scale dependence in physical results, 
the authors of~\cite{BL} have argued that the nucleon mass $m_N$ serves as
a ``natural'' scale in a relativistic baryon CHPT loop calculation and that
therefore, one should set $\lambda = m_N$ everywhere when using the infrared
regularization scheme. This was already suggested in \cite{BKKM} for the
framework of a relativistic theory with ordinary dimensional regularization.

\subsection{Power counting}
\label{sec:count}

To calculate any process to a given order, it is mandatory to have
a compact expression for the chiral power counting.\cite{wein79} \cite{ecker}
Consider first purely mesonic or single--baryon processes. To be
precise, I consider the heavy baryon approach, but these rules can be
extended to the relativistic treatment based on the IR regularization
method.
Any amplitude for a given physical process has a certain {\em chiral} {\em
dimension} $D$ which keeps track of the powers of external momenta and meson
masses, collectively labelled by the integer $d$. 
The building blocks to calculate this chiral dimension
from a general Feynman diagram in the CHPT loop expansion are 
(i) $I_M$ Goldstone boson (meson) propagators $\sim 1/(\ell^2 -M^2)$ 
(with $M=M_{\pi , K, \eta}$ the meson mass) of dimension $D= -2$, 
(ii) $I_B$ baryon propagators $\sim 1/ v \cdot \ell$ (in HBCHPT) with
$D= -1$, (iii) $N_d^M$ mesonic vertices with $d =2,4,6, \ldots$ and 
(iv) $N_d^{MB}$ meson--baryon vertices with $d = 1,2,3, \ldots$. 
Each loop integration introduces four powers of momenta. 
Putting pieces together, the chiral dimension $D$ of a given amplitude reads
\begin{equation}
D =4L - 2I_M - I_B + \sum_d d( \,  N_d^M + N_d^{MB} \, )
\end{equation}
with $L$ the number of loops.  For connected diagrams, one can use
the general topological relation
\begin{equation}
L = I_M + I_B -  \sum_d ( \,  N_d^M + N_d^{MB} \, ) + 1 
\end{equation}
to eliminate $I_M$~:
\begin{equation}
D =2L + 2 + I_B + \sum_d (d-2)  N_d^M + \sum_d (d-2) N_d^{MB} ~ .
\label{DLgen}
\end{equation}
Lorentz invariance and chiral symmetry demand $d \ge 2$ for mesonic 
interactions and thus the
term $\sum_d (d-2)  N_d^M$ is non--negative. Therefore, in the absence
of baryon fields, Eq.~(\ref{DLgen}) simplifies to~\cite{wein79}
\begin{equation}
D =2L + 2 + \sum_d (d-2)  N_d^M  \, \ge 2L + 2 ~ .
\label{DLmeson}
\end{equation}
To lowest order $p^2$, one has to deal with tree diagrams 
($L=0$) only. Loops are suppressed by powers of $p^{2L}$. 
Next, consider processes with a single baryon line running through 
the diagram (i.e., there is exactly one baryon in the in-- and one baryon 
in the out--state). In this case, the identity
\begin{equation}
\sum_d N_d^{MB} = I_B + 1 
\end{equation}
holds leading to~\cite{ecker}
\begin{equation}
D =2L + 1  + \sum_d (d-2)  N_d^M + \sum_d (d-1)  N_d^{MB}  \, \ge 2L + 1 ~ .
\label{DLMB}
\end{equation}
Therefore, tree diagrams start to contribute at order $p$ and one--loop
graphs at order $p^3$. 
Let me now consider diagrams with $N_\gamma$ external
photons.\footnote{These considerations can easily be extended to any
  type of external sources.}
Since gauge fields like the electromagnetic field appear
in covariant derivatives, their chiral dimension is obviously $D=1$.
One therefore writes the chiral dimension of a general amplitude
with $N_\gamma$ photons as
\begin{equation}
D = D_L + N_\gamma  ~,
\end{equation}
where $D_L$ is the degree of homogeneity of the (Feynman) amplitude $A$ as 
a function of external momenta ($p$) and meson masses ($M$) in the
following sense (see also \cite{rho}):
\begin{equation}
A(p,M;C_i^r(\lambda),\lambda/M) 
= M^{D_L} \, A ( p/M , 1;C_i^r(\lambda), \lambda/M )  ~ ,
\end{equation}
where $\lambda$ is an arbitrary renormalization scale and $C_i^r(\lambda)$
denote renormalized LECs. From now on, I will suppress the explicit dependence
on the renormalization scale and on the LECs. Since the total amplitude
is independent of the arbitrary scale $\lambda$, one may in particular
choose $\lambda = M$.
Note that $A(p,M)$ has also a certain physical dimension (which is 
of course independent of the number of loops and is therefore in
general different from $D_L$). The correct
physical dimension is ensured by appropriate factors of $F_\pi$ and $m$
in the denominators. Finally, consider a process with $E_n$ ($E_n = 4, 6, \ldots$) external
baryons (nucleons). The corresponding chiral dimension $D_n$ follows to be
\cite{weinnn}
\beq D_n = 2(L-C) + 4 - \frac{1}{2}E_n + \sum_i V_i \Delta_i \, ,
\quad \Delta_i = d_i + \frac{1}{2}n_i -2 \, \, , \eeq
where $C$ is  the number of connected pieces and one has $V_i$
vertices of type $i$ with $d_i$ derivatives and $n_i$ baryon fields
(these include the mesonic and meson--baryon vertices discussed before).
Chiral symmetry demands $\Delta_i \ge 0$. As before, loop diagrams
are suppressed by $p^{2L}$. Notice, however, that this chiral counting
only applies to the irreducible diagrams and not to the full S--matrix
since reducible diagrams can lead to IR pinch singularities and need
therefore a special treatment (for details, see Refs.40,41). 

\subsection{Phenomenological interpretation of the LECs}

In this section, we will be concerned with the phenomenological
interpretation of the values of the dimension two LECs $c_i$. For that, guided by
experience from the meson sector\cite{reso}, we use resonance
exchange. To introduce these concepts, I briefly discuss the 
Goldstone boson sector. In the meson sector at next--to-leading order, 
the effective three flavor
Lagrangian ${\cal L}_M^{(4)}$ contains ten LECs, called $L_i$.
These have been determined from data in Ref.11.
The actual values of the
$L_i$ can be understood in terms of resonance exchange.\cite{reso}
For that, one constructs the most general effective Lagrangian containing
besides the Goldstone bosons also resonance degrees of freedom.
Integrating out these heavy degrees of freedom from the EFT, one finds
that the renormalized $L_i^r (\mu=M_\rho)$ are practically saturated by 
resonance exchange ($S,P,V,A$). In some few cases, tensor mesons can
play a role.\cite{DT} This is sometimes called {\it chiral
  duality} because part of the excitation spectrum of QCD reveals itself
in the values of the LECs. Furthermore, whenever vector and axial
resonances can contribute, the $L_i^r (M_\rho)$ are completely
dominated by $V$ and $A$ exchange, called {\it chiral} {\it vector
  meson dominance (VMD)}.\cite{DRV}
As an example, consider the finite (and thus scale--independent) LEC
$L_9$. Its empirical value is $L_9 = (7.1 \pm 0.3) \cdot 10^{-3}$. The
well--known $\rho$--meson  exchange model for the pion form
factor (neglecting the width), $F_\pi^V (t) = M_\rho^2/ (M_\rho^2 -t) = 
1 + t / M_\rho^2 + {\cal O}(t^2)$ leads to
$L_9 = F_\pi^2 /(2 M_\rho^2) = 7.2 \cdot  10^{-3}$, by comparing to
the small momentum expansion of the pion form factor, $ F_\pi^V (q^2)
= 1 + \langle r^2 \rangle_\pi  \, t / 6 + {\cal O}(t^2)$.
The resonance exchange result is in good agreement with the empirical value.
Even in the symmetry breaking sector related to the quark masses, where
only scalar and (non--Goldstone) pseudoscalar mesons can contribute,
resonance exchange helps to understand why SU(3) breaking is generally
of the order of $25\%$, except for the Goldstone boson masses.
Consider now an effective Lagrangian with
resonances chirally coupled to the nucleons and pions. One can
generate local pion--nucleon operators of higher dimension with given
LECs by letting the resonance masses become
very large with fixed ratios of coupling constants to masses,
symbolically
\beq  
\tilde{{\cal L}}_{\rm eff}
 [U,M,N,N^\star] \to  {\cal L}_{\rm eff} [U,N]  \, \, , 
\eeq
where $M$ ($N^\star$) denotes meson (baryon) resonances.
This procedure amounts to decoupling the resonance degrees of freedom from
the effective field theory. However, the traces of these frozen
particles are encoded in the numerical values of certain LECs. In the
case at hand, we can have baryonic and mesonic excitations,
\beq \label{cirdef}
c_i = \sum_{N^*=\Delta,R,\ldots} c_i^{N^*} + \sum_{M=S,V,\ldots} c_i^M
\,\,\, ,
\eeq
where $R$ denotes the Roper $N^* (1440)$ resonance. 
Consider first scalar ($S$) meson exchange. The SU(2) $S\pi\pi$
interaction can be written as
\beq
{\cal L}_{\pi S} = S \, \Big[ \bar{c}_m \, {\rm Tr}(\chi_+) + 
\bar{c}_d \, {\rm Tr} (u_\mu u^\mu) \Big] \,\,\, .
\eeq
{}From this, one easily calculates the s--channel scalar meson
contribution to the invariant amplitude $A(s,t,u)$ for elastic 
$\pi \pi$ scattering,
\beqa
A^S (s,t,u) &=& \frac{4}{F_\pi^4 (M_S^2 -s)} \, [ 2 \bar{c}_m 
M_\pi^2 + \bar{c}_d (s- 2M_\pi^2) ]^2 \nonumber \\
&& \qquad \quad +{16 \bar c_m M_\pi^2 \over 3
F_\pi^4 M_S^2 } \Big[ \bar c_m M_\pi^2 + \bar c_d(3s-4M_\pi^2) \big]   \,\,\, .
\eeqa
Comparing to the SU(3) amplitude calculated in,\cite{bkmsu3} we
are able to relate the $\bar{c}_{m,d}$ to the  ${c}_{m,d}$ of
\cite{reso} (setting $M_{S_1} = M_{S_8} =M_S$ and using the large--$N_c$
relations $\tilde{c}_{m,d} = c_{m,d} / \sqrt{3}$ to express the
singlet couplings in terms of the octet ones),
$\bar{c}_{m,d} =   c_{m,d}/{\sqrt{2}}$,
with $|c_m| =42\,$MeV and $|c_d| = 32\,$MeV.\cite{reso} Assuming now
that $c_1$ is entirely due to scalar exchange, we get 
$c_1^S = - (g_S \, \bar{c}_{m})/M_S^2$.
Here, $g_S$ is the coupling constant of the scalar--isoscalar meson to
the nucleons, ${\cal L}_{SN} = - g_S \, \bar{N} N \, S$. What this 
scalar--isoscalar meson is essentially doing is to mock up the strong
pionic correlations coupled to nucleons. Such a phenomenon 
is also observed in the meson sector. The one loop description
of the scalar pion form factor fails beyond energies of 400 MeV,
well below the typical scale of chiral symmetry breaking,
$\Lambda_\chi \simeq 1\,$GeV. Higher loop effects are needed
to bring the chiral expansion in agreement with the data.\cite{game}
Effectively, one can simulate these higher loop effects by introducing
a scalar meson with a mass of about 600 MeV. This is exactly the line
of reasoning underlying the arguments used here (for a pedagogical 
discussion on this topic, see \cite{cnpp}). It does, however, not mean
that the range of applicability of the effective field theory is
bounded by this mass in general. In certain channels with strong pionic
correlations one simply has to work harder than in the channels
where the pions interact weakly and go beyond the one--loop approximation
which works well in most cases. For $c_1$ to be 
completely saturated by scalar exchange, $c_1 \equiv c_1^S$, we need
$M_S/\sqrt{g_S} = 180\,$MeV.
Here we made the assumption that such a scalar has the same couplings
to pseudoscalars as the real $a_0 (980)$ resonance.
It is interesting to note that the effective $\sigma$--meson in the 
Bonn one--boson--exchange potential \cite{bonn} with $M_S = 550\,$MeV and
$g_S^2/(4 \pi) = 7.1$ has $M_S / \sqrt{g_S} = 179\,$MeV. This number
is in stunning agreement with the  value demanded from scalar meson
saturation of the LEC $c_1$. With that, the scalar meson contribution
to $c_3$ is fixed including the sign, since $c_m c_d >0$,
$c_3^S = -2 g_s \, \bar{c}_d/M_S^2 = 2 c_d \, c_1 /c_m 
= -1.40\,$GeV$^{-1}$.
The isovector $\rho$--meson only contributes to $c_4$. Taking a universal
$\rho$--hadron coupling and using the KSFR relation, $M_\rho =
\sqrt{2} F_\pi g_\rho$,\cite{KFSR} we find
$c_4^\rho = \kappa_\rho/ (4m) = 1.63\,$GeV$^{-1}$,
using $\kappa_\rho = 6.1 \pm 0.4$ from the analysis of the nucleon
electromagnetic form factors, the process $\bar{N}N \to \pi \pi$
\cite{MMD} \cite{hoehpi} and the phenomenological one--boson--exchange
potential for the NN interaction.
I now turn to the baryon excitations. Here, the dominant one is the 
$\Delta (1232)$. Using the isobar model and the SU(4) coupling
constant relation, the $\Delta$ contribution to
the various LECs is
$c_2^\Delta = -c_3^\Delta = 2 c_4^\Delta = g_A^2 \, (m_\Delta
-m)/ (2 [(m_\Delta -m)^2 - M_\pi^2]) = 3.83\,$GeV$^{-1}$.
There is, however some sizeable uncertainty related to these, see 
Ref.37. 
One can deduce the following ranges: $c_2^\Delta = 
1.9 \ldots3.8, \, c_3^\Delta =-3.8 \ldots -3.0 , 
\, c_4^\Delta = 1.4 \ldots 2.0$ (in GeV$^{-1}$).  
The Roper $N^* (1440)$ resonance contributes only marginally,
see Ref.37.  
Putting pieces together, we have for $c_2$, $c_3$ and $c_4$ from
resonance exchange 
\beqa \label{cireso}
c_2^{\rm Res} &=& c_2^\Delta + c_2^R =  3.83 + 0.05 = 3.88 \,\, , \nonumber \\
c_3^{\rm Res} &=& c_3^\Delta + c_3^S + c_3^R 
= -3.83 -1.40 - 0.06  = -5.29 \,\, , \nonumber \\
c_4^{\rm Res} &=& c_4^\Delta + c_4^\rho + c_4^R 
= 1.92 + 1.63 + 0.12 = 3.67 \,\, ,
\eeqa
with all numbers given in units of GeV$^{-1}$.
Comparison with the empirical values listed in table~\ref{tab:LECs} 
shows that these LECs can be understood  from resonance
saturation, assuming only that $c_1$ is entirely given by scalar meson
exchange.  The LECs $\krig \kappa_s=-0.12$ and $\krig \kappa_v = 5.83$ 
can be estimated  from neutral vector meson exchange.
For the values from,\cite{MMD} $\kappa_\omega = -0.16 \pm
0.01$ and $\kappa_\rho = 6.1 \pm 0.4$, we see that the isoscalar and isovector
anomalous magnetic moments in the chiral limit can be well understood from 
$\omega$ and $\rho^0$ meson exchange. 
\renewcommand{\arraystretch}{1.2}
\begin{table}[t]
\caption{Values of the LECs $c_i$ in GeV$^{-1}$
for $i=1,\ldots,4$. Also given are the central values (cv) and the ranges
for the $c_i$ from resonance exchange. The $^*$ denotes an input
quantity. \label{tab:LECs}}
\begin{center}
\begin{tabular}{|l|r|r|r|c|}
   \hline
    $i$         & $c_i$~Ref.37 
                & $c_i$~Ref.20 
                & $c_i$~Ref.98 
                &  $ c_i^{\rm Res} \,\,$ cv [ranges]\\
    \hline
    1  &  $-0.93 \pm 0.10$  & $-1.23 \pm 0.16$ & $-0.81\pm 0.12$ & $-0.9^*$  [--] \\
    2  &  $3.34  \pm 0.20$  & $3.28  \pm 0.23$ & $8.43 \pm 56.9$ 
       & $3.9\,\,[2 \ldots 4]$
 \\    
    3  &  $-5.29 \pm 0.25$  & $5.94 \pm 0.09$& $-4.70\pm 1.16$ & $-5.3\,\, 
                                     [-4.5 \ldots -5.3]$ \\
    4  &  $3.63  \pm 0.10$   & $3.47 \pm 0.05$  & $3.40\pm 0.04$  & $3.7\,\, 
                                      [3.1 \ldots 3.7]$ \\    
    \hline
\end{tabular}
\end{center}
\end{table}
\noindent A systematic investigation
of dimension three or four LECs is not yet available. It has been found that
some dimension three LECs appearing in the chiral description of the 
nucleon electromagnetic form factors can be understood in terms of vector
meson ($\rho, \omega, \phi$) exchanges.

\subsection{Isospin violation and extension to virtual photons}
\label{sec:vir}

Up to now, I have mostly treated pure QCD in the isospin limit. We
know, however, that there are essentially two sources of isospin
symmetry violation (ISV). First, the quark mass difference $m_d -m_u$
leads to {\it strong} ISV. Second, switching on the {\it electromagnetic}
(em) interaction, charged particles are surrounded by a photon cloud
making them heavier than their neutral partners. An extreme case is
the pion, where the strong ISV is suppressed (see below) and the
photon cloud is almost entirely responsible for the charged
to neutral mass difference. Matters are different for the nucleon.
Here, pure electromagnetism would suggest the proton to be heavier than the
neutron by 0.8~MeV - at variance with the data. 
However, the quark mass (strong) contribution can be estimated to be
$(m_n-m_p)^{\rm str}= 2.1$~MeV. Combining these two numbers, one arrives at the
empirical value of $m_n - m_p = 1.3$~MeV. Let me now discuss how these
effects  arise in CHPT. 
Consider first the strong interactions. The symmetry
breaking part of the QCD Hamiltonian, i.e. the quark mass term,
 can be decomposed into an isoscalar and an isovector term
\beq
{\cal H}_{\rm QCD}^{\rm sb} = m_u \bar u u + m_d \bar d d =
{1\over 2} (m_u + m_d) ( \bar u u +  \bar d d ) +
{1\over 2} (m_u - m_d) ( \bar u u -  \bar d d ) \,\,\, .
\eeq
The quark mass ratios can easily be deduced from the ratios of the
(unmixed) Goldstone bosons, in particular, $m_d / m_u = 1.8\pm
 0.2$.\cite{leutw} Therefore, $(m_d-m_u)/(m_d+m_u) \simeq 0.3$ and
one could expect large isospin violating effects. However, the light
quark masses are only about $5...10$~MeV (at a renormalization
scale of 1~GeV) and the relevant scale to compare
to is of the order of the proton mass. This effectively suppresses the
effect of the sizeable light quark  mass difference in most cases,
as will be discussed below.
We notice that in the corresponding meson EFT, Eq.(\ref{pi eom}), 
the isoscalar term appears at leading order while the isovector one 
is suppressed. This is essentially the reason for the tiny quark mass 
contribution to the pion mass splitting. On the other hand, in the
pion--nucleon system, no symmetry breaking appears at lowest order
but at next--to--leading order, the isoscalar and the isovector terms
contribute. These are exactly the terms $\sim c_1$ and $\sim c_5$ in
Table~\ref{dim2}. In his seminal paper in 1977, Weinberg pointed out
that reactions involving nucleons and {\it neutral} pions might lead
to gross violations of isospin symmetry~\cite{weinmass} since the
leading terms from the dimension one Lagrangian are suppressed. In particular,
he argued that the mass difference of the up and down quarks can
produce a 30\% effect in the difference of the $\pi^0 p$ and $\pi^0 n$
S--wave scattering lengths while these would be equal in case of
isospin conservation. This was later reformulated in more modern
terminology.\cite{weinmit}
To arrive at the abovementioned result, Weinberg considered
Born terms and the dimension two symmetry breakers.
However, as shown in Ref.55,   
at this order there are other isospin--conserving terms which
make a precise prediction for the individual $\pi^0 p$ or $\pi^0 n$
scattering length very difficult. Furthermore, there is no way of
directly measuring these processes. On the other hand, there exists a huge
body of data for elastic charged pion--nucleon scattering ($\pi^\pm p
\to \pi^\pm p$) and charge exchange reactions ($\pi^- p \to \pi^0
n$).  In the framework of some models
it has been claimed that the presently available pion--nucleon data
basis exhibits strong isospin violation of the order of a few 
percent.\cite{gibbs} \cite{mats} What is, however, uncertain is to
what extent the methods used to separate the
electromagnetic from the strong ISV effects match. To really pin down isospin
breaking due to the light quark mass difference, one needs a machinery
that allows to {\it simultaneously} treat the electromagnetic and the
strong contributions. Here, CHPT comes into the game since one can
extend the effective Lagrangian to include virtual photons.
To be specific, consider now the photons as dynamical degrees of freedom.
To do this in a systematic fashion, one has to extend the power
counting. A very natural way to do this is to
assign to the electric charge the chiral dimension one, 
based on the observation that
\beq
\alpha = {e^2\over 4\pi} \sim {M_\pi^2 \over (4\pi F_\pi)^2} \sim {1 \over 100} \quad .
\eeq 
This is also a matter of consistency. While the neutral pion mass
is unaffected by the virtual photons, the charged pions acquire
a mass shift of the order $e^2$ from the photon cloud. If one would
assign the electric charge a chiral dimension zero, then the
power counting would be messed up. 
The extension of the meson Lagrangian is standard, I only give the
result here and refer to Refs.42,58,59,60
for all the details. Also, I  consider only the two flavor case,
\beq \label{L2meson}
{\cal L}^{(2)}_{\pi\pi} = - {1\over 4}F_{\mu\nu}F^{\mu\nu} - {\lambda \over 2} 
(\partial_\mu A^\mu )^2 + {F_\pi^2\over 4} \langle \nabla_\mu U \nabla^\mu
 U^\dagger + \chi U^\dagger + \chi^\dagger U \rangle + C \langle Q U
 Q U^\dagger \rangle \,\, , 
\eeq
with $F_{\mu\nu} = \partial_\mu A_\nu - \partial_\nu A_\mu$ the photon field
strength tensor and $\lambda$ the gauge--fixing parameter (from here on,
I use the Lorentz gauge $\lambda = 1$). Also, 
\beq
\nabla_\mu U = \partial_\mu U - i(v_\mu+a_\mu+QA_\mu)U
 +iU(v_\mu-a_\mu+QA_\mu) \,\, , 
\eeq
is the generalized pion covariant derivative containing the external
vector ($v_\mu$) and axial--vector ($a_\mu$)  sources.
It is important to stress that in
Refs.42,58,59   
$Q$ denotes the {\it quark} charge
matrix. To make use of the {\it nucleon} charge matrix commonly used 
in the pion--nucleon EFT, one can
perform a transformation of the type $Q \to Q + \alpha \, e\, \bf 1$, with
$\alpha$ a real parameter. One observes that $d \langle Q U Q U^\dagger \rangle
/ d\alpha \sim e^2 \, \bf 1$, i.e. to this order the difference between
the two charge matrices can completely be absorbed in an unobservable
constant term. The LEC $C$ can be
calculated from the neutral to charged pion mass difference via
$(\delta M^2_\pi)_{\rm em} = 2e^2C/F_\pi^2$. This gives
$C = 5.9\cdot 10^{-5}\,$GeV$^{4}$. Extending this unique lowest order
term in Eq.(\ref{L2meson}) to SU(3), one can easily derive Dashen's theorem.
To introduce virtual photons in the effective pion--nucleon field
theory, consider  the nucleon charge matrix $Q = e \,{\rm diag}(1,0)$.
Note that contrary to what was done before, the explicit factor
$e$ is subsumed in $Q$ so as to organize the power counting along the
lines discussed before. For the construction of chiral invariant 
operators, let me introduce the matrices
\beq
Q_{\pm} = \frac{1}{2} \, (u \, Q \,u^\dagger \pm u^\dagger \, Q \, u) \,\, ,
\quad
\widetilde{Q}_{\pm} = Q_\pm - {1 \over 2} \langle Q_\pm \rangle \,\, .
\eeq
Under chiral SU(2)$_L\times$SU(2)$_R$ symmetry, the $Q_\pm$ transform as 
\beq
Q_\pm  \to h(g,\Phi) \, Q_\pm  \,h^{-1} (g,\Phi)~,
\eeq
in terms of the compensator field introduced in section~\ref{sec:constru}. 
Furthermore, under parity ($P$) and charge conjugation ($C$)
transformations, one finds
\beq
P \, Q_\pm \, P^{-1} = \pm \, Q_\pm \,\,\, , \quad 
C \, Q_\pm \, C^{-1} = \pm \, Q_\pm^T \,\,\, . 
\eeq
For physical
processes, only quadratic combinations of the charge matrix $Q$
(or, equivalently, of the matrices $Q_\pm$) can appear. The following
relations are of use,
\beq
\langle Q_- \rangle = \langle Q_- \, Q_+ \rangle = 0 \,\, , \quad
\langle [i D_\mu , Q_\pm ] \rangle = 0 \,\,\, ,
\eeq
together with the SU(2) matrix relations discussed in 
section~\ref{sec:constru}.   It is now
straightforward to implement the (virtual) photons given in terms
of the gauge field $A_\mu$ in the effective pion--nucleon Lagrangian.
To be specific, I discuss here the heavy baryon approach.
Starting from the relativistic theory and decomposing the spinor
fields into light (denoted $N$) and heavy components (velocity
eigenstates), one can proceed as discussed before.
In particular, to lowest order (chiral dimension one),
\beq
{\cal L}_{\pi N}^{(1)} = \bar{N} \,\biggl( i v \cdot \bar{D} +
g_A \, S \cdot \bar{u} \, \biggr) \, N\,\,\, , 
\eeq
with 
\beq \label{covder}
\bar{D}_\mu = D_\mu -i \, Q_+ \, A_\mu \,\,\, , \quad 
\bar{u}_\mu = u_\mu - 2 \, Q_- \, A_\mu \,\,\, .
\eeq
At next order (chiral dimension two), we have 
\beq 
{\cal L}_{\pi N, {\rm em}}^{(2)} = \bar{N} \, F_\pi^2 \, \biggl\{ {f_1} \,
\langle Q^2_+ - Q^2_- \rangle + {f_2} \, \widetilde{Q}_+ \langle Q_+ \rangle
+ {f_3} \, \langle Q^2_+ + Q^2_- \rangle 
\, \biggr\} \, N \,\,\, .
\eeq
I have written down the
minimal number allowed by all symmetries. Note that the last  term
in ${\cal L}_{\pi N, {\rm em}}^{(2)}$ is proportional to $e^2 \bar{N}
N$. This means that it
only contributes to the electromagnetic nucleon mass in the chiral
limit and is thus not directly observable. However, this implies that
in the chiral two--flavor limit ($m_u =m_d=0$, $m_s$ fixed), the
proton is heavier than the neutron since it acquires an
electromagnetic mass shift. Only in pure QCD ($e^2=0$), this 
chiral limit mass is the same for both particles.
The numerical values of the electromagnetic LECs
$f_1$ and $f_2$ will be discussed below. The normalization factor of
$F_\pi^2$ in the electromagnetic pion--nucleon Lagrangian is
introduced so that the $f_i$ have the same dimension as the strong
LECs $c_i$. To go beyond tree level, we have to construct the
terms of order $p^3$ and $p^4$ (for a complete one--loop calculation).
The third order terms are given in Ref.61  
and the fourth order ones in Ref.62, 
together with a host of applications.
The interested reader is referred to these papers. A dimensional
analysis of the pertinent electromagnetic LECs is given in 
appendix~C.
I briefly return to the neutron--proton mass difference
as an instructive example. To order $p^3$, it is given
by a strong insertion $\sim c_5$ and an electromagnetic insertion
$\sim f_2$,
\beqa \label{delnp}
m_n - m_p &=& (m_n - m_p)_{\rm str} + (m_n - m_p)_{\rm em} 
\nonumber \\
&=& 4 \,c_5\, B \, (m_u - m_d) + 2 \, e^2 \, F_\pi^2 \, f_2  
+{\cal O}(p^4) \,\,\, .
\eeqa
Note that one--loop corrections only start at  order $p^4$.
This can be traced back to the fact that the
photonic self--energy diagram of the proton (on--shell) 
at order $p^3$ vanishes
since it is proportional to $\int d^dk \, [k^2 \, v\cdot k]^{-1}$.
Such an integral vanishes in dimensional regularization.  At chiral
dimension three, the  electromagnetic LEC $f_2$ can therefore be
fixed from the electromagnetic proton mass shift, $ (m_n - m_p)_{\rm
em} = -(0.7\pm 0.3)\,$MeV, i.e. $f_2 = -(0.45\pm 0.19)\,$GeV$^{-1}$.
The strong contribution has been used in~\cite{BKMLEC} to fix the LEC
$c_5 = -0.09 \pm 0.01\,$GeV$^{-1}$. Note that it is known that
one--loop graphs with an insertion $\sim m_d-m_u$ on the internal 
nucleon line, which in our counting appear at fourth order, can
contribute sizeably to the strong neutron-proton mass
difference.\cite{gl82} This calculation can be found in Ref.62.  
Also, the CHPT framework was used to systematically study ISV in
threshold $\pi$N scattering in all physical channels.\cite{fmsiso}
Some of these results will be discussed in later sections.

\subsection{The modern meaning of low--energy theorems}
\label{sec:LET}

In this section, I will briefly discuss the meaning of the so--called
low--energy theorems (LETs) in the context of CHPT, i.e. the
generalization of the current algebra LETs (which were
called ``soft--pion theorems'' in section~\ref{sec:dynamics}. I follow essentially
Ref.65. 
Let us first discuss a well--known example of a LET
 involving the electromagnetic current. Consider the
scattering of very soft photons on the proton, i.e., the Compton scattering
process $\gamma (k_1) + p(p_1) \to \gamma (k_2) + p(p_2)$ 
and denote by $\ve \, (\ve ')$
the polarization vector of the incoming (outgoing) photon. The transition
matrix element $T$ (normalized to $d\sigma / d\Omega = |T|^2$) can be 
expanded in a Taylor series in the small parameter $\delta =
|\vec{k_1}|/m$. 
In the forward  direction and in a gauge where
the polarization vectors have only space components, $T$ takes the form
\begin{equation}
T = a_0 \, \vec{\ve}\, ' \cdot \vec{\ve} + i \, a_1 \, \delta 
\, \vec{\sigma} \cdot (
\vec{\ve}\, ' \times \vec{\ve} \, ) + {\cal O}(\delta^2) ~ .
\label{Comp}
\end{equation}
The parameter $\delta$ can be made arbitrarily small in the laboratory so
that the first two terms in the Taylor expansion (\ref{Comp}) dominate. 
To be precise, the first one proportional to $a_0$ gives the low--energy 
limit for the spin--averaged Compton amplitude, while the second ($\sim a_1$) 
is of pure spin--flip type and can directly be detected in polarized 
photon proton scattering (to my knowledge, such a test has not yet
been performed). The pertinent LETs fix the values of $a_0$ and 
$a_1$ in terms of measurable quantities,\cite{low}
\begin{equation}
a_0 = - \frac{Z^2e^2}{4 \pi m} \, , \quad a_1 = 
- \frac{Z^2e^2 \kappa_p^2}{8 \pi m}
\label{c01}
\end{equation}
with $Z =1$ the  charge of the proton. To arrive at Eq.~(\ref{c01}), 
one only makes use of gauge 
invariance and the fact that the $T$--matrix can be written in terms of a
time--ordered product of two conserved vector currents sandwiched between
proton states. The derivation proceeds by showing that for small
enough photon energies the matrix element is determined by the electromagnetic
form factor of the proton at $q^2 = 0$.\cite{low}
Similar methods can be applied to other than the electromagnetic
currents. In strong interaction physics, a special role is played by the 
axial--vector currents. The associated symmetries are spontaneously
broken giving rise to the Goldstone matrix elements
\beq\label{GBM}
\langle 0|A^a_\mu(0)|\pi^b(p)\rangle = i \delta^{ab} F_\pi p_\mu~,
\eeq
where $a,b$ are isospin indices. In the chiral limit,
the massless pions play a similar role as the photon and many
LETs have been derived for ``soft pions".
In light of the previous discussion on Compton
scattering, the most obvious one is Weinberg's prediction for elastic 
$\pi p$ scattering.\cite{weinscatt} We only need the following translations~:
\begin{equation}
\langle p| T \, j_\mu^{\rm em} (x) j_\nu^{\rm em} (0)|p\rangle
 \, \, \to \, \,
\langle p| T \, A_\mu^{\pi^+} (x) A_\nu^{\pi^-} (0)|p \rangle ~, 
\end{equation}
\begin{equation}
\partial^\mu j_\mu^{\rm em} = 0 \, \, \to \, \,
\partial^\mu A_\mu^{\pi^-}  = 0  ~.
\end{equation}
In contrast to photons, pions are not massless in the real
world. It is therefore interesting to find out how the LETs for
soft pions are modified in the presence of non--zero pion masses
(due to non--vanishing quark masses). In the old days of current
algebra, a lot of emphasis was put on the PCAC relation,
consistent with the Goldstone matrix element, Eq.(\ref{GBM}),
\beq\label{PCAC}
\partial^\mu A^a_\mu = M_\pi^2 F_\pi \pi^a ~.
\eeq
Although the precise meaning of this formalism has
long been understood,\cite{col} it does not offer a systematic
method to calculate higher orders in the momentum and mass expansion
of LETs. The derivation of non--leading terms in the days of
current algebra and PCAC was more an art than a science, often
involving dangerous procedures like off--shell
extrapolations of amplitudes.
Current algebra can only provide the leading term in a systematic
expansion but can not be used to provide the corrections to these
lowest order statements. In the modern language, i.e. the EFT of
the Standard Model, these higher order corrections can be calculated 
unambiguously and one correspondingly defines a low--energy theorem
via:~{\bf L}(OW) {\bf E}(NERGY)  {\bf T}(HEOREM)  OF  ${\cal O}(p^n)
\equiv$  GENERAL  PREDICTION OF CHPT  TO ${\cal O}(p^n)$.
By general prediction I mean a
strict consequence of the SM depending on some LECs like 
$F_\pi, m, g_A, \kappa_p, \ldots$, but without any model assumption for these
parameters. This definition contains a
precise prescription how to obtain higher--order corrections to 
leading--order LETs. The soft--photon theorems, e.g. 
for Compton scattering,\cite{low}
involve the limit of small photon momenta, with all other momenta
remaining fixed. Therefore, they hold to all orders in the non--photonic
momenta and masses. In the low--energy expansion of CHPT, on the
other hand, the ratios of all small momenta and pseudoscalar meson
masses are held fixed. Of course, the soft--photon theorems are also
valid in CHPT as in any gauge invariant quantum field theory. 
To understand this difference  of low--energy limits,
I will now rederive and extend the LET for spin--averaged nucleon
Compton scattering in the framework of HBCHPT.\cite{BKKM} Consider the
spin--averaged Compton amplitude in forward direction (in the Coulomb
gauge $\ve \cdot v = 0$) 
\beq
e^2 \ve^\mu \ve^\nu
\frac{1}{4} {\rm Tr} \biggl[ (1 + \gamma_\lambda v^\lambda) T_{\mu \nu} (v,k)
\biggr] = e^2 \biggl[ \ve^2 U(\omega ) + (\ve \cdot k )^2
V(\omega ) \biggr]
\eeq
with $\omega = v \cdot k$ ($k$ is the photon momentum) and
\beq
T_{\mu \nu} (v,k) = \int d^4 k \, {\rm e}^{ik \cdot x} \, \langle
 N(v)| T j^{\rm em}_\mu (x) j_\nu^{\rm em} (0) |N(v)\rangle~.
\eeq
All dynamical information is
contained in the functions $U(\omega)$ and $V (\omega )$. We
only consider $U(\omega)$ here and refer to Ref.23  
for the calculation of both $U(\omega)$ and $V(\omega)$. In the
Thomson limit, only $U(0)$ contributes to the amplitude.
In the forward direction, the only quantities with non--zero chiral
dimension are $\omega$ and $M_\pi$. In order to make this dependence
explicit, we write $U(\omega,M_\pi)$ instead of $U(\omega)$. With
$N_\gamma = 2$ external photons, the degree of homogeneity $D_L$
for a given CHPT contribution to $U(\omega,M_\pi)$ follows from 
Eq.~(\ref{DLMB})~:
\beq
D_L =2L - 1  + \sum_d (d-2)  N_d^M + \sum_d (d-1)  N_d^{MB}  \, \ge - 1 ~ .
\label{DLC}
\eeq
Therefore, the chiral expansion of $U(\omega,M_\pi)$ takes the following
general form~:
\beq
U(\omega,M_\pi) = \sum_{D_L\ge -1} \omega^{D_L} f_{D_L}(\omega/M_\pi)~.
\label{Uce}
\eeq
The following arguments illuminate the difference and the
interplay between the soft--photon limit and the low--energy expansion
of CHPT. Let us consider first the leading terms in the chiral expansion
(\ref{Uce})~:
\beq
U(\omega,M_\pi) = {1\over \omega} f_{-1}(\omega/M_\pi) +
 f_0(\omega/M_\pi) + {\cal O}(p)~.
\eeq
Eq.~(\ref{DLC}) tells us that only tree diagrams can contribute
to the first two terms. However, the relevant tree diagrams
do not contain pion lines. Consequently, the functions
$f_{-1}$, $f_0$ cannot depend on $M_\pi$ and are therefore constants.
Since the soft--photon theorem \cite{low} requires $U(0,M_\pi)$
to be finite, $f_{-1}$ must actually vanish and the chiral
expansion of $U(\omega,M_\pi)$ can be written as
\beq
U(\omega,M_\pi) = f_0 + \sum_{D_L\ge 1} \omega^{D_L} f_{D_L}(\omega/M_\pi)~.
\label{Uce2}
\eeq
But the soft--photon theorem yields additional information~:
since the Compton amplitude is independent of $M_\pi$ in the Thomson
limit and since there is no term linear in $\omega$ in the spin--averaged
amplitude, we find
\beq
\lim_{\omega \to 0}~ \omega^{n-1} f_n(\omega/M_\pi) = 0 \qquad
(n\ge 1) \label{spl}
\eeq
implying in particular that the constant $f_0$ describes the
Thomson limit~:
\beq
U(0,M_\pi) = f_0~.
\eeq
Let me now verify these results by explicit calculation.
In the Coulomb gauge, there is no direct 
photon--nucleon coupling from the lowest--order effective Lagrangian 
${\cal L}_{\pi N}^{(1)}$ since it is proportional to $\ve \cdot v$. 
Consequently, the corresponding Born diagrams  vanish so that indeed
$f_{-1}=0$. On the other hand, the heavy mass expansion of the
relativistic $\pi N$ Dirac Lagrangian 
leads to a Feynman insertion of the form:
\beq
i \frac{e^2}{m} \frac{1}{2} (1 + \tau_3 ) \biggl[ \ve^2 -( \ve
\cdot v)^2 \biggr] = i \frac{e^2 Z^2}{m} \,  \ve^2
\eeq
producing the desired result $f_0 = Z^2 / m$, the Thomson limit.
At the next order in the chiral expansion, ${\cal O}(p^3)$ ($D_L = 1$), 
the function $f_1(\omega/M_\pi)$ is given by the
finite sum of 9 one--loop diagrams.\cite{BKM} \cite{BKKM} According
to Eq.~(\ref{spl}), $f_1$ vanishes for $\omega \to 0$. The term linear
in $\omega/M_\pi$ yields the leading contribution to the sum of the
electric and magnetic polarizabilities of the nucleon, defined by
the second--order Taylor coefficient in the expansion of $U(\omega,M_\pi)$
in $\omega$~:
\beq
f_1(\omega/M_\pi) = - {11 g_A^2 \omega\over 192 \pi F_\pi^2 M_\pi}
+ {\cal O}(\omega^2)~.
\eeq
The $1/M_\pi$ behaviour should not come as a surprise
-- in the chiral limit the  pion cloud becomes long--ranged (instead of being
Yukawa--suppressed) so that the polarizabilities explode.
This behaviour is specific to the leading contribution
of ${\cal O}(p^3)$. In fact, from the general form (\ref{Uce2}) one
immediately derives that the contribution of ${\cal O}(p^n)$
($D_L = n - 2$) to the polarizabilities is of the form $c_n M_\pi^{n-4}$
($n\ge 3$), where $c_n$ is a constant that may be zero.
One can perform a similar analysis for the amplitude 
$V(\omega)$ and for the spin--flip amplitude. 
Next, let us consider the processes
$\gamma N \to \pi^0 N$ $(N=p,n)$
at threshold, i.e., for vanishing three--momentum of the pion
in the nucleon rest frame. At
threshold, only the electric dipole amplitude $E_{0+}$ survives and
the only quantity with non--zero chiral dimension is $M_\pi$.
In the usual conventions,
$E_{0+}$ has physical dimension $-1$ and it can therefore be
written as
\beq
E_{0+} = {e g_A \over F}~A\left( {M_\pi\over m},{M_\pi\over F}\right)~.
\eeq
The dimensionless amplitude $A$ will be expressed as a power
series in $M_\pi$. The various parts are characterized by the
degree of homogeneity (in $M_\pi$) $D_L$ according to the chiral
expansion. Since $N_\gamma=1$ in the present case, we obtain from
Eq.~(\ref{DLMB})
\beq
D_L = D-1 = 2L + \sum_d (d-2)  N_d^M + \sum_d (d-1)  N_d^{MB} ~ .
\eeq
For the LET of ${\cal O}(p^3)$ in question, only lowest--order mesonic vertices
($d=2$) will appear. Therefore, in this case the general formula for $D_L$ 
takes the simpler form 
\beq
D_L = 2L + \sum_d (d-1)  N_d^{MB} ~ .
\label{DLphoto}
\eeq
I will not discuss the chiral expansion of $E_{0+}$ step by step, but
rather refer
to the literature \cite{BGKM} \cite{BKM1} \cite{BKKM} for the actual 
calculation and for more details to the Comment.\cite{gerulf} 
Up--to--and--including order $\mu^2 = (M_\pi / m)^2$, one has to
consider contributions with $D_L = 0, 1$ and $2$. In fact, for neutral
pion photoproduction, there is no term with $D_L=0$ since the
time--honored Kroll--Ruderman contact term
\cite{KR} where both the pion and the photon emanate from the same
vertex, only exists for charged pions.
For $D_L = 2$
there is a one--loop
contribution ($L=1$) with leading--order vertices only
($N_d^{MB}=0 ~(d>1)$).  It is considerably easier to work out the
relevant diagrams in HBCHPT \cite{BKKM} than in the original 
derivation.\cite{BGKM}~\cite{BKM1} In fact, at threshold only the so--called triangle
diagram (and its crossed partner) survive out
of some 60 diagrams. The main reason for the enormous simplification
in HBCHPT is that one can choose a gauge without a direct $\gamma NN$
coupling of lowest order and that there is no direct coupling of
the produced $\pi^0$ to the nucleon at threshold. 
Notice that the loop contributions are finite and they are identical for proton
and neutron. They were omitted in the original version of the LET
\cite{VZ} \cite{deB} and in many later rederivations.
The full LETs of ${\cal O}(p^3)$ are given by \cite{BGKM} 
\beqa
E_{0+}(\pi^0 p) &=&  - \dfrac{e g_A}{8\pi F_\pi}\biggl[ \biggr.
\dfrac{M_\pi}{m} - \dfrac{M_\pi^2}{2 m^2}~(3+\kappa_p)  -
\dfrac{M_\pi^2}{16 F_\pi^2} + \biggl.{\cal O}(M_\pi^3)\biggr]~,
\\
E_{0+}(\pi^0 n) &=&  - \dfrac{e g_A}{8\pi F_\pi}\biggl[ \biggr. 
 \dfrac{M_\pi^2}{2 m^2}~\kappa_n  -
\dfrac{M_\pi^2}{16 F_\pi^2} + \biggl. {\cal O}(M_\pi^3)\biggr]~. \eeqa
We note that the electric dipole amplitude for neutral pion production
vanishes in the chiral limit. By now, even the terms of order
$M_\pi^3$ have been worked out, see Ref.\cite{bkmpi0}
The derivation of LETs sketched above is based on a well--defined
quantum field theory where each step can be checked explicitly. Nevertheless, 
the corrected LETs have been questioned by several authors. A detailed
discussions of the various assumptions like e.g. analyticity in the
pion mass, which do not hold in QCD, can be found in Ref.65.
Even better, by now the data support the
CHPT predictions, see section~\ref{sec:photo}. Consequently, a LET to
a certain order as defined here is always strict but might not always be
useful (in case of large higher order corrections).

\subsection{Inclusion of resonances}
\label{sec:SSE}
This section departs a bit from the main line of reasoning performed so far.
Up to now, any QCD state of higher mass  (nucleon or meson resonances) 
appeared only implicitly, in the values of certain LECs. In addition, there is a decoupling
theorem,\cite{GaZe} which states that in the chiral limit all S--matrix elements
and transition currents to leading order are given in terms of the Goldstone
bosons and the ground state baryon octet.  Nevertheless, under certain
circumstances one has to extend the effective field theory to include
some of these resonances explicitly. In particular,
when going to higher energies, the usefulness of the chiral
expansion is limited by the appearance of the nucleon (or meson) resonances, 
the most prominent and important of these being the $\Delta (1232)$ with spin and isospin
3/2.\footnote{The role of t--channel meson excitations will be discussed later.} Its
implications for hadronic and nuclear physics are well established. Consequently,
one would like to have a consistent and systematic framework to include this
important degree of freedom in baryon chiral perturbation theory, as first
stressed by Jenkins and Manohar~\cite{JMdel}
and only recently formalized by Hemmert, Holstein and
Kambor~\cite{HHK} (see also~\cite{rhodel} for a general discussion of
including the delta). This is possible because one can count
the nucleon--delta mass splitting as an additional small
parameter. Doing that, one arrives at the so--called small scale expansion. 
It has already
been established that most of the low--energy constants appearing in the effective chiral
pion--nucleon Lagrangian are saturated by the delta and thus the
resummation of such terms underlying the small scale expansion (SSE) lets one expect a better
convergence as compared to the chiral expansion. In addition, the radius of convergence
is clearly enlarged when the delta is included as an explicit degree of freedom.
This of course increases the complexity of the approach since the pertinent effective
Lagrangian contains more structures consistent with all symmetries. 
Here, I will briefly discuss the structure of the effective Lagrangian underlying
the SSE. All this  is found in much more detail in the work of Hemmert et al.~\cite{HHK}
and I refer the interested reader to that paper to fill in 
the details omitted here.  The explicit
inclusion of the delta into an effective pion--nucleon field theory is
motivated by the fact that in certain observables this resonance plays a
prominent role already at low energies. The reason for this is twofold.
First, the delta--nucleon mass splitting is small,
\be
\Delta \equiv m_\Delta -m_N = 294~{\rm MeV} \simeq 3F_\pi~
\eeq
or stated differently, $\Delta = {\cal O}(M_\pi)$. However, in the chiral
limit of vanishing quark masses, neither $\Delta$ nor $F_\pi$ vanish.
Therefore, such an extended EFT does not have the same chiral limit
as QCD, as it is well--known since long.\cite{GaZe} Second, the delta
couples very strongly to the $\pi N \gamma$ system, e.g. the strong
$\Delta N \gamma$ M1 transition plays a prominent role in charged pion
photoproduction or nucleon Compton scattering. One can set up a
consistent power counting by
using the well--known heavy baryon techniques~\cite{JM,BKKM} and by treating 
the mass splitting $\Delta$ as an additional small parameter besides 
the external momenta and quark (meson) masses. Therefore, any matrix element
or transition current has a low energy expansion of the form
\be
{\cal M} = \ve^{n} \, {\cal M}_1 + \ve^{n+1} \, {\cal M}_2 +
\ve^{n+2} \, {\cal M}_3 + {\cal O}(\ve^{n+3})~,
\ee
where the power $n$ depends on the process under consideration 
(for example in case of elastic pion--nucleon scattering, $n$ equals one)
and $\ve$ collects the three different small parameters,
\be
\ve \in \left\{ {p\over \Lambda_\chi},  {M_\pi\over \Lambda_\chi},
 {\Delta \over \Lambda_\chi} \right\}~,
\ee
with $\Lambda_\chi \simeq 1\,$GeV the scale of chiral symmetry
breaking. This power
counting scheme is often called $\ve$-- or small scale expansion (SSE).
As it is common in heavy baryon approaches,
the expansion in the inverse of the heavy mass and with respect to the
chiral symmetry breaking scale are treated simultaneously (because $m_N \sim
m_\Delta \sim \Lambda_\chi$). The inclusion of the delta is more
complicated because as a spin--3/2 field, one has to eliminate not
only the ``small'' spin--3/2 component but also the four intrinsic
(``off--shell'') spin--1/2 components. The technology to do that
is spelled out in.\cite{HHK} The resulting effective Lagrangian has the following 
low--energy expansion
\be
{\cal L}_{\rm eff} = {\cal L}^{(1)} + {\cal L}^{(2)} + {\cal L}^{(3)} + \ldots
\ee
where each of the terms ${\cal L}^{(n)}$ decomposes into a pure nucleon
($\pi N$), a nucleon--delta ($\pi N \Delta$) and  a pure delta part
($\pi\Delta$),
\be
 {\cal L}^{(n)} =   {\cal L}^{(n)}_{\pi N} + {\cal L}^{(n)}_{\pi N\Delta} +
  {\cal L}^{(n)}_{\pi\Delta}~, \quad n= 1,2,3, \ldots~.
\ee
I will not discuss these terms in detail but only make some general
remarks. First, while the terms with nucleon fields only have the
same structure as in the pion--nucleon EFT, the presence of the
explicit delta modifies of course the numerical values of most LECs.
The $N\Delta$ terms are to be understood symmetrically and the pure
delta part is mostly needed inside loop integrals. The coupling of external fields
is done by standard methods. So far, only some reactions have been
investigated systematically in this framework, these are threshold pion
photoproduction,\cite{HHK1} Compton scattering,\cite{HHK2,HHKK}
electroweak nucleon and nucleon--delta transition form factors~\cite{BFHM,george} 
or pion--nucleon scattering.\cite{FMdel} Some of the pertinent results
will be discussed below.

\smallskip\noindent
It is also well established that in many reations vector mesons play
a prominent role. A formal study of chiral Lagrangians involving
baryons, Goldstone bosons and spin--1 fields is given in.\cite{BoMe}
At present, there is no generalization of CHPT which fully includes the
effects of vector mesons as intermediate states to arbitrary loop orders. As the masses
of the vector mesons do not vanish in the chiral limit, they introduce a new mass scale
which, when appearing inside loop integrals, potentially spoils chiral power counting 
in a similar manner as the nucleon mass in a ``naive'' relativistic baryon CHPT approach. 
In analogy to the heavy fermion theories, 
``heavy meson effective theory'' has been used to investigate vector meson properties
like masses and decay constants when coupling these to the Goldstone
bosons in a chirally invariant fashion 
(see~\cite{BGT} or for some special processes~\cite{JMW}).
This approach does not work with the heavy particle number not conserved 
as in processes with these resonances being virtual intermediate states. 
However, these difficulties do not arise as long as there is no loop integration 
over intermediate vector meson momenta. Such special cases will be discussed below.

\section{Two flavors: The structure of the nucleon}
\label{sec:SU2}

\subsection{Two--point functions I: The nucleon mass}
It is instructive to consider the chiral expansion of the nucleon mass
in heavy baryon CHPT. Here, I will be concerned with the chiral SU(2) limit, i.e.
$m_u=m_d=0$ and $m_s$ fixed at its physical value. That means that
possible effects due to strangeness admixture into the nucleon
wave function are subsumed in the value of the nucleon mass in the
chiral limit, which is denoted by $\krig{m}$. One expects  that
the corrections at $n^{\rm th}$ order are given by $c_n
M_\pi^n/\Lambda_\chi^{n-1}$, with $c_n$ of order one. To fourth order in small
momenta, the chiral expansion of the nucleon mass
reads:\cite{FMSwf,FM}
\beqa\label{Nmass}
m &=& \krig{m} -4 M^2 c_1 -\frac{3 g_A^2 M^3}{32 \pi F^2} \nonumber\\
&& \quad -  M^4 (16\bar{e}_{38} + 2\bar{e}_{115}+ \frac{1}{2}\bar{e}_{116}) 
+ \frac{3 M^4 c_2}{128 \pi^2 F^2} - \frac{3 g_A^2 M^4}{64 \pi^2 m
  F^2}~.
\eeqa
The LECs $c_{1,2}$ can in principle be determined from the analysis of
pion--nucleon scattering discussed in section~\ref{sec:piN}.
Note that while the term $\sim \bar{e}_{38}$ is nothing but a quark mass
renormalization of the operator $\sim c_1$, the LECs
$\bar{e}_{115,116}$ cannot be determined from pion--nucleon scattering. Setting the fourth
order LECs to zero, one finds that the chiral expansion of the nucleon
mass is well--behaved and follows the scaling arguments given above
\beq
\delta m^{(4)} = ( 72.5 - 15.1 + 0.4 - 0.4 )~{\rm MeV}~,
\eeq
using the LECs as determined in.\cite{BKMLEC} The fifth
order contribution has been determined in 
Ref.89,   
they find that all loop contributions cancel and that one is simply left with a
$1/m$ correction to the third order piece, which is numerically tiny.
This is a good starting point for further investigations of $n$--point
functions in HBCHPT. The effects due to virtual photons have also
been worked out to fourth order.\cite{MMiso} At third order, one only
has an effect due to the pion mass difference, which is a 7\%
correction to the strong result given in Eq.(\ref{Nmass}). The fourth
order correction depends on the em LEC $f_1$ and other known
LECs.\cite{MMiso} Setting $f_1 = \pm 1/(4\pi)$ as expected from
dimensional analysis (see app.~C), the numerical value of the em fourth
order mass shift is tiny, for the proton one
finds $\delta m^{(4, \rm em)}_{p}
= -0.10 \ldots -0.01\,$MeV and for the neutron $\delta m^{(4, \rm em)}_{n}
= -0.06 \ldots +0.03\,$MeV. This is completely negligible and comparable
to the strong fourth order mass shift. This shows that the effects of
virtual photons can be calculated with their precise values depending
on the electromagnetic LECs. Most of these can at present only be
estimated using dimensional analysis (as discussed before).

\subsection{Three--point functions I: Scalar form factor and sigma term}
\label{sec:sigma}

Pion--nucleon scattering data allow to extract information on the size
of the pion--nucleon  $\sigma$--term, $\sigma (0)$, which measures 
the explicit chiral symmetry breaking in QCD due to the up-- and
down--quark masses. A venerable (current algebra) low--energy theorem  due to 
Brown, Pardee and Peccei~\cite{bpp} relates $\sigma (0)$ to the isoscalar
$\pi$N scattering amplitude (with the pseudovector Born term subtracted) via 
\begin{equation} 
F_\pi^2 \, \bar D^+(0,2M_\pi^2) - \sigma(2M_\pi^2) = 
F_\pi^2 \, \bar D^+(0,2M_\pi^2) -\Delta \sigma - \sigma (0) = \Delta_R
= M_\pi^4 \, C_R~, 
\end{equation}
with  $\Delta \sigma = \sigma(2M_\pi^2) - \sigma(0)$. 
The crucial statement of the low--energy theorem is that the 
remainder $\Delta_R$ grows quadratically with the light quark mass,
because $C_R$ is a quark mass independent constant. 
At the Cheng--Dashen point $\nu = 0,\, s=m^2,\, t = 2M_\pi^2$,\cite{cd} 
\begin{equation} 
\bar D^+(0,2M_\pi^2) = A^+(m^2,2M_\pi^2) - {g_{\pi N}^2 \over m} \,\,
\, ,
\end{equation}
in terms of the standard isospin even $\pi$N scattering amplitude $A^+
(\nu,t)$ and the last term is due to the subtraction of the pseudovector tree amplitude.
Furthermore, the nucleon scalar form factor $\sigma(t)$ is given by the matrix
element  
\begin{equation}
\langle N(p') |\hat m(\bar u u+\bar dd)|N(p)\rangle 
= \bar u(p') u(p)\, \sigma(t)~,\quad 
t = (p-p')^2 \,\, \, ,
\end{equation}
with $\hat m = (m_u+m_d)/2$ the average light quark mass.
Although the Cheng--Dashen point is not in the
physical region of the $\pi$N scattering process, it lies well
within the Lehmann ellipse and thus $\bar D^+(0,2M_\pi^2)$ can 
 be obtained by analytic
continuation, i.e. using dispersion relations. Employing the only available 
coherent phase shift analyses, KA80~\cite{KA80} and KA85,\cite{KA85}
leads to $F_\pi^2 \,\bar D^+(0,2M_\pi^2) = 60 \, (62) \,
$MeV.\cite{gls} 
For a discussion of the 
uncertainties (typically $\pm 8$ MeV) and previous determinations, 
I refer to.\cite{gls}  There is also on--going debate about the size of this
quantity and I refer to Ref.95 
for the various opinions.
The leading non--analytic contribution to the
scalar form factor difference, $\Delta \sigma$, is $3g_{\pi N}^2
M_\pi^3 /(64 \pi m^2) $ and gives about 8 MeV.\cite{pp} \cite{GSS} 
A  detailed dispersive analysis~\cite{gls}, with $\pi \pi$ and
$\pi$N  information consistent with chiral symmetry, yields $\Delta \sigma =
15.2 \pm 0.4 \,$MeV. 
The remainder $\Delta_R$ is not fixed by chiral
symmetry. It has to be known, however, in order to extract information on the
$\sigma$--term, i.e. $\sigma(0)$, and from it the strangeness content of the
proton, i.e. the matrix element $\langle p|\, \bar s s \,|p\rangle$,
via the so--called strangeness fraction
\beq
y = \frac{2 \langle p|\, \bar s s \,|p\rangle }
{\langle p|\, \bar uu + \bar d d \,|p\rangle}~.
\eeq 
The strangeness fraction can also be related to the baryon mass
spectrum via $\sigma = \hat m \, (m_\Xi -m_\Sigma -2m_N)  /(m_s-\hat m)/
(1-y) +{\cal O}(m_q^{3/2})$.
The remainder $\Delta_R$ has been evaluated to order $q^4$ in HBCHPT in.\cite{BKMcd}
It was proven that $C_R$ has $no$ chiral logarithm and thus it is finite 
in the chiral limit. To this order, only local contact terms
contribute to the remainder. The result
based on the complete $q^4$ CHPT calculation with the low energy constants of
the pertinent counterterms saturated via resonance exchange is
\begin{equation} 
\Delta_R \approx 2 \, {\rm MeV} \,\,\, ,
\end{equation}
which should be considered an  upper bound. As conjectured in Ref.94  
the remainder $\Delta_R$ indeed does not play any role in the extraction of the
$\sigma$--term from the $\pi$N data considering the present status of accuracy
of these data in the threshold region. Thus, the analysis of
Ref.94  
leads to
\beq
\sigma (0) \simeq 45\,{\rm MeV}~, \quad \Delta \sigma \simeq 15\,{\rm
  MeV}~, \quad y \simeq 0.2~.
\eeq
This value for the sigma term has been confirmed using a different
method in.\cite{paul} Of course, the KA80 and KA85 phase shift
analyses are based on some data which are considered inconsistent
right now. All modern data are included in the partial wave analysis
of the GW/VPI group and reevaluations of the sigma term based on that
should be available in the near future. However, it is mandatory that
the same precise machinery, i.e. combining dispersion relations with
chiral symmetry constraints, is used to obtain reliable results. This
point is particularly stressed in Ref.98.   
In the case of isospin violation, one has of course
to differentiate between the proton and the neutron $\sigma$--terms, as
detailed in Ref.61. 
There it was shown that the third order
effects can shift the proton $\sigma$--term by about 8\% and have a smaller
influence on the shift to the Cheng--Dashen  point. The
isospin violating corrections to this shift to fourth
order were evaluated in Ref.62 
and were found to be
very small. However, it was also shown that there is an isospin
conserving electromagnetic contribution to the shift which can be 
as large as $\pm 2\,$MeV, which is a substantial electromagnetic effect.
What is still missing is a complete reanalysis of the LET~\cite{bpp}
in the presence of strong and electromagnetic isospin violation.

\subsection{Three--point functions II: Electromagnetic form factors}
\label{sec:ff}

\begin{figure}[tbh]
\hskip .5in
\epsfysize=2.7in
\epsffile{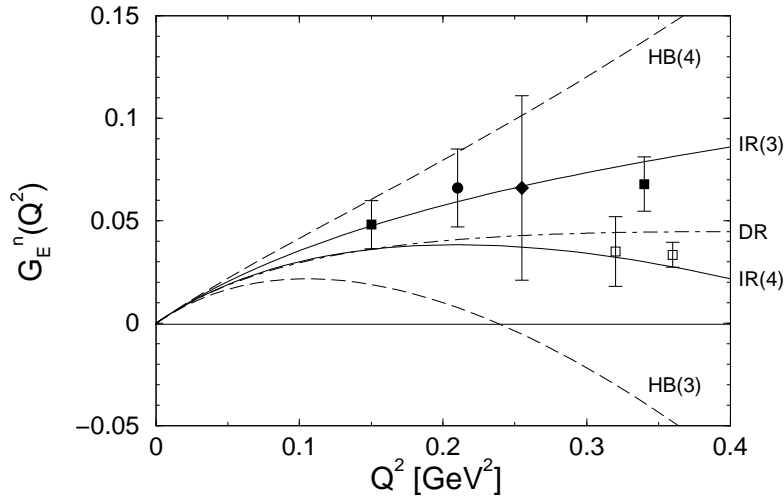}
\vspace{-0.1cm}
 \caption{Electric form factor of  the neutron.
          The dashed lines refer to the HBCHPT and the
          solid ones to the IR regularized relativistic calculation. The
          order is indicated by the number on each line. The dot--dashed
          line is the dispersion theoretical analysis of Ref.50. 
          The data are from MAMI, BATES and NIKHEF involving polarized $^3$He
          targets or polarization transfer on deuterium.
          \label{fig:GE} }
\end{figure}

\begin{figure}[htb]
\hskip .5in
\epsfysize=2.7in
\epsffile{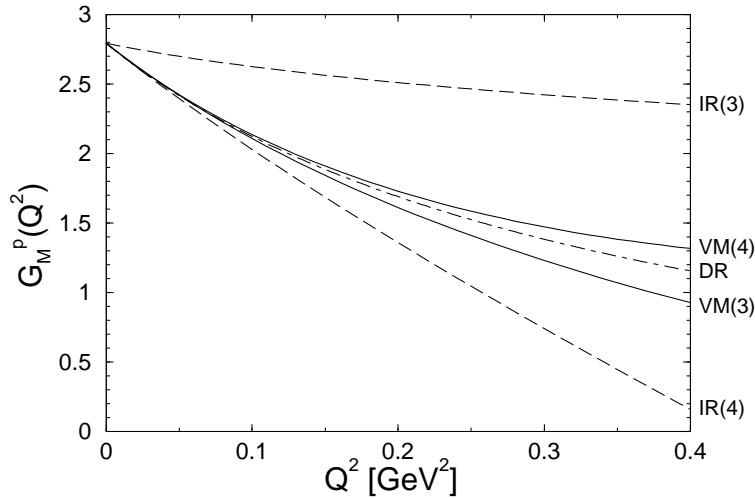}
\vspace{-0.1cm}
\caption{Magnetic form factor of the proton.
         The dashed lines refer to the 
         IR regularized relativistic calculation. The solid lines
         are obtained by adding in vector mesons in a chirally invariant
         way and readjusting the LECs to the radius. The chiral order
         is indicated by the number on each line. The dot--dashed
         line is the dispersion theoretical analysis of Ref.50. 
          \label{fig:GM} }
\end{figure}

The structure of the nucleon  as probed by virtual photons
is parameterized in terms of four form factors (here, $N$ is the fully
relativistic spin--1/2 field),
\beq
\langle N(p')\, | \, {\cal J}_\mu \,  | \, N(p) \rangle
= e \,  \bar{u}(p') \, \biggl\{  \gamma_\mu F_1^{N} (t)
+ \frac{i \sigma_{\mu \nu} k^\nu}{2 m} F_2^{N} (t) \biggr\} 
\,  u(p) \,, \quad N=p,n \,,
\eeq
with $t = k_\mu k^\mu = (p'-p)^2$ the invariant momentum 
transfer squared and ${\cal J}_\mu$ 
the em current related to the photon field.
In electron scattering, $t$ is negative and it is thus convenient 
to define the positive
quantity $Q^2 = -t > 0$. $F_1$ and $F_2$ are called the Pauli and the Dirac
form factor (ff), respectively, with the normalizations $F_1^p (0) =1$,
$F_1^n (0) =0$, $F_2^p (0) =\kappa_p$ and $F_2^n (0) =\kappa_n$. 
Also used are the electric and magnetic Sachs ffs,
\beq
G_E = F_1 - \tau F_2 \, , \quad G_M = F_1 + F_2 \, ,
 \quad \tau =Q^2/4m^2 \,.
\eeq
In the Breit--frame, $G_E$ and $G_M$ are nothing but the Fourier--transforms
of the charge and the magnetization distribution, respectively. These are
the natural quantities for the heavy nucleon scheme.\cite{BKKM}
There exists already a large body of data
for the proton and also for the neutron. 
In the latter case, one has to perform
some model--dependent extractions to go from the deuteron
or $^3$He to the neutron. 
There are also data in the time--like region from the reactions $e^+ e^-
\to p \bar{p} , n \bar{n}$ and from annihilation $p \bar{p} \to
e^+ e^-$, for $t \ge 4m_N^2$. A largely model--independent analysis of this
body of data is based on dispersion theory using
an unsubtracted dispersion relation for $F(t)$ (which
is a generic symbol for any one of the four ff's),
\beq
F(t) = \frac{1}{\pi} \int_{t_0}^\infty \, dt' \, \frac{{\rm Im} \, F(t)}{t'-t}
\, \, , \eeq
with $t_0$ the two (three) pion threshold for the isovector (isoscalar) ffs.
Im~$F(t)$ is called the {\it spectral}
{\it  function}. It is advantageous to work in
the isospin basis, $F_i^{s,v} = F_i^p \pm F_i^n$, since the photon
has an isoscalar ($I=s$) and an isovector ($I=v$) component. These spectral 
functions are the natural meeting ground for theory and experiment, 
like e.g. the partial
wave amplitudes in $\pi N$ scattering. In general, the spectral functions 
can be thought  of as a superposition of vector meson poles and some 
continua, related to n-particle thresholds, like
e.g. $2\pi$, $3\pi$, $K \bar{K}$, $N\bar{N}$ and so on. 
For example, in the Vector Meson
Dominance (VMD) picture one simply retains a set of poles.
It is important to realize that  there are some
powerful constraints which the spectral functions have to obey. 
Consider the spectral functions just above threshold.
Here, {\it unitarity} plays a central role. As pointed out by Frazer and 
Fulco  \cite{frafu}
long time ago, extended unitarity leads to a drastic enhancement 
of the isovector spectral functions on the left wing of the $\rho$
resonance. This is due to a logarithmic singularity on the second
Riemann sheet at the {\it anomalous} threshold $t_c = 4M_\pi^2 - 
(M_\pi^4 / m^2) = 3.98 M_\pi^2$,
very close to the normal threshold $t_0 = 4M_\pi^2$. Leaving out this
contribution from the two--pion cut leads to a gross underestimation of the
isovector charge and magnetic radii. This very fundamental constraint is 
often overlooked. In the framework of chiral pertubation theory, this 
enhancement is also present at the one--loop level as first shown in
Ref.21. 
It is also correctly given when using the method of infrared 
regularization,\cite{KuMe} but distorted in the strict heavy baryon approach
as detailed in Ref.101.   
The question whether a similar phenomenon appears in
the isoscalar spectral function has been answered in HBCHPT.\cite{bkmff} 
For that, one has to consider certain two--loop graphs.
Although the analysis of Landau equations reveals a branch point on 
the second Riemann sheet close to the threshold at $t_0 = 9M_\pi^2$,
the three--body  phase factors suppress its influence
in the physical region. Consequently, the spectral functions rise smoothly
up to the $\omega$ pole and the common practice of simply retaining vector
meson poles at low $t$ in the isoscalar channel is justified.
The ffs have been worked out to third order in Refs.23,82 
and to fourth order in\cite{HGKun} as shown by the dashed lines
for the electric ffs in Fig.\ref{fig:GE}.  
At third order, one has
two electric counterterms which can be fixed from fitting the proton and the
neutron charge radius and two counterterms which are fixed from the
magnetic moments of the proton and the neutron. To that order, the 
momentum dependence of the magnetic ffs is free of
counterterms. At fourth order, one has two additional magnetic counterterms,
which allow to fix the proton and neutron magnetic radius.
Still,  only for momentum
transfer $Q^2 < 0.2\,$GeV$^2$ one gets a decent description for the
large ffs ($G_E^p$, $G_M^p$ and $G_M^n$) whereas the relative
magnitude of the fourth order
corrections to the neutron electric ff becomes large at even
smaller $Q^2$. As can be seen from the figure, the use of  IR 
regularization (which has no additional counterterms) clearly 
improves the situation for $G_E^n$, not only is the prediction much
closer to the data up to $Q^2 \simeq  0.4\,$GeV$^2$ but also the
difference between the third and fourth order is sizeably reduced
up to higher $Q^2$. However, the predictions for the large 
(dipole--like) ffs are not
very much affected. The coupling of vector mesons
in a chirally invariant way allows to fix this problems without any
new parameters (these can be taken from existing determinations based
on dispersion theory~\cite{MMD,hmd}), as
shown in Fig.\ref{fig:GM} for the magnetic ff of the proton. 
Very similar results are found for the electric proton and the magnetic
neutron form factors. On the other hand, the good prediction for the
neutron electric form factor is not destroyed (because the vector
meson contributions largely cancel in this case). To be more precise,
one can formulate an effective chiral Lagrangian including spin--1 fields,\cite{BoMe}
which then allows to substitute the local contact terms by explicit vector meson
dynamics and a remainder, which is refitted as in the case of the pure
pion--nucleon Lagrangian and turns out to be small for most but not
all operators (chiral VMD).
For details, the reader should consult Ref.100.
Clearly, the pion cloud plays an important role in a
precise description of these fundamental nucleon structure observables
but needs to be supplemented by vector meson dynamics.

\begin{figure}[htb]
\hspace{2.2cm}
\psfig{figure=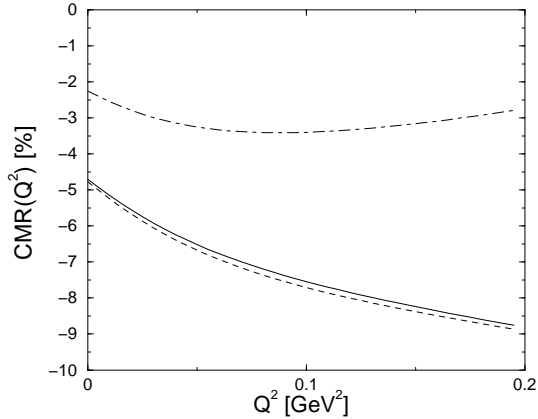,width=7.cm}
\vspace{-0.2cm}
\caption{The ratio $C_2/M_1$ versus $Q^2$. The solid, dashed and
  dot--dashed line refer to input from the Mainz, RPI and VPI
  multipole analysis, in order.}
\label{fig:cmr}
\end{figure}

\medskip\noindent
Within the framework of the SSE, discussed in section~\ref{sec:SSE},
one can also calculate form factors related to the delta or to
nucleon--delta transitions. In Ref.83, 
the isovector $N\Delta$--transition was
calculated to third order in the SSE. The transition matrix element
is parametrized in terms of three form factors $G_{1,2,3} (Q^2)$.
To that order, one has only two non--vanishing and finite loop
diagrams and all counterterms are momentum--independent. That means
that the $Q^2$--dependence of the transition ffs is predicted in a
parameter--free way and thus is a good testing ground for chiral dynamics.
Since the intermediate $\pi N$ state can go on mass--shell, these
ffs are complex, even at $Q^2=0$. That is not accounted for in most
models. These ffs can be mapped uniquely onto the multipole
amplitudes $M_1 (Q^2)$, $E_2 (Q^2)$ and $C_2 (Q^2)$,
measurable in pion photo/electroproduction. Consequently,
the $Q^2$--dependence of the so--called EMR $E_2/M_1$ and of the 
so--called CMR $C_2/M_1$
can be predicted. In Fig.\ref{fig:cmr} the momentum dependence of the
CMR is shown (for a similar plot for the EMR, see~\cite{george}). 
The various lines refer to the multipole analysis of the
Mainz, RPI and VPI groups, which are used to pin down the LECs at
$Q^2=0$. A more detailed discussion of these topics can be found
in.\cite{george} It will be important to confront the recent
measurements from ELSA and BATES with these predictions.
\noindent

\subsection{Three--point functions III: Axial form factors}

\begin{figure}[hbt]
\centerline{
\psfig{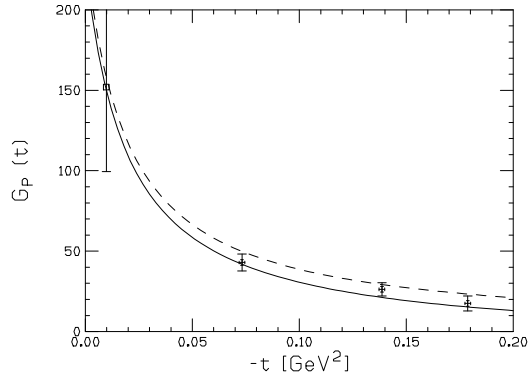}
}
\caption{
         The ``world data'' for the induced pseudoscalar form factor
         $G_P(t)$.  Dashed curve: Pion--pole prediction. Solid curve:
         Full chiral prediction. The pion electroproduction data
         (crosses) are from Ref.108. 
         Also shown is the OMC result from Ref.107   
         (open square).
\label{fig:gp}
}
\end{figure}

The structure of the nucleon as probed by charged axial currents is encoded in
two form factors -- the axial and the induced pseudoscalar ones,
denoted by $G_A$ and $G_P$, respectively. To be specific,
consider the matrix element of the isovector axial quark current, $A_\mu^a =
\bar{q}\gamma_\mu \gamma_5 (\tau^a/2)q$, between nucleon states,
\begin{equation}
\langle N(p')|A_\mu^a|N(p)\rangle = \bar{u}(p')\left[G_A (t)
\gamma_\mu +\frac{G_P(t)}{2m_N}(p'-p)_\mu \right] \gamma_5 {\tau^a \over 2}
u(p)~, \label{Acorr} 
\end{equation}
where $t=(p'-p)^2$ is the invariant momentum transfer squared. 
In the axial matrix element  one has assumed the absence
of second class currents (in agreement with the data). While $G_A (t)$ can be
extracted from (anti)neutrino--proton scattering or charged pion
electroproduction data, $G_P(t)$ is harder to pin down (from muon capture
or pion electroproduction) and in fact
constitutes the least known nucleon form factor.
Consider first the axial form factor. To one loop accuracy, it is given
by $G_A(t) = g_A (1  + t \langle r_A^2 \rangle /6)$, with the axial
radius containing one LEC. Still, CHPT has been used to make an interesting
prediction, namely that this radius as extracted from charged pion electroproduction
is different from the one obtained by analyzing neutrino--proton scattering
data by a computable and unique loop correction,\cite{BKMax}
\beq
\Delta \langle r_A^2 \rangle = \frac{3}{64F_\pi^2}\left(1-\frac{12}{\pi^2}
\right) = -0.046\,{\rm fm}^2~,
\eeq
subject to  small fourth order corrections.\cite{bkmrev} Translated into a dipole
mass via $ \langle r_A^2 \rangle = 12/M_A^2$, this amounts to a shift in the
axial mass by $\Delta M_A = 0.056\,$GeV. This prediction  has recently been 
verified at MAMI,\cite{MAMIax} they find  
$\Delta M_A = (0.043 \pm 0.026)\,$GeV (for more details, I
refer to that paper). Consider now $G_P(t)$.
Based on the chiral Ward identities of QCD, a very precise prediction for
the induced pseudoscalar form factor was given in~\cite{BKMgp} (see that
paper for a brief discussion on the history of that quantity)
\begin{eqnarray}
G_P(t) &=& \frac {4 m_N g_{\pi N} F_\pi}{M_\pi^2 - t }-\frac{2}{3}
g_A m_N^2 r_A^2\; , \label{gpff}
\end{eqnarray}
which leads to a very accurate prediction for the induced pseudoscalar
coupling constant, $g_P = (M_\mu/2m)G_P(t=-0.88M_\mu^2) = 8.44 \pm 0.23$,
with $M_\mu$ the muon mass. The theoretical uncertainty is almost entirely
due to our poor knowledge of the pion--nucleon coupling constant, $g_{\pi N}
= 13.05 \ldots 13.40$.
The first term on the right hand side of Eq.(\ref{gpff}) is nothing but the
pion pole (current algebra) result. This prediction agrees with the much
less precise experimental determination extracted from ordinary muon capture 
(OMC).\cite{baldin} It has, however, been challenged by the recent
TRIUMF result deduced from radiative muon capture (which is based on
a Born term model~\cite{BF1} to extract $g_P$ from the photon spectrum), 
$g_P^{\rm RMC} = 12.35 \pm 0.88 \pm 0.38 = 1.46 \, g_P^{\rm CHPT}$.
This has spurred some sizeable theoretical activity, see
e.g. Refs.111,112,113. 
Although this puzzle is not yet
resolved, the most promising ideas towards its solution are spelled out 
in.\cite{BHMmc} Much like in the case of the pion--nucleon sigma term,
a sum of various small effects seems to make up this sizeable discrepancy.
In any case, the Born model analysis underlying the extraction is certainly
very doubtful. Next, let me discuss the momentum dependence of $G_P$.
In Fig.\ref{fig:gp} the difference between the usual pion-pole parameterization for
$G_P(t)$ and the full chiral structure of the form
factor is displayed.  The data shown in that figure are from
one pion electroproduction experiment~\cite{choi} and one data point is from 
OMC.\cite{baldin} In this kinematical regime
the structure proportional to $r_A^2$ produces the biggest effect and a new
dedicated experiment should be able to verify it. In fact,
at the Mainz Microtron MAMI-B an experiment has been
proposed to measure the axial and the induced pseudoscalar form
factors by means of charged pion electroproduction at low momentum
transfer.\cite{MAMIgp}

\subsection{Four--point functions I: Pion--nucleon scattering}
\label{sec:piN}
\begin{figure}[p]
\centerline{
\psfig{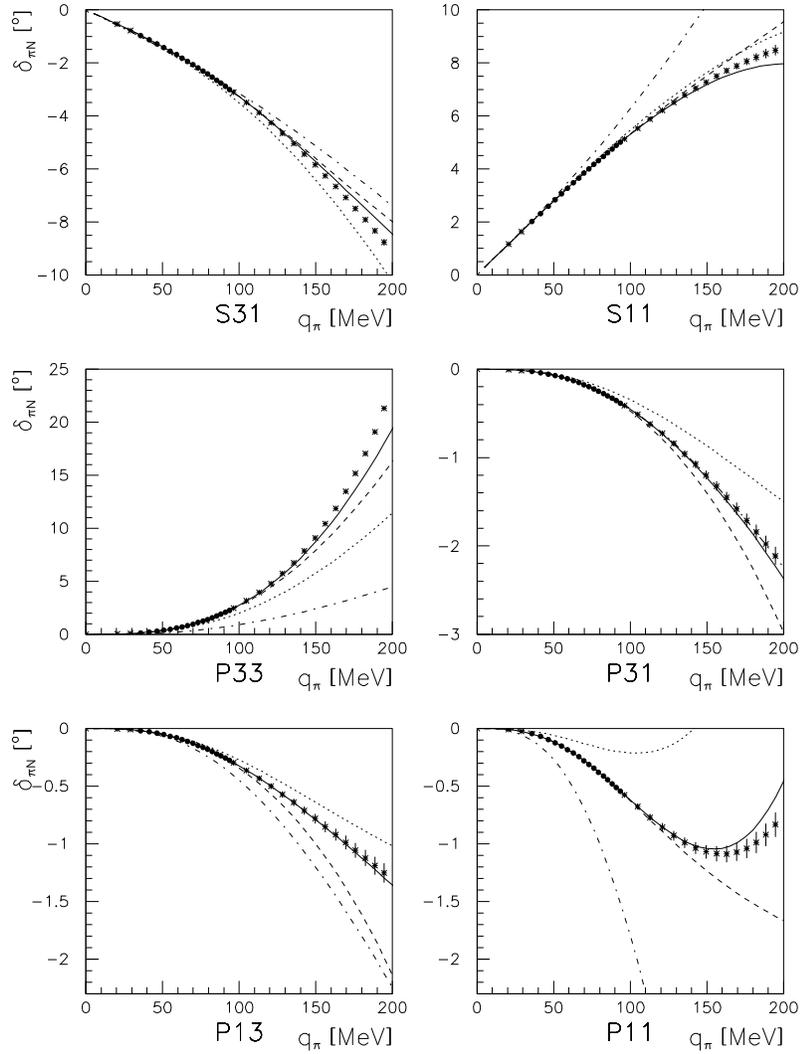}
}
\caption{HBCHPT description of the S-- and P--wave phase shifts as
  described in the text. The data are from Ref.118. 
\label{fig:pin}}
\end{figure}

In the following, we consider elastic $\pi N$ scattering.
In the center-of-mass system (cms)  the amplitude for the process
$\pi^a(q_1) + N(p_1) \to \pi^b(q_2) + N(p_2)$ takes the
following form (in the isospin basis): 
\beqa 
T^{ba}_{\pi N} &=& 
\biggl(\frac{E+m}{2m}\biggr) \, \biggl\lbrace \delta^{ba} 
\Big[ g^+(\omega,t)+ i \vec
\sigma \cdot(\vec{q}_2\times \vec{q}_1\,) \, h^+(\omega,t) \Big]
\nonumber \\ && \qquad\quad
+i \, \epsilon^{bac}
\tau^c \Big[ g^-(\omega,t)+ i \vec \sigma \cdot(\vec{q}_2 \times \vec{q}_1\,) \,
h^-(\omega,t) \Big] \biggr\rbrace ~,
\eeqa
with $\omega = v\cdot q_1 = v\cdot q_2$ the pion cms energy, $t=(q_1-q_2)^2$ 
the invariant momentum transfer squared, $E_1 = E_2 \equiv E = ( \vec{q\,}^
2 + m^2)^{1/2}$ the nucleon energy and
$\vec{q\,}_1^2 = \vec{q\,}_2^2 \equiv
\vec{q\,}^2 = ((s-M^2-m^2)^2 -4m^2M^2)/(4s)$.
Furthermore, $g^\pm(\omega,t)$ refers to the
isoscalar/isovector non--spin--flip amplitude and $h^\pm(\omega,t)$ to the
isoscalar/isovector spin--flip amplitude. This form is most suitable
for the HBCHPT calculation since it is already defined in a
two--component framework.\footnote{These amplitudes should not be
called ``non-relativistic''. They are, however, more suitable for our
discussion than the amplitudes $A^\pm (s,t) ,B^\pm (s,t)$ which arise from a
manifest Lorentz--invariant formulation of the $\pi N$ scattering
amplitude.}  These amplitudes can then be projected onto partial waves
by standard methods and at threshold lead to the pertinent scattering 
lengths $(a^\pm_{l\pm}$) and effective ranges $(b^\pm_{l\pm}$), with
$l$ the orbital angular momentum and the total angular momentum
$j=l\pm 1/2$. 
Pion--nucleon scattering has played a very important role in
establishing chiral symmetry in hadron physics as discussed before.
The first corrections to Weinberg's CA prediction for the S--wave
scattering lengths were considered in Ref.55 
including third order loop graphs and contact interactions. The LECs were
estimated by resonance saturation. In Ref.116 
it was further shown that the dominant correction to the Weinberg--Tomozawa result
for the isovector S--wave scattering length is a pure loop effect.
More systematic third order studies were performed in
Refs.20,37,117. 
Finally, in Ref.88   
the complete fourth--order one--loop results were presented. At this order, one has 14 LECs,
which were determined by a fit to the existing S-- and P--wave phases from
various partial wave analyses for pion momenta between 50 and 100~MeV.
This allows to make predictions for the threshold parameters and the
phase shifts at higher energies. A typical result is shown in
Fig.\ref{fig:pin}, where the dotted lines give the first order result
and the dot-dashed, dashed and solid lines refer to the best fit at
second (4 LECs), third (9 LECs) and fourth (14 LECs) order. Clearly,
one observes convergence as long as the pion momenta stay below 200~MeV.
Of particular interest are the predictions for the the S--wave scattering 
lengths  $a^\pm_{0+}$. We find $-1.0 \le a^+_{0+} \le 0.6$ and $8.3
\le a^-_{0+} \le 9.3$ in units of $10^{-2}/M_\pi$. The  uncertainties  are 
mostly due to the input, i.e. to the various phase shift
analyses in the physical region, and not to the  theory. 
It is instructive to study the convergence, using e.g. the KA85 phases as
input, one finds
\beq
\begin{array}{c|rrrr}
         & {\cal O}(p) & {\cal O}(p^2) & {\cal O}(p^3) & {\cal O}(p^4)
         \\ \hline
a^+_{0+} & 0.0         & 0.46          &  -1.00        &  -0.96 \\
a^-_{0+} & 7.90        & 7.90          &   9.05        &   9.03 \\
\end{array}
\eeq
where ${\cal O}(p^n)$ means that all terms up--to--and--including order
$n$ are included. Note that the fourth order correction for the isoscalar scattering
length is not large and that the small difference between the third
and fourth order results for the isovector one is due to the
readjustment of a LEC. Translated  into the physical channels, we have
\beq
\begin{array}{lll}
a_{\pi^- p \to \pi^0 n} = & -0.131 \ldots -0.117 \, M_{\pi}^{-1} \quad
                            &
                            [(-0.128 \pm 0.006) \, M_{\pi}^{-1}]~,\\
a_{\pi^- p \to \pi^- p} = & $\,\,\,\,\,$ 0.073 \ldots $\,\,\,\,\,$
                            0.093 \, M_{\pi}^{-1} \quad & 
                            [(0.0883 \pm 0.0008) \, M_{\pi}^{-1}]~,\\
\end{array}
\eeq
where the experimental numbers (in the brackets) are taken from
Ref.119. 
It is worth emphasizing that the theoretical
prediction is of similar precision as  the experimental one.
\noindent
Higher threshold parameters can be predicted without any free
parameters, these are collected in table~\ref{tab:piNDF}. The
agreement with the determination from dispersion theory is mostly
satisfactory. However, the HBCHPT representation of the amplitude is not precise enough
to also give stable predictions for the so--called subthreshold
parameters when the LECs are determined in the physical region. 
On the other hand, using dispersion theory one can
determine the LECs from a fit within the interior of the Mandelstam
triangle.\cite{paul} In particular, the LEC $c_1$, which governs the 
size of the sigma term, comes out very different to the determinations
from fitting the phase shifts. It is hoped that the approach based on
the IR regularization will give better results in this context.
\renewcommand{\arraystretch}{1.2}
\begin{table}[htb]
\caption{\label{tab:piNDF}
D-- and F--wave threshold parameters predicted by 
CHPT\protect \cite{BKMLEC}. The order
of the prediction is also given together with the "experimental
values", which  come from the Karlsruhe--Helsinki dispersive 
analysis.\protect\cite{hoeh}}
\begin{center}
\begin{tabular}{|c|c|c|c|c|}
    \hline  
    Obs. & CHPT &  Order & Exp. value & Units \\
    \hline
$a^+_{2+}$ &  $-1.83$ & $p^3$  & $ -1.8 \pm 0.3$  & $10^{-3}\, M_\pi^{-5} $ \\
$a^+_{2-}$ &  $2.38$  & $p^3$  & $ 2.20 \pm 0.33$ & $10^{-3}\, M_\pi^{-5} $ \\
$a^-_{2+}$ &  $3.21$  & $p^3$  & $ 3.20 \pm 0.13$ & $10^{-3}\, M_\pi^{-5} $ \\
$a^-_{2-}$ &  $-0.21$ & $p^3$  & $ 0.10 \pm 0.15$ & $10^{-3}\, M_\pi^{-5} $ \\
   \hline
$a^+_{3+}$ &  $0.29$  & $p^3$  & $ 0.43 $          & $10^{-3}\, M_\pi^{-7} $ \\
$a^+_{3-}$ &  $0.06$  & $p^3$  & $ 0.15 \pm 0.12$  & $10^{-3}\, M_\pi^{-7} $ \\
$a^-_{3+}$ &  $-0.20$ & $p^3$  & $ -0.25 \pm 0.02$ & $10^{-3}\, M_\pi^{-7} $ \\
$a^-_{3-}$ &  $0.06$  & $p^3$  & $ 0.10 \pm 0.02$  & $10^{-3}\, M_\pi^{-7} $ \\
   \hline   
  \end{tabular} \end{center}
\end{table}

\subsection{Four--point functions II: Real and virtual Compton scattering}

As discussed before, the low--energy Compton scattering amplitude
can be decomposed into a spin--independent, called $f_1$, and a spin--dependent part,
denoted by $f_2$. In forward direction, the T--matrix takes the form
\beq
T(\gamma N \to \gamma N) = f_1 (\omega) \, {\vec\epsilon \, '}{}^* \cdot \vec\epsilon
+ f_2 (\omega) \, \omega \, i \vec{\sigma} \cdot ({\vec\epsilon \, '}{}^* \times \vec\epsilon )
\eeq
in terms of the photon polarization four--vectors ($\vec\epsilon, {\vec\epsilon \, '}$),
the photon energy $\omega$ and the nucleon spin matrix $\vec\sigma$.
In the spin--independent case, the nucleon structure is encoded in the so--called
electromagnetic polarizabilities, denoted $\bar\alpha$ (electric)
and $\bar\beta$ (magnetic),
\beq
f_1 (\omega ) = {e^2 Z^2\over m} + (\bar\alpha + \bar\beta) \omega^2 + {\cal O}
(\omega^4)~,
\eeq 
with the first term being the Thomsom amplitude discussed before.
These polarizabilities have been measured over the years at
Illinois, Mainz, Moscow, Oak Ridge and Saskatoon. 
Calculations have been performed to fourth order with some LECs
determined from data and others from resonance saturation. 
The leading third order one--loop result, which is free of any LEC and
solely due to finite loop graphs,
has been given a decade ago,\cite{bkmpola,BKKM}
\beq
\bar\alpha_p = \bar\alpha_n = 10\bar\beta_p = 10\bar\beta_n = 
{5e^2 g_A^2 \over 384 \pi^2 F_\pi^2 M_\pi} =
12.5\cdot 10^{-4}~{\rm  fm}^3~,
\eeq
which describes the trend of the data, namely that the electric
polarizability is considerably larger than the magnetic one and also
that the sum of both spin--independent polarizabilities
 is approximately the same for the proton and the neutron.
This latter statement is based on  the so--called Baldin sum rule, which relates the
sum of the electric and magnetic polarizabilities to an integral
over the total photoabsorption cross section. Older and more recent 
evaluations\cite{Petrunkin,Mato,Lev} give
$(\bar{\alpha} +\bar{\beta})_p = (14.2\pm 0.5),(13.69\pm 0.14),(14.0\pm 0.5)
\cdot 10^{-4}$~fm$^3$ and
$(\bar{\alpha} +\bar{\beta})_n = (15.8\pm 0.5),(14.40 \pm 0.66),(15.2\pm 0.5)
\cdot 10^{-4}$~fm$^3$.
The third order investigation has then be supplemented by a fourth order calculation,
mostly triggered by the suspicion that the result for the magnetic
polarizabilities, which are believed to receive a large order $p^4$ contribution
from the $\Delta$ resonance, is accidental. A summary of the predictions as compared
to the data for the proton reads
\beq
\begin{array}{crr} 
\bar{\alpha}_p = & (10.5\pm 2.0)\cdot 10^{-4} \, {\rm fm}^3
               & [(12.1\pm0.8\pm0.5) \cdot 10^{-4} \, {\rm fm}^3]~, \\ 
\bar{\beta}_p = & (3.5\pm 3.6) \cdot 10^{-4} \, {\rm fm}^3  
               & [(2.1\mp0.8\mp0.5)  \cdot 10^{-4} \, {\rm fm}^3]~, \\
\end{array}
\eeq
The experimental numbers in the square brackets are taken from Ref.125.
While in case of the
proton a consistent picture is emerging,\cite{bkms} the published empirical
values deduced from scattering slow neutrons on heavy atoms have
recently been put into question.\cite{koes} Furthermore, recent measurements
of elastic and of quasi--free Compton scattering from the deuteron at SAL
lead to contradictory results for the neutron polarizabilities.\cite{Horn,Kolb}
Therefore, I refrain here from comparing the chiral predictions to
the data. This experimental problem remains to be sorted out. 
The most promising result is the almost complete 
cancellation of a negative  non--analytic pion loop contribution with the large
positive contribution of tree level $\Delta$--exchange in case of the
proton magnetic polarizability.\cite{bkms} This is the first time that a
consistent picture of the para-- and diamagnetic contributions to
$\bar{\beta}_p$ has been found and it underlines the importance of chiral,
i.e. pion loop, physics in understanding the nucleon structure as 
revealed in Compton scattering. 

\medskip\noindent
Furthermore, there exist chiral predictions for the four
spin--dependent polarizabilities, denoted $\gamma_{1,2,3,4}$.
 As of today, no direct measurements
exist but some indirect information based on multipole analysis points
towards the important role of the $\Delta (1232)$ as an active degree
of freedom to
understand some of these quantities. In table~\ref{tab:copo}, the
present status of the theoretical precdictions compared with the 
{\it indirect} determinations is given. For a cleaner separation of the
effects of third and fourth order, the isosaclar and isovector
contributions are given. These are related to the ones of the proton
and neutron via $\gamma_i^{(p)}=\gamma_i^{(s)}+\gamma_i^{(v)}$, and
$\gamma_i^{(n)}=\gamma_i^{(s)}-\gamma_i^{(v)}$. Note, however, that with the
advent of polarized targets and new sources with a high
flux of polarized photons, the case of polarized Compton scattering off
the proton, $\vec{\gamma}\,\vec{p}\rightarrow\gamma p$, has come close to
experimental feasibility. It thus remains an active field to further
sharpen these theoretical predictions.\cite{GHM} Also, as can be seen from
table~\ref{tab:copo}, there is large $\Delta$ contribution to the
spin--polarizabilities $\gamma_{2,4}$. In the conventional power
counting such effects only start to appear via counterterms at fifth
order. Thus, in such a case the explicit inclusion of the delta is
mandatory. Other  calculations of the spin polarizabilities can be
found in Refs.131,132.   
\renewcommand{\arraystretch}{1.1}
\begin{table}[htb]
\caption{\label{tab:copo}
Predictions for the spin-polarizabilities in HBChPT in comparison with the
dispersion
analyses of Refs.133,134,135 
and the ${\cal O}(\epsilon^3)$ results of the
small scale expansion \protect \cite{HHKK} (SSE1).
All results are given in the units of $10^{-4}\;{\rm fm}^4$.}
\medskip
\begin{center}
\begin{tabular}{|c||cc|c||cccc|}
\hline
 & ${\cal O}(p^3)$ & ${\cal O}(p^4)$ & Sum  & Mainz1 & Mainz2 & BGLMN & SSE1  \\
\hline
$\gamma_{1}^{(s)}$ & $+ 4.6$ & $-2.1$ & $+2.5$ & $+5.6$ &$+5.7$ &  $+4.7$
&$+4.4$\\
$\gamma_{2}^{(s)}$ & $+ 2.3$ & $-0.6$ & $+1.7$ & $-1.0$ &$-0.7$ &  $-0.9$
&$-0.4$\\
$\gamma_{3}^{(s)}$ & $+ 1.1$ & $-0.5$ & $+0.6$ & $-0.6$ &$-0.5$ &  $-0.2$
&$+1.0$\\
$\gamma_{4}^{(s)}$ & $- 1.1$ & $+1.5$ & $+0.4$ & $+3.4$ &$+3.4$ &  $+3.3$
&$+1.4$\\
\hline
$\gamma_{1}^{(v)}$ & - & $-1.3$ & $-1.3$ & $-0.5$ & $-1.3$ & $-1.6$ & - \\
$\gamma_{2}^{(v)}$ & - & $-0.2$ & $-0.2$ & $-0.2$ & $+0.0$ & $+0.1$ & - \\
$\gamma_{3}^{(v)}$ & - & $+0.1$ & $+0.1$ & $-0.0$ & $+0.5$ & $+0.5$ & - \\
$\gamma_{4}^{(v)}$ & - & $+0.0$ & $+0.0$ & $+0.0$ & $-0.5$ & $-0.6$ & - \\
\hline
\end{tabular}
\end{center}
\end{table}

\begin{figure}[htb]
\centerline{
\psfig{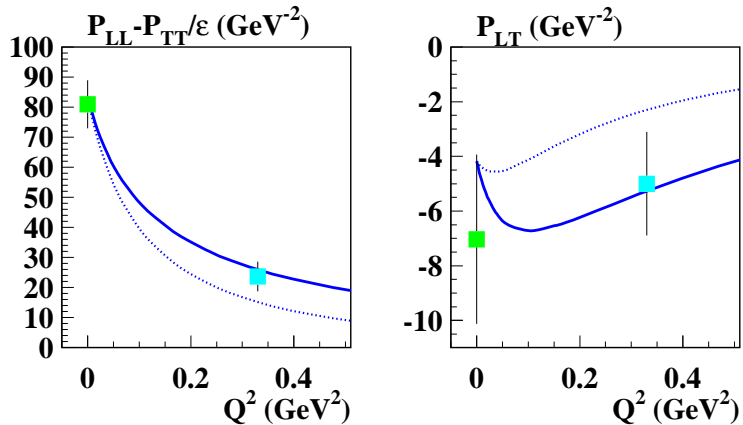}
}
\caption{The influence of the generalized polarizabilities on unpolarized
         VCS (solid line) in comparison to the MAMI measurement
         at $Q^2 = 0.33\,$GeV$^2$. Dashed line: Structure functions obtained
         by switching off the spin--dependent polarizabilities.
\label{fig:vcs}}
\end{figure}

\medskip\noindent
Virtual Compton scattering (VCS), in which one of the photons has a
non--vanishing virtuality, offers additional information about the
nucleon structure. The VCS amplitude can be parameterized in terms of
six so--called generalized polarizabilities (for a precise definition
I refer to Refs.136,137).   
In unpolarized electron scattering, one can extract two of these generalized
polarizabilities from the cross section after properly subtracting the
dominant contributions from the Bethe--Heitler process and nucleon
form factors. That this is indeed possible was recently demonstrated
at MAMI.\cite{MAMIVCS} Unpolarized VCS was measured at $Q^2 =
0.33\,$GeV$^2$. As shown in Fig.~\ref{fig:vcs}, the measured
structure functions compare favorably with the  third order heavy
baryon prediction of Ref.139.  
Clearly, the fourth order
corrections have to be calculated to solidify this result, eventually
also using the previously described relativistic formalism. Since
this field is very active both theoretically and experimentally
(MIT-Bates, MAMI, Jefferson Lab), we are looking forward to many
interesting developments. 
\noindent

\subsection{Four--point functions III: Pion photo-- and electroproduction}
\label{sec:photo}
\begin{figure}[htb]
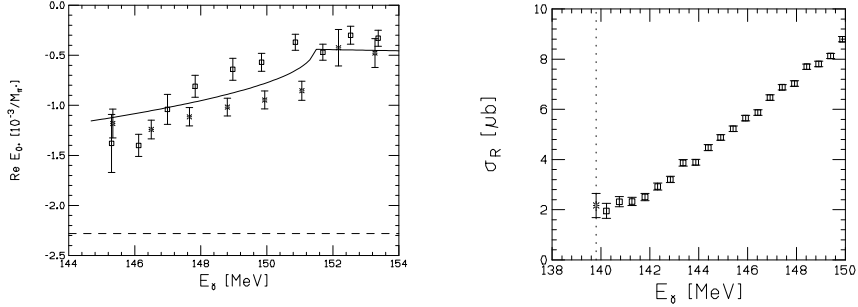

\vspace{6.cm}
\includegraphics{pi03.epsi}
\hspace{4.8cm}
\includegraphics{jack.epsi}
\vspace{-2.2cm}
\caption{Left panel: CHPT prediction (solid line) 
for the electric dipole amplitude
in   $\gamma p \to \pi^0 p$ in comparison to the  data  from MAMI and SAL.
The dashed line  gives the prediction of the incorrect ancient LET
discussed in section~\ref{sec:LET}. 
Right panel: Reduced total cross section for $\gamma d \to \pi^0 d$. The
data from SAL are depicted by the boxes, the CHPT threshold prediction is the
star on the the dotted line (indicating the threshold photon energy).
\label{fig:photo}} 
\end{figure}

\noindent Charged pion photoproduction at threshold is well described
in terms of the Kroll--Ruderman contact term, which is non--vanishing
in the chiral limit. All chiral corrections including the
third order in the pion mass have been calculated.\cite{bkmcp} The
chiral series is quickly converging and the theoretical error on the
CHPT predictions is rather small, see table~\ref{tab:char}. The LECs have been
determined from resonance exchange. 
Notice that these uncertainties do not account for the variations in 
the pion--nucleon coupling constant.
The available threshold data are quite  old, with
the exception of the recent TRIUMF experiment on the inverse reaction
$\pi^- p \to \gamma n$ and the SAL measurement for $\gamma p \to \pi^+ n$.
While the overall agreement is quite good for 
the $\pi^+ n$ channel, in the $\pi^- p$ channel the CHPT prediction
is on the large side of the data. Clearly, we need more precise data
to draw a final conclusion. It is, however, remarkable to have
predictions with an error of only 2\%$\,$.
\renewcommand{\arraystretch}{1.3}
\begin{table}[htb]
\begin{center}
\caption{Predictions and data for the charged pion 
electric dipole amplitudes in $10^{-3}/M_{\pi^+}$.\label{tab:char}}
\medskip
\begin{tabular}{|l|c|c|}
\hline
                                &  ${\cal O}(p^4)$ & Experiment  \\
\hline
$E_{0+}^{\rm thr} (\pi^+ n)$    & $28.2 \pm 0.6$   
                                & $27.9\pm 0.5$ \protect{\cite{burg}},
                                $28.8 \pm 0.7$ \protect{\cite{adam}},
                                $27.6 \pm 0.6$ \protect{\cite{salp}}   \\
$E_{0+}^{\rm thr} (\pi^- p)$    & $-32.7 \pm 0.6$   & $-31.4
                                \pm 1.3$ \protect{\cite{burg}}, 
                               $-32.2 \pm 1.2$ \protect{\cite{gold}}, 
                               $-31.5\pm 0.8$ \protect{\cite{triumf}}  \\
\hline
\end{tabular}
\end{center}
\end{table}
\noindent The threshold production of neutral pions is much more
subtle since the corresponding electric dipole amplitudes vanish in
the chiral limit. Space does not allow to tell the tale of the
experimental and theoretical developments concerning the electric
dipole amplitude for neutral pion production off protons, for details
see.\cite{ulf95} Even so the convergence for this particular
observable is slow, a CHPT calculation to order $p^4$ does allow to
understand the energy dependence of $E_{0+}$ in the threshold region
once three LECs are fitted to the total and differential cross section
data,\cite{bkme0p} see the left panel of Fig.\ref{fig:photo}. The threshold
value agrees with the data, see table~\ref{tab:e0plus0}.
Even more interesting is the
case of the neutron. Here, CHPT predicts a sizeably larger $E_{0+}$
than for the proton (in magnitude).
The CHPT predictions for $E_{0+} (\pi^0 p,\pi^0 n)$ in the threshold
region  clearly
exhibit the unitary cusp due to the opening of the secondary
threshold, $\gamma p \to \pi^+ n \to \pi^0 p$ and $\gamma n \to \pi^-
p \to \pi^0 n$, respectively. Note, however, that while $E_{0+} (\pi^0
p)$ is almost vanishing after the secondary threshold, the neutron
electric dipole amplitude is sizeable ($-0.4$ compared to $2.8$ in
units of $10^{-3}/M_{\pi^+}$).
The question arises how to measure the neutron amplitude?
The natural neutron target is the deuteron. The corresponding electric
dipole $E_d$ amplitude has been calculated to   order $p^4$ in Ref.148.
It was shown that the next--to--leading order
three--body corrections and the possible four--fermion contact terms
do not induce any new unknown LEC and one therefore can calculate
$E_d$ in parameter--free manner. Furthermore, the leading order
three--body terms are dominant, but one finds a good convergence for
these corrections and also a sizeable sensitivity to the elementary
neutron amplitude. Note also that neutral pion production off
deuterium has recently been measured at SAL. The CHPT prediction in
comaprison to the data for the reduced cross section of coherent
neutral pion production off deuterium is shown in the right panel of
Fig.\ref{fig:photo}.
\noindent
Still, there is more interesting physics in these channels.
Quite in contrast to what was believed for a long
time, there exist a set of LETs for the slopes of the P--waves
$P_{1,2} = 3E_{1+} \pm M_{1+} \mp M_{1-}$ at threshold, for example
\beq \frac{1}{|\vec q \,|} P_{1, {\rm thr}}^{\pi^0 p} = \frac{e g_{\pi
  N}}{8 \pi m^2} \left\lbrace 1 + \kappa_p + \mu \left[ -1 - 
\frac{\kappa_p}{2} + \frac{g_{\pi N}^2}{48 \pi}(10 -3\pi) \right]
+ {\cal O}(\mu^2) \right\rbrace \, \, . \eeq
Numerically, this translates into 
\beq \frac{1}{|\vec q \,|} P_{1, {\rm thr}}^{\pi^0 p} = 0.512 \, ( 1
-0.062) \, {\rm GeV}^{-2} = 0.480 \, {\rm GeV}^{-2} \, \, , \eeq
which is given in table~\ref{tab:e0plus0} 
in units which are more used in the literature. The agreement with
the data is stunning. A theoretical uncertainty on this order $p^3$ 
calculation can only be given when the next order has been calculated.
That calculation is underway. Soon, there will also be an experimental value
for $P_2$ at threshold once the MAMI data on $\vec{\gamma} p \to \pi^0
p$ have been analyzed.
\renewcommand{\arraystretch}{1.4}
\begin{table}[hbt]
\caption{Predictions and data for  neutral pion S-- and P--wave
         multipoles. Units are $10^{-3}/M_{\pi^+}$ and
         $|q||k|10^{-3}/M_{\pi^+}^3$ for the S--and P--waves,
         respectively. \label{tab:e0plus0}}
\begin{center}
\begin{tabular}{|l|c|c|c|}
\hline
                        & CHPT & Order & Experiment   \\
\hline
$E_{0+}^{\rm thr} (\pi^0 p)$    & $-1.16$ \protect{\cite{bkme0p}} & $p^4$ 
                                & $-1.31\pm 0.08$ \protect{\cite{fuchs}},
                                $-1.32\pm 0.05$ \protect{\cite{berg}}
                                \\
$P_{1}^{\rm thr} (\pi^0 p)$     & $10.3$ \protect{\cite{bkmcp}}  & $p^3$ 
                                & $10.02\pm 0.15$ \protect{\cite{fuchs}},
                                $10.26\pm 0.1$ \protect{\cite{berg}}
                                \\
$E_{0+}^{\rm thr} (\pi^0 d)$    & $-1.8\pm0.2$ \protect{\cite{bblmvk}}  
                                & $p^4$ & 
                                $-1.7 \pm 0.2$ \protect{\cite{argan}}, 
                                $-1.5 \pm 0.1$ \protect{\cite{sald}} 
                                 \\
 \hline
\end{tabular}
\end{center}
\end{table}

\begin{figure}[hbt]
\centerline{
\psfig{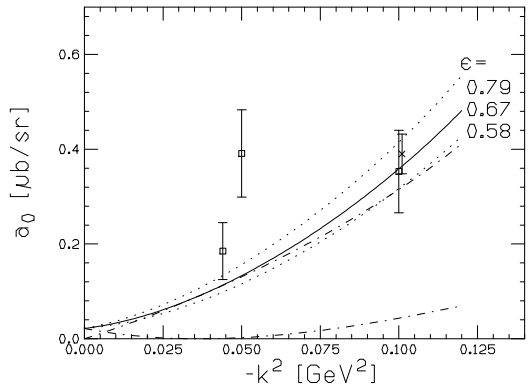}
}
\caption{The S--wave cross section $a_0$ compared to the data for
          various photon polrizations. The upper (lower) dash--dotted
          line gives the contribution of the longitudinal (electric)
          dipole amplitude to $a_0$ for $\epsilon = 0.79$ (solid
          line).\label{fig:a0}  }
\end{figure}

\medskip\noindent
Producing the pion with virtual photons offers further insight
since one can extract the longitudinal S--wave multipole $L_{0+}$ and
also novel P--wave multipoles. Data have been taken at
NIKHEF~\cite{welch} \cite{benno} and MAMI~\cite{distler} for
photon virtuality of $k^2 = -0.1$~GeV$^2$. In fact, it has been argued
previously that such photon four--momenta are already too large for
CHPT tests since the loop corrections are large.\cite{bklm} However,
these calculations were performed in relativistic baryon CHPT and thus
it was necessary to redo them in the heavy fermion formalism. This was
done in.\cite{bkmel} The abovementioned data for differential cross
sections were used to determine the three novel S--wave LECs. I should
mention that one of the operators used is of dimension five, i.e. one
order higher than the calculation was done. This can not be
circumvented since it was shown that the two S--waves are
overconstrained by a LET valid up to order $p^4$. The resulting
S--wave cross section $a_0 = |E_{0+}|^2 + \ve_L \,|L_{0+}|^2$ shown in  
Fig.~\ref{fig:a0} is in fair agreement with the data. Note also that it is
dominated completely by the $L_{0+}$ multipole (upper dot-dashed line)
since $E_{0+}$ passes through zero at $k^2 \simeq
-0.04$~GeV$^2$. However, in agreement with the older (and less precise)
calculations, the one loop corrections are large so one should compare at
lower photon virtualities. In Ref.157,  
many predictions for $k^2 \simeq -0.05$~GeV$^2$ are given. At MAMI, data have been taken in
this range of $k^2$ and we are looking forward to their analysis, in
particular it will be interesting to nail down the zero--crossing of
the electric dipole amplitude and to test the novel electroproduction P--wave 
LETs.\cite{bkmprl} I remark that the preliminary analysis of thes data at 
lower virtuality seems to indicate some inconsistency with the chiral
prediction if one uses the data at $k^2 = -0.1$~GeV$^2$ to pin down the LECs.

\subsection{Five--point functions I: Electromagnetic two--pion production}

\begin{figure}[hbt]
\centerline{
\psfig{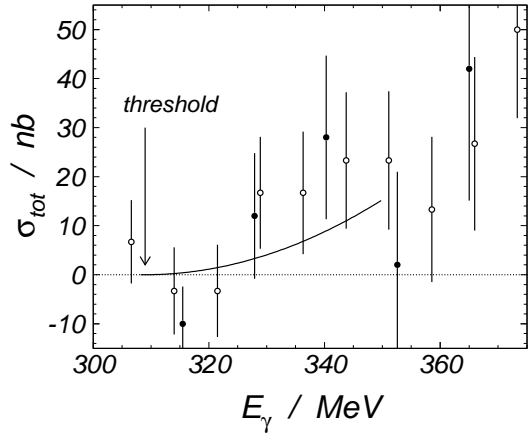}
}
\caption{Total cross section for the process $\gamma p \to \pi^0 \pi^0 p$
         as measured by the TAPS collaboration at MAMI.\protect\cite{TAPS2p}
         The solid line is the fourth order HBCHPT prediction. I am grateful
         to the TAPS collaboration for providing me with this figure.         
          \label{fig:tpi0}  }
\end{figure}

Electromagnetic production of two pions off a nucleon can be used to
study the excitation of certain nucleon resonances, in particular the
$\Delta (1232)$, the Roper $N^* (1440)$ or the $D_{15} (1520)$. However,
close to threshold one observes an interesting effect due to the chiral pion
loops of QCD. To be specific, consider the reaction $\gamma p \to \pi\pi N$,
where the two pions in the final state can both be charged, both neutral or
one charged and one neutral. To leading order in the chiral expansion, the
production of two neutral pions is strictly suppressed. However, at
next--to--leading order, due to finite chiral loops the production cross
section for final states with two neutral pions is considerably
enhanced.\cite{bkms2p}  Also, in a small window above threshold, the
potentially large contribution from double--delta excitation is strongly
suppressed, leaving a window in which one can detect much more neutrals than
expected. This prediction was further sharpened in Ref.160,  
where all fourth order corrections including the excitation of the Roper and its
successive decay into two neutral pions here considered. The predicted near
threshold cross section is
\beq
\sigma_{\rm tot} (E_\gamma ) \leq 0.91~{\rm nb}~\biggl({ E_\gamma -
  E_\gamma^{\rm thr} \over 10~{\rm MeV}} \biggr)^2~,
\eeq
with $E_\gamma^{\rm thr} = 308.8\,$MeV. This prediction can only be applied
for the first 30 MeV above threshold. A recent measurement by the TAPS 
collaboration~\cite{TAPS2p} has shown that such an enhancement of the
the 2$\pi^0$ production indeed happens, see Fig.\ref{fig:tpi0}, proving once
again the importance of pion loop effects, which can lead to rather unexpected 
predictions and results.

\subsection{Five--point functions I: $\pi N \to \pi \pi N$}

\begin{figure}[p]
\centerline{
\psfig{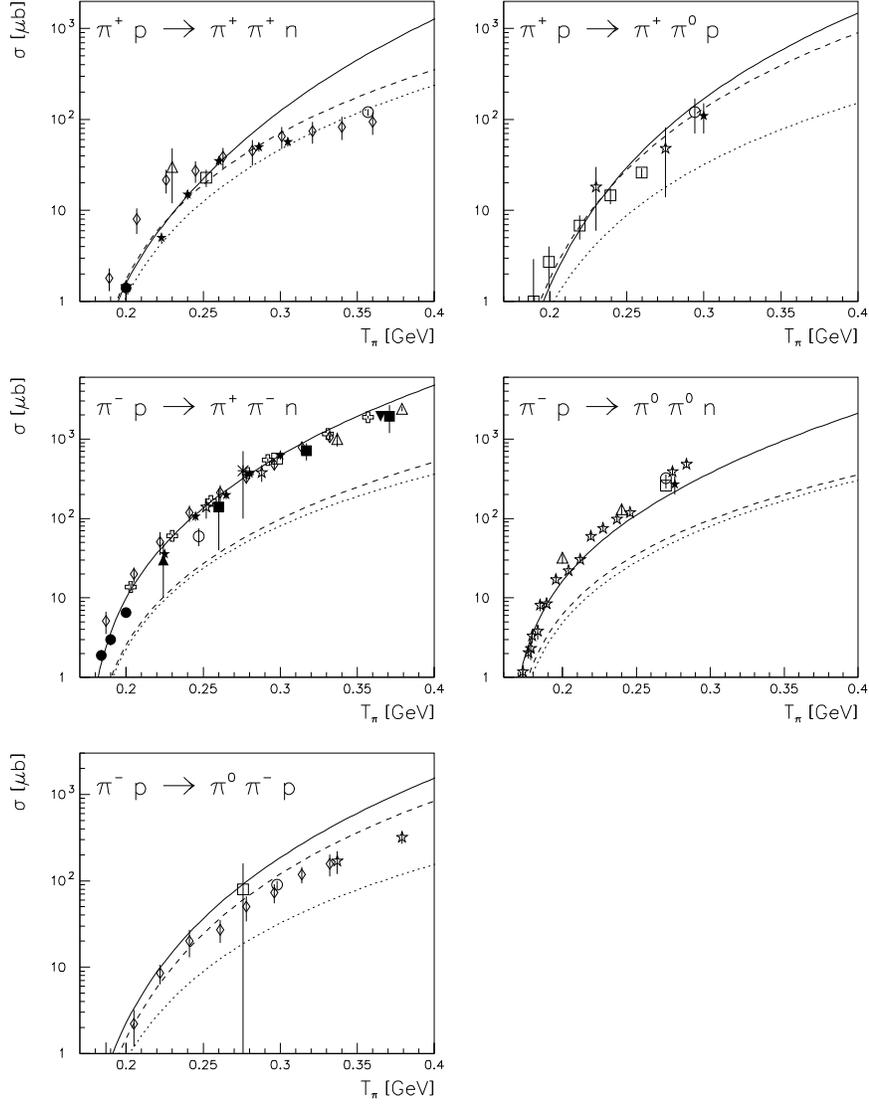}
}
\caption{Predictions for the total cross sections up to incoming pion
kinetic energy of $T_\pi = 400\,$MeV (solid lines). Fitted in each channel
are the data below  $T_\pi = 250\,$MeV. The dashed and dotted lines refer
to the second and first order contributions, respectively.
\label{fig:pipin}}
\end{figure}

Single pion production off nucleons has been at the center of numerous
experimental and theoretical investigations since many years. One of
the original motivations of these works was the observation that
the elusive pion--pion threshold S--wave interaction could be deduced
from the pion--pole graph contribution. A whole series of precision
experiments at PSI, TRIUMF, CERN and other laboratories 
has been performed over the last
decade and there is still on--going activity. On the theoretical side,
chiral perturbation theory has been used to consider these processes.
Beringer considered the reaction $\pi N \to \pi
\pi N$ to lowest order in chiral perturbation theory.\cite{bering} 
Low--energy theorems for the threshold amplitudes $D_1$
and $D_2$ were derived in,\cite{bkmplb}
\beq D_1 = \frac{g_A}{8 F_\pi^3} \left( 1 +\frac{7 M_\pi}{2m} \right)
+ {\cal O}(M_\pi^2)  = 2.4 \, \, {\rm fm}^3 \,\, , \label{lppn1}\eeq 
\beq D_2 = -\frac{g_A}{8 F_\pi^3} \left( 3 +\frac{17 M_\pi}{2m} \right)
+ {\cal O}(M_\pi^2)  = -6.8 \, \, {\rm fm}^3 \,\,  . \label{lppn2} \eeq
These are free of unknown parameters and not sensitive to the $\pi
\pi$--interaction beyond tree level. 
A direct comparison with the threshold data for
the channel $\pi^+ p \to \pi^+ \pi^+ n$, which is only sensitive to $D_1$,
leads to a very satisfactory description whereas in case of the process
$\pi^- p \to \pi^0 \pi^0 n$, which is only sensitive to $D_2$, sizeable 
deviations are found for the total cross sections near threshold. These were 
originally attributed to the strong pionic final--state interactions in the
$I_{\pi\pi}=0$ channel. However, this conjecture turned out to be incorrect 
when a complete higher order calculation of the threshold amplitudes $D_{1,2}$
was performed.\cite{bkmppn} In that paper, the relation
between the threshold amplitudes $D_1$ and $D_2$ for the reaction $\pi N \to
\pi \pi N$ and the $\pi\pi$ S--wave scattering lengths $a_0^0$ and $a_0^2$
in the framework of HBCHPT to second order in the pion mass was worked
out. Notice that the pion loop and pionic counterterm 
corrections  only start  contributing to the $\pi\pi N$ threshold amplitudes at
second order. One of these counterterms, proportional to the low--energy
constant $\ell_3$, eventually allows to
measure the scalar quark condensate, i.e. the strength of the spontaneous 
chiral symmetry breaking in QCD. However, at that order, the largest 
contributions to $D_{1,2}$ stem indeed from insertions of the dimension two
chiral pion--nucleon Lagrangian, which is characterized by the LECs
$c_i$.  In particular this is the case for the amplitude
$D_2$. To be  specific, consider the threshold  amplitudes $D_{1,2}$ calculated
from the relativistic Born graphs (with lowest order vertices) and the
relativistic $c_i$--terms expanded  to second order in the pion mass. This 
gives 
$D_1^{{\rm Born}} + D_1^{c_i} = (2.33 + 0.24 \pm 0.10) \, {\rm
    fm}^3 = (2.57 \pm 0.10) \, {\rm  fm}^3 \,$, 
$D_2^{{\rm Born}} + D_2^{c_i} = (-6.61 - 2.85 \pm 0.06) \, {\rm
    fm}^3 = (-9.46 \pm 0.06) \, {\rm  fm}^3 \,$,
which are within 14\% and 5\% off the empirical values, 
$D_1^{\exp} =(2.26 \pm 0.10)\,{\rm fm}^3$ and $D_2^{\exp} 
= (-9.05 \pm 0.36)\, {\rm  fm}^3 \,$,
respectively.  It appears therefore natural to extend the same
calculation above threshold and to compare to the large body of data
for the various reaction channels that exist.\cite{bkmppn2} 
It was already shown by
Beringer \cite{bering} that taking into account only the relativistic Born terms
does indeed not suffice to describe the total cross section data for
incoming pion energies up to 400~MeV in most channels. Such a failure 
can be expected from the threshold expansion of $D_2$, where the Born terms 
only amount to 73\% of the empirical value. We therefore expect that the
inclusion of the dimension two operators, which clearly improves the
prediction for $D_2$ at threshold, will lead to a better description of
the above threshold data. In particular, it will tell to which extent
loop effects are necessary (and thus testing the sensitivity to the
pion--pion interaction beyond tree level) and to which extent one has to
incorporate explicit resonance degrees of freedom like the Roper and the
$\Delta$--isobar as well as heavier mesons ($\sigma, \rho,\omega $) 
as dynamical degrees of freedom (as it is done in
many models, see e.g.~\cite{oset}~\cite{jaeck}). 
Since the LECs $c_i$ have previously been determined, 
all  results to this order\cite{bkmppn2}  are based on a 
truly parameter--free calculation.
One finds that (a) for pion energies up to $T_\pi = 250\,$MeV, in all but
one case the inclusion of the contribution $\sim c_i$ clearly improves
the description of the total cross sections. Up to $T_\pi =
400\,$MeV, the trend of the data can be described although some
discrepancies particularly towards the higher energies persist,
and  (b) double differential cross sections for 
$\pi^- p \to \pi^+ \pi^- n$ at incident pion energies
below $T_\pi = 250\,$MeV are well described. 
These findings were further strengthened and extended by the results of 
a complete third order HBCHPT calculation presented in.\cite{BFM}
It was shown that the effect of unitarity corrections (pion loops)
is indeed very small, justifying the use of tree level models,
and that the available threshold data are not precise enough to
pin down the LEC $\ell_3$. The predictions for the total cross sections
of all five  physical channels are shown in Fig.\ref{fig:pipin}
in comparison to the available data for incoming pion momenta up to 400~MeV. Despite the
large momentum transfers involved, the chiral description works 
fairly well, in particular in the $\pi^- p \to \pi^+ \pi^- n$ channel.
On the other hand, for the two reactions $\pi^+ p \to \pi^+ \pi^+ n$
and $\pi^- p \to \pi^0 \pi^- p$ the chiral prediction overshoots the
data for energies larger than 300~MeV. For further results and discussion,
see.\cite{BFM}
\noindent

\section{Three flavors: Hyperons, cascades and the nucleon revisited}
\label{sec:SU3}

\subsection{General remarks}
\label{sec:strange}
Chiral baryon dynamics with strange quarks is a much less developed
field than the two--flavor (pion--nucleon) case discussed so far, 
for two reasons. First, the strange quark mass is sizeably larger
so that it is not a priori clear that it can be treated perturbatively.
Second, the appearance of close to or even subthreshold resonances
(the most prominent example being the $\Lambda (1405)$)
in some channels requires some non--perturbative resummation. Ultimately,
one is after a marriage of such techniques with chiral symmetry constraints.
Although some pionering work has been done, see e.g. the 
papers and reviews,\cite{Munrev,Valrev} I will focus here on the results obtained
in``plain'' chiral perturbation theory.

\subsection{Two--point functions I: From baryon to quark masses}
\label{sec:masses}

The scalar sector of baryon CHPT is particularly interesting since
it is sensitive to scalar--isoscalar operators and thus directly to the
symmetry breaking of QCD. This is most obvious for the pion-- and 
kaon--nucleon $\sigma$--terms, which measure the strength of the scalar
quark condensates $\bar q q$ in the proton. Here, $q$ is a generic symbol for 
any one of the light quarks $u$, $d$ and $s$. Furthermore, the quark mass
expansion of the baryon masses allows to give bounds on the ratios of the
light quark masses. 
The quark mass expansion of the octet baryon masses takes the form
\begin{equation}
m = \, \, \krig{m} + \sum_q \, B_q \, m_q + \sum_q \, C_q \, m_q^{3/2} + 
\sum_q \, D_q \, m_q^2  + \ldots
\label{massform}
\end{equation}
modulo logs. Here, $\krig{m}$ is the octet mass in the chiral limit of
vanishing quark masses and the coefficients $B_q, C_q, D_q$ are 
state--dependent. Furthermore, they include contributions proportional
to some LECs which appear beyond leading order in the effective Lagrangian.
If one retains only the terms linear in the quark masses, one obtains
the Gell-Mann--Okubo relation $m_\Sigma + 3m_\Lambda = 2(m_N +m_\Xi)$
(which is fulfilled within 0.6 percent in nature) for the octet and
the equal spacing rule for the decuplet, $m_\Omega - m_{\Xi^*}=
m_{\Xi^*} - m_{\Sigma^*} = m_{\Sigma^*} - m_\Delta$ (experimentally,
one has 142:145:153~MeV). At  third order, the leading
non-analytic terms $\sim m_q^{3/2}$ are generated from finite loop graphs.
In contrast, the order $p^4$ loops are  no longer finite, thus
requiring renormalization.
A complete fourth order calculation (in the isospin limit $m_u = m_d$)
was presented in.\cite{bo:mei} There are 10 LECs related to symmetry
breaking, which can be estimated from a generalized resonance saturation
hypothesis, i.e. from loop graphs with intermediate baryon excitations.
The analytic pieces of such diagrams can be mapped onto higher order
symmetry breaking terms, for details see Ref.171.  
Within this scheme, one finds for the octet baryon mass in the chiral limit
$\krig{m} = (770\pm 110)\, {\rm MeV}$. The quark mass expansion of the baryon 
masses,  in the notation of Eq.(\ref{massform}), reads
\begin{eqnarray}
&& m_N  = \,  \krig{m}  \, ( 1 + 0.34 - 0.35 + 0.24 \, ) \, \, ,
\nonumber \\
&& m_\Lambda  = \,  \krig{m} \, ( 1 + 0.69 - 0.77 + 0.54 \, ) \, \, ,
\nonumber \\
&& m_\Sigma  = \,  \krig{m} \, ( 1 + 0.81 - 0.70 + 0.44 \, ) \, \, ,
\nonumber \\
&& m_\Xi  = \,  \krig{m} \, ( 1 + 1.10 - 1.16 + 0.78 \, ) \, \, .   
\label{mexpand}
\end{eqnarray}
One observes that there are large cancellations between the second order
and the leading non--analytic terms of order $p^3$, a well--known effect.
The fourth order contribution to the nucleon mass is fairly small, whereas it
is sizeable for the $\Lambda$, the $\Sigma$ and the $\Xi$. This is
partly due to the small value of $\krig{m}$, e.g. for the $\Xi$ the
leading term in the quark mass expansion gives only about 55\% of the
physical mass and the second and third order terms cancel almost completely.
From the chiral expansions exhibited in Eq.(\ref{mexpand}) one can not    
yet draw a final conclusion about the rate of convergence in the
three--flavor sector of baryon CHPT. Certainly, the breakdown of CHPT
claimed by~\cite{juerg} is not observed. On the other hand, the
conjecture by~\cite{jm2} that only the leading non--analytic corrections (LNAC)
$\sim m_q^{3/2}$ are large and that further terms like the ones $\sim m_q^2$
are moderately small (of the order of 100 MeV) is not supported. 
Based on a third order calculation and estimating the fourth order, it
was shown in~\cite{EllTo} that using relavistic baryon fields and
employing  infrared regularization, the
magnitude of the loop integrals is reduced so that an improved
convergence of the chiral series emerges. These calculations are,
however, not yet precise enough to draw conclusions about the quark mass ratios
from the baryon masses. It was suggested a long time ago by
Gasser\cite{juerg} to use a static source plus meson cloud model to
tame the large chiral corrections. Such an approach is less systematic
than the strict chiral expansion and has a certain cut--off
dependence. Including isospin breaking terms $\sim (m_d-m_u)$ and
electromagnetic corrections to the baryon masses, this approach was
used to deduce the quark mass ratio
\beq
R = {m_s - \hat{m} \over m_d -m_u} = 43.5 \pm 3.2
\eeq
from the mass splittings $m_p -m_n$, $m_{\Sigma^+} - m_{\Sigma^-}$
and $m_{\Xi^0}- m_{\Xi^-}$. This result is further supported by the
so--called ``flavor perturbation theory'' of Ref.175, 
$R \ge 38 \pm 11$.

\subsection{Three--point functions I: Magnetic moments}

\begin{figure}[htb]
\hskip 1.4in
\epsfxsize=1.9in
\epsffile{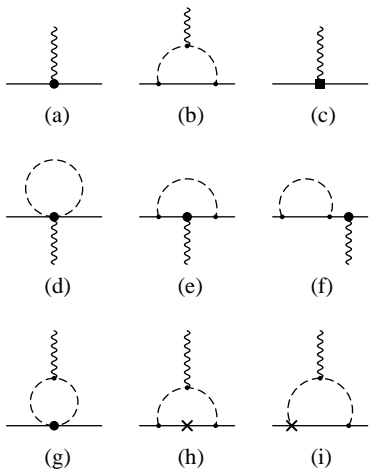}
\caption{Chiral expansion of the magnetic moments
to order $p^2$ (a), $p^3$ (b) and $p^4$ (c-i).
The cross denotes either a $1/m$ insertion with a fixed LEC or
one related to scalar symmetry breaking. The dot is
the leading dimension two insertion and the box depicts the
counterterms at fourth order.\label{fig:mm}} 
\end{figure}

The magnetic moments of the octet baryons  have been measured with
very high precision. On the theoretical side, SU(3)
flavor symmetry was first used by Coleman and Glashow\cite{CG} to  
predict seven relations between the eight moments of the $p$, $n$, $\Lambda$,
$\Sigma^\pm$, $\Sigma^0$, $\Xi^-$, $\Xi^0$ and the $\mu_{\Lambda \Sigma^0}$ 
transition moment in terms of two parameters.  One of these relations is
in fact a consequence of isospin symmetry alone. In modern language, 
this was a tree level calculation employing the lowest order effective
Lagrangian of dimension two, compare Fig.\ref{fig:mm}a.
Given the simplicity of this approach, these relations work remarkably well,
truely a benchmark success of SU(3).
The first loop corrections arise at order $p^3$ in 
the chiral counting, see Ref.177 
(cf. Fig.\ref{fig:mm}b). 
They are given entirely in 
terms of the lowest order parameters from the dimension one (two)
meson--baryon (meson) Lagrangian. It was
found that these loop corrections are large for standard values of the
two axial couplings $F$ and $D$. Caldi and Pagels~\cite{CP} derived
three relations independent of these coupling constants.
These are, in fact, in good agreement with the data. However, the deviations
from the Coleman--Glashow relations worsen considerably. This fact has
some times been taken as an indication for the breakdown of SU(3) CHPT.
To draw any such conclusion, a calculation to fourth order  is mandatory.
This was attempted in Ref.178,   
however, not all terms were accounted for.
To be precise, in that calculation the contribution from the
graphs corresponding to Fig.\ref{fig:mm}c-f were worked out. As pointed out 
in Ref.179, 
there are additional one--loop graphs at ${\cal O}{(p^4)}$, namely tadpole
graphs with double--derivative meson--baryon vertices (Fig.\ref{fig:mm}g) 
and diagrams with fixed $1/m$ and symmetry breaking insertions $\sim b_{D,F}$
from the dimension two Lagrangian, see
Fig.\ref{fig:mm}h,i. Some (but not all) of these contributions were implicitely
contained in the work of Ref.178   
as becomes obvious when one expands the graphs with intermediate decuplet states. 
However, apart form these decuplet contributions to the LECs, there are 
also important t--channel vector meson
exchanges which are not accounted for if one includes the spin--3/2
decuplet in the effective theory and calculates the corresponding
tree and loop graphs. In total, there
are seven LECs related to symmetry breaking and three related to scattering
processes (the once appearing in the class of graphs Fig.\ref{fig:mm}g). 
These latter LECs
can be estimated with some accuracy from resonance exchange. The strategy
of Ref.179  
was to leave the others as free parameters and fit the magnetic
moments. One is thus able to investigate the chiral expansion of the
magnetic moments and to predict the ${\Lambda \Sigma^0}$ transition moment.
The chiral expansion of the various magnetic moments thus takes the
form
\beq 
\label{mubform}
\mu_B = \mu_B^{(2)} + \mu_B^{(3)} + \mu_B^{(4)}
= \mu_B^{(2)} \,( \, 1 + \varepsilon^{(3)} + \varepsilon^{(4)} \, ) \quad ,
\eeq
with the result (all numbers in nuclear magnetons)
\begin{eqnarray} \label{conv}
\begin{array}{llrllr}
\mu_p                 & = & 4.15  & (1 - 0.53 + 0.20) & = & 2.79 \, , \\
\mu_n                 & = & -2.53 & (1 - 0.34 + 0.09) & = & -1.91 \, , \\
\mu_{\Sigma^+}        & = & 4.15  & (1 - 0.67 + 0.26) & = & 2.46 \, , \\
\mu_{\Sigma^-}        & = & -1.63 & (1 - 0.38 - 0.09) & = & -1.16 \, , \\
\mu_{\Sigma^0}        & = & 1.26  & (1 - 0.85 + 0.37) & = & 0.65 \, , \\
\mu_\Lambda           & = & -1.26 & (1 - 0.85 + 0.34) & = & -0.61 \, , \\
\mu_{\Xi^0}           & = & -2.53 & (1 - 0.87 + 0.37) & = & -1.25 \, , \\
\mu_{\Xi^-}           & = & -1.63 & (1 - 0.80 + 0.20) & = & -0.65 \, , \\
\mu_{\Lambda\Sigma^0} & = & 2.19  & (1 - 0.52 + 0.15) & = & 1.37 \, , \\
\end{array} 
\end{eqnarray}
setting here the scale of dimensional regularization $\lambda = 1\,$GeV.
In all cases the ${\cal O}(p^4)$ contribution is smaller than the one 
from order $p^3$ by at least a factor of two, in most cases even by 
a factor of three. 
Like in the case of the baryon masses, one finds sizeable
cancellations between the leading and next--to--leading order terms
making a {\it precise} calculation of the ${\cal O}(p^4)$ terms absolutely
necessary. In particular, the previously omitted terms with a
symmetry-breaking insertion  turn out to be very important. One can
predict the transition moment to be $\mu_{\Lambda \Sigma^0} 
= (1.40 \pm 0.01) \mu_N$ in fair agreement with the lattice gauge theory 
result of,\cite{lwd} $\mu_{\Lambda \Sigma^0} = (1.54 \pm 0.09) \mu_N$. 
Furthermore, in~\cite{jlms}  one relation amongst the magnetic moments
independent of the axial couplings was derived. This relation does not hold any
more in the complete ${\cal O}(p^4)$ calculation. To be specific, the
graphs of Fig.\ref{fig:mm}h with a scalar symmetry breaking insertion do not respect
this relation. That the convergence is improved in a
lorentz--invariant formulation can be seen as follows: In such an
approach, the diagrams h) and i) of Fig.\ref{fig:mm} would appear at
second order leading to 
\begin{eqnarray} \label{convBL}
\begin{array}{llrllr}
\mu_p                 & = & 4.15  & (1 - 0.45 + 0.12) & = & 2.79 \, , \\
\mu_n                 & = & -2.53 & (1 - 0.27 + 0.03) & = & -1.91 \, , \\
\end{array} 
\end{eqnarray}
and similarly for the other members of the octet. In all cases, the
fourth order contribution are at most one quarter of the leading one.
This is expected by dimensional arguments since $(M_K/m_B)^2 \simeq 1/4$.
Certainly, a complete one--loop calculation using IR regularization
should be done.

\subsection{Three--point functions II: Electromagnetic form factors}

\begin{figure}[hbt]
\centerline{
\psfig{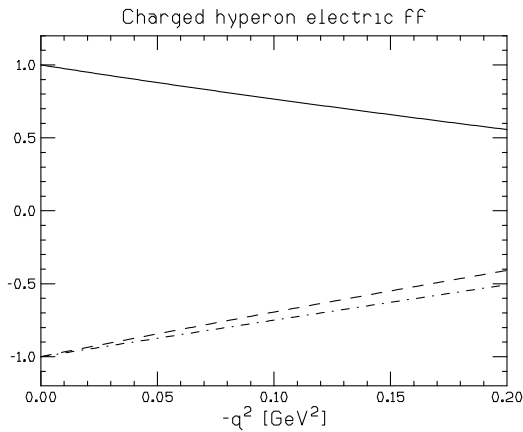}
}
\caption{
The electric form factors of the charged hyperons calculated in
three flavor baryon CHPT.
Solid, dashed, dot--dashed line: $\Sigma^+$, $\Sigma^-$, $\Xi^-$, in order.
\label{fig:FFhyp}}
\end{figure}

The electromagnetic form factors (ffs) of the nucleon were discussed
to one loop accuracy in section~\ref{sec:ff}. Let us recall the
salient features of the leading one loop, i.e. third order, analysis
in HBCHPT. At that order, one has to deal with two
counterterms in the electric  and two in the magnetic ffs. 
Using e.g. the proton and neutron electric radii and magnetic moments as input, the
ffs are fully determined to that order. In particular, no counterterms appear
in the momentum expansion of the magnetic ffs. To this order in the chiral
expansion, the ffs are fairly described for momentum transfer squared
up to $Q^2 \simeq 0.2\,$GeV$^2$. It appears therefore natural
to extend such an investigation to the three flavor case as done in Ref.181.    
This investigation was triggered by the recent results on the
$\Sigma^-$ radius
reported by the WA89 collaboration at CERN~\cite{WA89},
$\langle r^2_{\Sigma^-} \rangle_{\rm exp} = 0.92\pm 0.32\pm 0.40~{\rm
  fm}^2$, and by the SELEX collaboration at
FNAL~\cite{SELEX}, $\langle r^2_{\Sigma^-} \rangle_{\rm exp} 
= 0.60\pm 0.08\pm 0.08~{\rm fm}^2$
 (note that the SELEX results are still preliminary),
obtained by scattering a highly boosted hyperon beam off the electronic
cloud of a heavy
atom (elastic hadron--electron scattering). The pattern of the charge radii
embodies information on SU(3) breaking and the structure of the groundstate octet.
In a CHPT calculation of the corresponding ffs, the baryon structure is to some 
part given by the meson (pion and kaon) cloud and in part by short distance physics 
parameterized in terms of local contact interactions. In the general case, 
such a splitting depends on the regulator scheme and scale one chooses. Here, we work in 
standard dimensional
regularization and set $\lambda =1\,$GeV throughout (since this is the natural
hadronic scale). If one performs the SU(3) calculation to third order, one
has no new counterterms as compared to the SU(2) case. Therefore, fixing
the LECs from proton and neutron properties allows one
to make parameter--free predictions for the hyperons. As an added bonus, kaon
loops induce a momentum dependence in the isoscalar magnetic form factor of the
nucleon, as first pointed out in Ref.184,  
whereas in the pure SU(2) calculation, $G_M^{I=0} (Q^2)$ is simply constant.  
Consider now the hyperons. The electric ffs of the charged hyperons
are given in Fig.\ref{fig:FFhyp}. The corresponding radii are (a more detailed discussion
also of the neutral particles and magnetic radii is given in~\cite{khm})
\begin{equation}
\langle r^2_{\Sigma^+}\rangle = 0.64\ldots0.66~{\rm fm}^2~,
\langle r^2_{\Sigma^-}\rangle = 0.77\ldots0.80~{\rm fm}^2~,
\langle r^2_{\Xi^-}\rangle = 0.61\ldots0.65~{\rm fm}^2~.
\end{equation}
The given uncertainty does not reflect the contribution from higher orders,
which should be calculated. 
\noindent
The prediction for the $\Sigma^-$ is in fair
agreement with the recent measurements.
The result for the $\Sigma$ radii is at variance
with quenched lattice QCD calculations, which give 0.56(5)~fm$^2$ and
0.72(6)~fm$^2$ for the negative 
and positive $\Sigma$, respectively.\cite{lwd} However, quenched lattice calculations
should be taken with a grain of salt (the true error due to the quenching is
only known for very few quantities, certainly not for the radii). In the CHPT approach, 
the difference of the radii is due to some short distance physics
encoded in the LEC $d_{102}^0$~\cite{mue:mei} and to the Foldy term. 
It is also important to note that some of these hyperon form factors can
eventually be extracted from kaon electroproduction data obtained at JLab.
Clearly, a fourth order calculation is called for to further quantify these results.
In addition, a calculation using IR regularization and including
explicit vector mesons as done for the nucleon form factors is
necessary to gain further insight.

\subsection{Four--point functions I: Kaon photo-- and electroproduction}

\begin{figure}[hbt]
\centerline{
\psfig{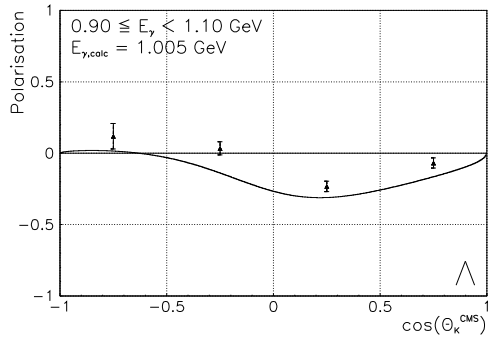}
}
\caption{
Recoil polarisation in $\gamma p \to K^+ \Lambda$. The third order HBCHPT
prediction is compared to the data from ELSA.
\label{fig:kaon}}
\end{figure}

At ELSA (Bonn) and JLab ample kaon photoproduction data
have been taken over a wide energy range.\cite{bock,Elsa1,Elsa2,JLAB}
It therefore is interesting  to study the reactions $\gamma p \to
\Sigma^+ K^0$, $\Lambda K^0$ and $\Sigma^0 K^+$ in the framework of CHPT. 
This has been done in an exploratory study published in~\cite{ste:mei}
to third order.  The threshold energies for these three processes are
$1046$, $1048$ and $911\,$MeV, in order, which is already rather high.
Concerning  the loop diagrams, two remarks are in order. First, the SU(3) 
calculation allows one to investigate the effect of kaon loops on the SU(2) predictions.
As expected, it is found that these effects are small, in agreement with
the decoupling theorem. In the chiral SU(2) limit, that is for a fixed
strange quark mass, kaon loop effects must decouple, which means that they
are suppressed by inverse powers or logs of the heavy mass, here $ M_K$. 
Second, the loop graphs give rise to the imaginary part of the transistion
amplitude. In particular, one has intermediate $\pi N$ states at high
energy which leads to too large imaginary parts. Therefore, a safer
way of calculating these imaginary parts is to use the
Fermi--Watson final state theorem and taking the pertinent $\pi N$ photoproduction
multipoles from the existing multipole analysis. 
In addition, there are 13 various operators with unknown LECs, which 
are known or can be determined from differential cross section data.
Most of the total cross sections for energies up to 100~MeV above
threshold are well described, which is rather surprising.
The most interesting result is the prediction for the
recoil polarization $P$ at $E_\gamma = 1$~GeV in the $K^+ \Lambda$ channel. 
Amazingly, the shape and magnitude of the data is well described for forward angles,
but comes out on the small side for backward angles, see Fig.\ref{fig:kaon}.  
Most isobar models, that give a descent description of the total and differential 
cross sections also at higher energies, fail to explain this angular
dependence of the recoil polarization. A similar observation is made
for the recoil polarization in the $K^+ \Sigma^0$ channel.

\noindent Clearly, these results 
should only be considered indicative since one should include (a) higher 
order effects, (b) higher partial waves and (c) has to  get a better handle
on the ranges of the various coupling constants. In addition, one would
also need more data closer to threshold, i.e. in a region where the
method is applicable. However, the results presented are encouraging 
enough  to pursue a more detailed study of these reactions (for real 
and virtual photons) in the framework of "strict" chiral perturbation theory.

\section{Summary}
In this essay, I  have tried to show that the role of the
Goldstone boson cloud in describing the baryon structure precisely and
in a controlled fashion has become evident over the last years.
I have also provided some tools to better understand and use the 
underlying machinery.
Despite all the interesting progress and results presented in the
previous sections,  chiral baryon dynamics is still faced with many challenges.
In particular, in the strange quark sector, we still do not know whether
a perturbative treatment in the strange quark mass is justified or has
to be supplemented by some resummation technique in harmony with chiral
symmetry. Also, the extension of hadron effective field theory to higher energies
is a topic of numerous present investigations, in particular the matching to the
perturbative QCD regime is of current interest. Ultimately, these investigations
should help us to understand the mechanism behind the chiral symmetry breaking of QCD 
and also give additional bounds on the ratios of the light quark masses, thus
leading to some insight about some of the fundamental parameters of QCD.
There is also considerable experimental effort which is mandatory to further sharpen 
our understanding of chiral symmetry breaking and its consequences for the
structure, dynamics and interactions of the low--lying baryons.

\section*{Acknowledgements}
I would like to thank Misha Shifman for giving me this opportunity.
I am also grateful to my collaborators, friends and colleagues for sharing
their insight into chiral dynamics, in particular 
Silas Beane, V\'eronique Bernard,
Bugra Borasoy, Paul B\"uttiker, Gerhard Ecker, Evgeni Epelbaum, 
Nadia Fettes, J\"urg Gasser, George Gellas, Barry Holstein,
Norbert Kaiser, Marc Knecht, Hermann Krebs, Bastian Kubis, 
Heiri Leutwyler, Thomas Hemmert,  Guido M\"uller, Jos\'e Oller, 
Sven Steininger, Jan Stern and Markus Walzl.

\section*{Appendix A: Generating functional and effective Lagrangian}

In this appendix, I briefly discuss the path integral approach to
the effective meson--baryon Lagrangian (for the three flavor case). The
generating functional for Green functions of quark currents 
between single baryon states, $Z[j,\eta, \bar \eta$], is defined via
\beq
\exp  \bigl\{ i  Z[j,\eta, \bar \eta] \bigr\} 
= {\cal N} \int [dU] [dB] [d{\bar B}]  \exp 
 i \biggl[ S_M + S_{MB} + \int d^4x \langle\bar \eta B\rangle 
+\langle\bar B \eta\rangle
\biggr] \,  ,
\label{defgenfun}
\eeq
with $S_M$ and $S_{MB}$ denoting the meson and the meson--baryon
effective action, respectively, to be discussed below. $\eta$ and
$\bar \eta$ are fermionic sources coupled to the baryons and $j$
collectively denotes the external fields of vector ($v_\mu$), 
axial--vector ($a_\mu$), scalar ($s$) and pseudoscalar ($p$) type. 
These are coupled in the standard chiral
invariant manner. The Goldstone bosons are collecetd in matrix $U$ 
and the baryons in the matrix $B$ as discussed in section~2. The action follows
from the Lagrangian via $S = \int d^4x {\cal L}$ and thus admits the
same type of low--energy expansion. Consider now the heavy baryon
approach. For that, define velocity--dependent spin--1/2 fields 
by a particular choice of Lorentz frame and decompose the fields into their velocity
eigenstates (called 'light' and 'heavy' components),
$H_v (x) = \exp \{ i m_B v \cdot x \} \, P_v^+ \, B(x)$,
$h_v (x)  =  \exp \{ i m_B v \cdot x \} \, P_v^- \, B(x)$, in terms of
velocity projection operators $P_v^\pm =  (1 \pm v \!\!\!/)/2\,$.
In this basis, the three flavor meson--baryon action takes the form 
\beq
S_{MB} = \int d^4x \, \biggl\{ \bar{H}_v^a \, A^{ab} H^b_v 
- \bar{h}_v^a \, C^{ab} h^b_v + \bar{h}_v^a \, B^{ab} H^b_v
+ \bar{H}_v^a \, \gamma_0 \,  {B^{ab}}^\dagger \, \gamma_0 \, h^b_v 
\biggr\}\,\,\, ,
\eeq
with $a,b= 1, \ldots ,8$ flavor indices. The 8$\times$8 matrices $A$,
$B$ and $C$ admit low energy expansions,
\beq  
X^{ab} = X^{ab,(1)} + X^{ab,(2)} + X^{ab,(3)}+ \ldots,
\quad X\in \{A,B,C\}~. 
\eeq
Similarly, one splits the
baryon source fields $\eta (x)$ into velocity eigenstates,
\beq
R_v (x) = \exp \{ i m_B v \cdot x \} \, P_v^+ \, \eta(x) \,\, , \quad
\rho_v (x) = \exp \{ i m_B v \cdot x \} \, P_v^- \, \eta(x) \,\, , 
\label{sourceheavy}
\eeq
and shift variables
\beq
h_v^{a'} = h_v^a - (C^{ac})^{-1} \, ( B^{cd} \, H_v^d + \rho_v^c \, )
\,\, \, ,
\eeq
so that the generating functional takes the form
\beq
\exp[iZ] = {\cal N} \, \Delta_h \, \int [dU][dH_v][d\bar{H}_v] \, \exp \bigl\{
iS_M + i S_{MB}' \, \bigr\}
\label{Zinter}
\eeq
in terms of the meson--baryon action $S_{MB}'$,
\beqa
S_{MB}' &=& \int d^4x \, \bar{H}_v^a \bigl( A^{ab} + \gamma_0
[B^{ac}]^\dagger \gamma_0 [C^{cd}]^{-1} B^{db} \, \bigr) H_v^b
\\
&+& \bar{H}_v^a \bigl( R_v^a + \gamma_0 [B^{ac}]^\dagger \gamma_0 
[C^{cd}]^{-1} \rho_v^d \bigr) + \bigr( \bar{R}_v^a + \bar{\rho}_v^c
[C^{cb}]^{-1} B^{ba} \bigr) H_v^a \, \, .\nonumber 
\eeqa        
The determinant $\Delta_h$ related to the 'heavy' components is
identical to one. The generating
functional is thus entirely expressed in terms of the Goldstone bosons
and the 'light' components of the spin--1/2 fields. The action is,
however, highly non--local due to the appearance of the inverse of 
the matrix $C$. To render it local, one now expands $C^{-1}$ in powers
of $1/m_B$, i.e. in terms of increasing chiral dimension,
\beqa
[C^{ab}]^{-1} &=& 
\frac{\delta^{ab}}{2m_B} - \frac{1}{(2m_B)^2}\biggl\{
\langle {\lambda^a}^\dagger [i v \cdot \nabla  , \lambda^b ]\rangle
+ D \langle  {\lambda^a}^\dagger  \{S \cdot u, \lambda^b\} \rangle
\nonumber \\
&& \qquad\qquad
 + F \langle  {\lambda^a}^\dagger  [ S \cdot u, \lambda^b ] \rangle
\biggr\} + {\cal O}(p^2) \,\,\, ,
\label{Cinvexp}
\eeqa
with $S_\mu$ the Pauli--Lubanski spin vector and $u_\mu \sim i
\partial_\mu \phi/F_0 + \ldots \,$.
To any finite power in $1/m_B$, one can now perform the integration of
the 'light' baryon field components $N_v$ by again completing the
square,
\beqa
H_v^{a'} &=& [T^{ac}]^{-1} \, \bigl( R_v^c + \gamma_0 \, [B^{cd}]^\dagger
\, \gamma_0 \, [C^{db}]^{-1} \, \rho_v^b \, \bigr) \nonumber \\
T^{ab} &=& A^{ab} + \gamma_0 \, [B^{ac}]^\dagger \, \gamma_0 \,
[C^{cd}]^{-1} \, B^{db} \,\,\, .
\eeqa
Notice that the second term in the expression for $T^{ab}$ only starts
to contribute at chiral dimension two. In this manner,
one can construct the effective meson--baryon Lagrangian with the
added virtue that the $1/m_B$ corrections related to the Lorentz
invariance of the underlying relativistic theory are correctly given.
Another method to do that is discussed in the next appendix.

\section*{Appendix B: On reparametrization invariance}
\label{app:repara}
It was first shown by Luke and Manohar~\cite{LM} that a simple
reparametrization invariance relates the coupling constants of terms
of different orders in the $1/m$ expansion.  This invariance also
restricts the forms of operators which can appear in the chiral
Lagrangian for heavy particles. This method allows one to stay
entirely within the heavy particle EFT but is more tedious to apply
than using the path integral based on the relativistic version of the
theory to construct the terms with fixed coefficients. Still, the idea
is very neat and deserves discussion. The heavy particles in the
effective theory are described by velocity--dependent fields with
velocity $v$, residual momentum $k$, and total momentum $p =mv+k$.
Now, there is an ambiguity in assigning a velocity and a momentum to a
particle when one considers $1/m$ corrections. To see that, note that
the same physical momentum may be parametrized by
\beq\label{repa}
(v,k) \leftrightarrow \left( v +\frac{q}{m}, k-q\right)~, 
\quad v^2 = \left( v + \frac{q}{m}\right)^2 = 1~,
\eeq
where $q$ is an arbitrary four--vector which  satisfies $(v+q/m)^2=1$.
This is simply a consequence of the energy--momentum relation for a
particle with mass $m$. Consider as an example a colored scalar field
coupled to gluons,
\beq\label{Ltoy}
{\cal L}_{\rm toy} = D_\mu \phi^* D^\mu \phi - m^2 \phi^* \phi~.
\eeq 
The low--energy effective Lagrangian can be formulated in terms of
a velocity--dependent effective field
\beq
\phi_v (x) = \sqrt{2m} \, {\rm e}^{im v\cdot x} \phi(x)~,
\eeq
with $v$ a velocity four--vector of unit length. Note that I
explicitely keep the subscript ``$v$''. The field $\phi_v (x)$
creates and annihilates scalars with definite velocity $v$, which is
a good quantum number in the limit that the mass becomes infinite.
The effective Lagrangian which describes the low--energy dynamics of
the full theory, Eq.(\ref{Ltoy}), is
\beq
{\cal L}_{\rm eff} = \sum_v \phi_v^* (i v\cdot D)\phi_v + {\cal
  O}(1/m)~.
\eeq
The reparametrization transformation corresponding to Eq.(\ref{repa})
for the velocity--dependent fields is
\beq\label{refe}
\phi_w (x) = {\rm e}^{i q\cdot x} \phi_v (x)~, \quad w = v +
\frac{q}{m}~,
\eeq
under which the effective Lagrangian must remain invariant. Let us
work out explicitely the consequences of the reparametrization
invariance to order $1/m$. The most general effective Lagrangian for 
a scalar field theory coupled to gluons up to terms of order $1/m$ is
\beq\label{Ltoyeff}
{\cal L}_{\rm eff} = \sum_v \phi_v^* (i v\cdot D)\phi_v -
\frac{A}{2m} \phi_v^* D^2 \phi_v + {\cal O}(1/m^2)~,
\eeq
where $A$ is a constant and the lowest order equation of motion was
used to eliminate a term of the form $\phi_v^* (v\cdot D)^2 \phi_v$.
In terms of the reparametrized fields, Eq.(\ref{refe}), the effective
Lagrangian takes the form
\beq\label{Lw}
{\cal L}_{\rm eff} = \sum_v \phi_w^* [v\cdot (i D + q)]\phi_w -
\frac{A}{2m} \phi_w^* (D-iq)^2 \phi_w + {\cal O}(1/m^2)~.
\eeq
Relabelling now the dummy variable $w$ as $v$ and $v$ as $v-q/m$ in
Eq.(\ref{Lw}) and expanding to first order in the small momentum $q$
gives the change in ${\cal L}_{\rm eff}$,
\beq
\delta{\cal L}_{\rm eff} = (A-1) \, \sum_v \phi_v^* \frac{q\cdot D}{m}
\phi_v + {\cal O}(q^2, 1/m^2)~,
\eeq
using $v\cdot q = {\cal O}(q^2/m)$ from Eq.(\ref{repa}). Therefore,
the Lagrangian given in  Eq.(\ref{Ltoyeff}) is invariant under
velocity reparametrization up to
order $1/m$ only if $A=1$. Thus reparametrization invariance has fixed
the coefficient of one of the $1/m$ terms in the effective Lagrangian.
This is, of course, nothing but the expansion of the full kinetic
energy in powers of $1/m$ which we encountered already a couple of
times. The method is, however, more general and leads to much richer
constraints in less trivial applications.

\section*{Appendix C: Dimensional analysis of the electromagnetic LECs}
\label{app:LECdim}
Since very little empirical information exists to pin down the
em LECs $f_i$, $g_i$ and $h_i$ appearing at second, third and
fourth order (I follow the notation of Ref.62), 
one has to resort to dimensional analysis to get
some idea about their values. The corresponding operators contain an
even number of charge matrices. It is convenient to use the 
following normalization. Each power
of a charge matrix $Q$ appearing in any monomial should be accompanied by
a factor of $F_\pi$ so that the corresponding LECs have the same
mass dimension as their strong  counterparts. Therefore, the
$f_i$, $h_{6\ldots 90}$, $g_i$ and $h_{1\ldots 5}$ scale as 
[mass$^{-1}$], [mass$^{-1}$],  [mass$^{-2}$] and [mass$^{-3}$], in
order. Here, $g_i$ denotes the third order em  LECs and the ones
from fourth order, $h_i$, contain either four ($i=1,\ldots,5$) or
two ($i=6,\ldots,90$) powers of $Q$.   The physical origin
of the em LECs is the integration of hard photon loops. Therefore, each
power in $e^2$, as it is the case in QED, is really a power in the
fine structure constant $\alpha =e^2/4\pi$. Thus, since the natural
scale of chiral symmetry breaking is $\Lambda_\chi \sim m_N \sim M_\rho
\sim 1\,$GeV, we can deduce the following estimates for the renormalized 
em LECs  at the typical hadronic scale, which we chose as $M_\rho$ (naturalness
conditions)
\beq\label{diman}
f_i  = \frac{\tilde{f}_i}{4\pi}~, \quad
g_i^r (  M_\rho) = \frac{\tilde{g}_i}{4\pi}~, \quad
h_{1\ldots5}^r (  M_\rho) = \frac{\tilde{h}_{1\ldots5}}{(4\pi)^2}~, \quad
h_{6\ldots90}^r ( M_\rho) = \frac{\tilde{h}_{6\ldots90}}{4\pi}~,
\eeq
with the $\tilde{f}_i$, $\tilde{g}_i$ and $\tilde{h}_i$ are numbers
of order one,
\beq
\tilde{f}_i \sim \tilde{g}_i \sim \tilde{h}_i = {\cal O}(1)~.
\eeq
The $f_i$ are finite and scale--independent since loops only start at third order.
Of course, such type of analysis does not allow to fix the signs of the LECs. 
One should also remember that the numbers of
order one appearing in Eq.(\ref{diman}) can be sizeably smaller or
larger than one. From the determination of $f_2$ from the the neutron--proton
mass difference to third order, we conclude e.g. that $\tilde{f}_2 = -5.65$.
It would be interesting to develop some model
which would allow one to calculate or estimate these LECs.

\section*{References}

\end{document}